\documentclass[%
reprint,
superscriptaddress,
nofootinbib,
 amsmath,amssymb,
 aps,
 showkeys,
floatfix,
]{revtex4-2}

\usepackage{dcolumn}
\usepackage{bm}
\usepackage[utf8]{inputenc}
\usepackage{amsmath, amssymb, amsthm}
\usepackage{braket}
\usepackage[urlcolor=blue, colorlinks=true,citecolor=blue,linkcolor=black,pagebackref=true]{hyperref}
\usepackage{cleveref}
\usepackage{graphicx}
\usepackage[caption=false]{subfig}
\usepackage{float}
\usepackage{tabularx}
\usepackage{amsfonts}
\usepackage{xcolor}
\usepackage{dsfont}
\usepackage{tikz}
\usepackage{booktabs}
\usepackage{comment}

\usepackage[ruled]{algorithm2e}
\usepackage{algpseudocode}
\crefname{algocf}{alg.}{algs.}
\Crefname{algocf}{Algorithm}{Algorithms}

\usepackage{bookmark}

\usepackage{physics}
\usepackage{tikz}

\usepackage{tikz}
  \usetikzlibrary{positioning}
  \usetikzlibrary{quantikz}

\newcommand{\obs}{O}
\newcommand{\Var}{\text{Var}}
\newcommand{\E}{\mathbb{E}}

\renewcommand*\backref[1]{\ifx#1\relax \else (Cited on #1) \fi}

\begin{document}

\title{Robustness of quantum reinforcement learning under hardware errors}

\author{Andrea Skolik$^*$}
\affiliation{Leiden University, Niels Bohrweg 1, 2333 CA Leiden, The Netherlands}
\affiliation{Volkswagen AG, Ungererstra\ss e 69, 80805 Munich, Germany}
\email[]{Corresponding author a.skolik@liacs.leidenuniv.nl}
 
\author{Stefano Mangini}
\affiliation{Dipartimento di Fisica, Università di Pavia, Via Bassi 6, I-27100, Pavia, Italy}
\affiliation{INFN Sezione di Pavia, Via Bassi 6, I-27100, Pavia, Italy}

\author{Thomas Bäck}
\affiliation{Leiden University, Niels Bohrweg 1, 2333 CA Leiden, The Netherlands}

\author{Chiara Macchiavello}
\affiliation{Dipartimento di Fisica, Università di Pavia, Via Bassi 6, I-27100, Pavia, Italy}
\affiliation{INFN Sezione di Pavia, Via Bassi 6, I-27100, Pavia, Italy}
\affiliation{CNR-INO - Largo E. Fermi 6, I-50125, Firenze, Italy}
    
\author{Vedran Dunjko}
\affiliation{Leiden University, Niels Bohrweg 1, 2333 CA Leiden, The Netherlands}

\date{\today}

\begin{abstract}
Variational quantum machine learning algorithms have become the focus of recent research on how to utilize near-term quantum devices for machine learning tasks. They are considered suitable for this as the circuits that are run can be tailored to the device, and a big part of the computation is delegated to the classical optimizer. It has also been hypothesized that they may be more robust to hardware noise than conventional algorithms due to their hybrid nature. However, the effect of training quantum machine learning models under the influence of hardware-induced noise has not yet been extensively studied. In this work, we address this question for a specific type of learning, namely variational reinforcement learning, by studying its performance in the presence of various noise sources: shot noise, coherent and incoherent errors. We analytically and empirically investigate how the presence of noise during training and evaluation of variational quantum reinforcement learning algorithms affect the performance of the agents and robustness of the learned policies. Furthermore, we provide a method to reduce the number of measurements required to train Q-learning agents, using the inherent structure of the algorithm.
\end{abstract}

\keywords{variational quantum algorithms, quantum machine learning, quantum hardware noise}

\maketitle

\section{Introduction \label{sec:intro}}

Quantum machine learning (QML) is advertised as one of the most promising candidates for a near-term advantage in quantum computing~\cite{Bharti2021NIQSReview}. The variational quantum algorithms (VQAs) that are used for this are trained in a hybrid fashion, where a classical optimizer is used to tune the parameters of a quantum circuit~\cite{CerezoPQC2021Review, ManginiQNN}. It is hypothesized that the hybrid training scheme along with the freedom of adjusting the parameters appropriately, makes these algorithms inherently robust to quantum hardware noise to some extent~\cite{GentiniNoiseVQA_2020, CerezoPQC2021Review}. This hypothesis is also inspired by classical neural networks, which are robust under certain types of noise. In the classical setting, one can broadly distinguish between two types of noise: benign noise that does not severely impact the training procedure or can even improve generalization~\cite{jim1996analysis,noh2017regularizing,graves2011practical,graves2013speech}, and adversarial noise which is deliberately constructed to study where neural networks fail~\cite{balda2020adversarial,xie2017mitigating,gilmer2019adversarial,jaeckle2021generating}. Furthermore, we can distinguish between noise that is present during training, and noise that is present when using the trained model. Adversarial noise is usually of the latter case, where a trained neural network can produce completely wrong outputs due to small perturbations of the input data~\cite{goodfellow2014explaining}. The benign type of noise mentioned above on the other hand is usually present at training time in form of perturbations of the input data, activation functions, weights or structure of the neural network, and has even been established as a method to combat overfitting in the classical literature~\cite{jim1996analysis,noh2017regularizing,graves2011practical,graves2013speech,srivastava2014dropout}. These results inspired the hypothesis that variational quantum algorithms possess a similar robustness to certain types of noise and may even benefit from its presence when trained on a quantum device. However, thorough investigations that confirm such robustness of VQAs against hardware-related noise, or even a beneficial effect from it, are still lacking. In terms of negative results for the trainability of VQAs under noise, it has been shown that optimization landscapes of noisy quantum circuits become increasingly flat at a rate that scales exponentially with the number of qubits under local Pauli noise when the circuit depth grows linearly with the number of qubits~\cite{WangNoiseinducedBarrenPlateaus2021}. In the case of the variational quantum eigensolver, where the goal is to find the ground state of a given Hamiltonian, the presence of noise has been shown to lead to increasing deviation from the ideal energy~\cite{zeng2021simulating}. Similar effects have been studied in the context of the quantum approximate optimization algorithm (QAOA)~\cite{farhi2014quantum}, where the goal is to find the ground state of a Hamiltonian that represents the solution to a combinatorial optimization problem~\cite{alam2019analysis,harrigan2021quantum}. 

\begin{figure*}[ht]
    \centering
    \includegraphics[width=\textwidth]{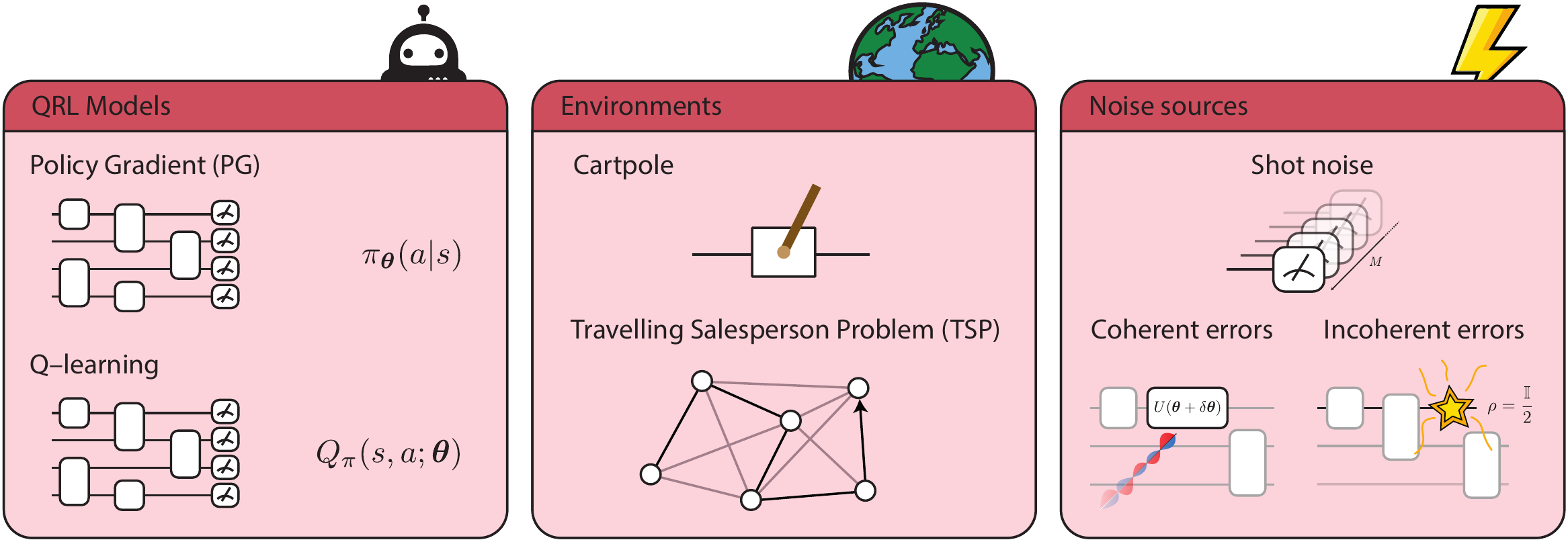}
    \caption{Summary of the scenarios analysed in the present work. We consider two models for quantum reinforcement learning (QRL) agents and test their performances on two environments, CartPole and the Travelling Salesperson Problem (TSP). We analyse the performances of the agents when these are trained and used in the presence of most common noise sources found on real quantum hardware, namely statistical fluctuations due to shot noise, coherent errors due to imperfect control or calibration of the device, and finally incoherent errors coming from the unavoidable interaction of the quantum hardware with its environment.}
    \label{fig:summary_figure}
\end{figure*}

When it comes to QML, in-depth studies on the effect of noise on the trainability and performance of VQAs are scarce. Apart from the work mentioned above on noise-induced barren plateaus~\cite{WangNoiseinducedBarrenPlateaus2021}, the authors of~\cite{LaRose2020Robust} provided first insights into how the data encoding method used in a quantum classifier influences its resilience to varying types of noise. As for the potential benefit of noise, the authors of~\cite{liu2022noise} show that the stochasticity induced by measurements in a QML model can help the optimizer to escape saddle points. The above results show that, on the one hand, too much noise will make the model untrainable, while on the other hand, modest amounts of noise can even improve trainability~\cite{liu2022noise}. However, it remains unclear how large the gap is between tolerable and harmful amounts of noise~\cite{GentiniNoiseVQA_2020}, and it is not expected that this can be answered in a general way for all different types of learning algorithms and noise sources. 

In this work, we shed light on this question from the angle of variational quantum reinforcement learning (QRL). Classical reinforcement learning (RL) models have been shown to be sensitive to noise, either during training~\cite{wang2020reinforcement} or in the form of adversarial samples~\cite{huang2017adversarial,kos2017delving}. Additionally, it is known that a bottleneck of RL algorithms is their sample inefficiency, i.e., many interactions with an environment are needed for training~\cite{yu2018towards}. Still, RL resembles human-type learning most closely among the main branches of modern ML, and therefore motivates further studies in this area. Among these studies, RL with VQAs has been proposed and extensively investigated in the noise-free setting over the past few years~\cite{chen2020variational, lockwood2020reinforcement, jerbi2021parametrized,skolik2022quantum, lan2021variational, wu2020quantum, sequeira2022variational, lockwood2021playing, franz2022uncovering}. These results provide promising perspectives, as quantum models have empirically been shown to perform similarly to neural networks on small classical benchmark tasks~\cite{skolik2022quantum, sequeira2022variational}, while at the same time an exponential separation between classical and quantum learners can be proven for specific contrived environments based on classically hard tasks~\cite{jerbi2021parametrized, skolik2022quantum}. These results motivate further studies on how large the above-mentioned gap between tolerable and too much noise is in the case of variational RL algorithms, and how close the algorithm performance can get to the noise-free setting for various types of noise that can be present on near-term devices.

We investigate this for two types of variational RL algorithms, Q-learning and the policy gradient method, by performing extensive numerical experiments for both types of algorithms with two different environments, CartPole and the Travelling Salesperson Problem, and under the effect of a wide class of noise sources, namely shot noise, coherent and incoherent errors. In Figure~\ref{fig:summary_figure} we summarise the approach of the present work showing the QRL models, environments and noise sources considered in the analysis. We start by considering the trade-off between the number of measurement shots taken for each circuit evaluation and the performance of variational agents. As the number of shots required by a QML algorithm can be a bottleneck on near-term devices and RL is known to require many interactions with the environment to learn, we propose a method for Q-learning to reduce the number of overall measurements by taking advantage of the structure of the underlying RL algorithm. Second, we model coherent errors with a random Gaussian perturbation of the variational parameters, and analytically study the effect of these perturbations on the output of parameterised quantum circuits, similarly to~\cite{ito2021universal}. We provide an upper bound on the perturbation induced by such Gaussian coherent noise based on the Hessian matrix of the circuit, and theoretically and numerically show that hardware-efficient ansätze may be particularly resilient against this type of error due to small second derivatives~\cite{CerezoHigherOrder2021}. Finally, we analyse the performance of the above algorithms under the action of incoherent errors coming from the unavoidable interaction of the qubits with the environment which we have no control over. To study this type of noise, we start by investigating the effect of single-qubit depolarization channels. In addition, we consider a custom noise model that combines various types of errors present on hardware, and study the effect of this noise model with error probabilities that are present in currently available superconducting quantum hardware. Our results show that both policy gradient methods and Q-learning exhibit a robustness to noise that may enable successfully running them on near-term devices. This motivates further study in the quest to find a real-world problem of interest where a quantum advantage for variational RL could be possible.

\section{Reinforcement Learning}

In this section, we will provide a brief introduction to RL that contains the basics necessary to understand this work. For a more in-depth introduction to the topic we refer the reader to~\cite{sutton2018reinforcement}.

In RL, an agent learns to perform a specific task by trial and error through interacting with an environment. In contrast to supervised learning, this means that there is no necessity for a preexisting training dataset made of pairs of inputs and corresponding correct labels. Instead, the learning task is specified in terms of an \textit{environment} and a \textit{reward function}. The environment is defined in terms of its state space $\mathcal{S}$ and its action space $\mathcal{A}$, as well as a transition function $P^a_{ss'} = P(s'|s, a)$ that specifies the probability of transitioning to state $s'$, given that the previous state is $s$ and action $a$ is taken. The agent can use the actions $a \in \mathcal{A}$ to move across states $s \in \mathcal{S}$ of the environment, and receives a reward $r$ that informs it about the quality of the chosen action. The agent chooses its actions based on a policy $\pi(a|s)$ which specifies the probability of taking actions given states, and its goal is to maximize the rewards. This is formally defined as a quantity called the \textit{expected return}, which is the random variable $G_t$,
\begin{equation}\label{eq:expected_return}
    G_t = \sum_{k=0}^{\infty} \gamma^k \cdot r_{t+1+k},
\end{equation}
where $\gamma \in [0, 1)$ is a discount factor that controls the significance of delayed rewards, $t$ is the current time step and $r_t$ represents the reward at the given time step. Typically we work in \textit{episodic} environments with a fixed time \textit{horizon} $H$, so that the sum in~\Cref{eq:expected_return} runs until $H$ instead of infinity. We can then quantify the agent's performance in terms of a \textit{value function},
\begin{equation}\label{eq:value_function}
    V_{\pi}(s) = \underset{\pi}{\mathbb{E}} \left[\sum_{k=0}^{H-1} \gamma^k \cdot r_{t+1+k} | s_t = s \right],
\end{equation}
which is the expected return when following a given policy $\pi$ from an initial state $s$. There are many different approaches to maximize the expected return, and we focus on the two main paradigms used in state-of-the-art RL: value-based and policy gradient methods. We will now introduce both of these in more detail.

\subsection{Value-based methods}\label{sec:value_methods}
One approach to maximizing the expected return is to parameterize and train the value function in \Cref{eq:value_function} directly with a function approximator. This function approximator can be implemented for example as a neural network (NN)~\cite{mnih2015human} or a parameterised quantum circuit (PQC)~\cite{chen2020variational,lockwood2020reinforcement,skolik2022quantum}. The value-based method that we focus on in this work is called \textit{Q-learning}. While the value function in \Cref{eq:value_function} is called the \textit{state-value} function as it only depends on the state, in Q-learning we try to approximate the \textit{action-value} function that additionally depends on the action,
\begin{equation}
    Q_{\pi}(s, a) = \underset{\pi}{\mathbb{E}} \left[\sum_{k=0}^{H-1} \gamma^k \cdot r_{t+1+k} | s_t=s, a_t=a \right].
\end{equation}

For a parametrized Q-function $Q_{\pi}(s, a; \boldsymbol \theta)$ the goal is then to approximate the optimal Q-function $Q^*$ as closely as possible, where the optimal Q-function is the one that leads to the optimal policy. The actions are chosen such that in each time step, the agent prefers to take the action that has the highest expected return, i.e.,
\begin{equation}\label{eq:argmax_qval}
    a_t = \mathrm{argmax}_a Q_{\pi}(s_t, a; \boldsymbol \theta).
\end{equation}
Due to this choice being deterministic, a Q-learning agent may never visit certain states of the environment and therefore not explore the state space sufficiently to find a good policy. In order to facilitate exploration, in practice a so-called $\epsilon$-greedy policy is used, where the agent selects a random action instead of that corresponding to the largest Q-value with probability $\epsilon$. Typically, $\epsilon$ is chosen large at the beginning and decreased over the course of training. In each training step, the Q-values are updated as follows,
\begin{equation}\label{eq:q_update}
    Q_{\pi}(s_t, a_t; \boldsymbol \theta) \leftarrow r_{t+1} + \gamma \max_a Q_{\pi}(s_{t+1}, a; \boldsymbol \theta).
\end{equation}

In order to train a function approximator like a NN or a PQC, the right-hand side of \Cref{eq:q_update} is used as a label in a supervised learning setting. This means that the function approximator is updated based on its own predictions about the expected return under the current parametrization, in addition to the reward given by the environment. Consequently, the agent needs to learn a moving target, which can lead to instability of training and delayed convergence. Additionally, updates are always based on the latest observed rewards, so the agent can ``forget'' previously learned behaviour even when it was beneficial. 

To stabilize training, two components have been added to the algorithm: a second model to compute the Q-values on the right-hand side of~\Cref{eq:q_update}, called the \textit{target model}, which is identical to the Q-function approximator but with parameters that are updated with a copy of the Q-function approximator's parameters only at fixed intervals. This decreases the rate of change in the prediction of the expected return used for parameter updates, and can therefore make learning more stable. Additionally, past interactions with the environment are stored in a memory and then randomly sampled to perform parameter updates to remove temporal correlations between transitions. For more detail on Q-learning with function approximators, also referred to as \textit{deep Q-learning} in classical literature, we refer the reader to the seminal work~\cite{mnih2015human}.

\subsection{Policy gradient method}\label{sec:pg_methods}

As described above, a RL agent chooses its actions based on a policy $\pi(a|s)$, which is the conditional probability distribution of actions given states. To maximize the expected return, the agent needs to find the optimal policy $\pi^*$. In policy gradient training, the agent is implemented in form of a parametrized policy $\pi_{\boldsymbol \theta}$, and the goal of the algorithm is to find the parameters $\boldsymbol \theta^*$ that produce the optimal policy. The quality of the policy is measured by a quantity $J(\boldsymbol \theta)$, that in the fixed-horizon setting is equal to the value function~\eqref{eq:value_function},
\begin{equation}
    J(\boldsymbol \theta) := V_{\pi_{\boldsymbol \theta}}(s).
\end{equation}
In a gradient-based optimization procedure the parameters are updated according to
\begin{equation}
    \boldsymbol \theta_{t+1} = \boldsymbol \theta_{t} + \alpha \nabla J(\boldsymbol \theta_t),
\end{equation}
with a learning rate $\alpha$, i.e., we perform gradient ascent on the parameters to maximize the expected return. 
The policy gradient theorem~\cite{sutton2018reinforcement} then states that the gradient of our performance measure can be written as
\begin{align}\label{eq:pg_theorem}
    \nabla J(\boldsymbol \theta) &= \nabla V_{\pi_{\boldsymbol \theta}}(s)\nonumber\\
    &\propto \sum_s \mu(s) \sum_a \nabla \pi_{\boldsymbol \theta}(a|s) Q_{\pi}(s, a)\nonumber\\
    &= \underset{\pi_{\boldsymbol \theta}}{\mathbb{E}}\left[\sum_a \nabla \pi_{\boldsymbol \theta}(a|S_t) Q_{\pi}(S_t, a) \right],
\end{align}
where $\mu(s)$ is the on-policy distribution under the current policy, which depends on the time spent in each state, and $S_t$ in the third line of \Cref{eq:pg_theorem} are states sampled under the policy $\pi$. Using this, we can now derive the REINFORCE algorithm, that is the basis of policy gradient based training.

Our goal is to perform gradient ascent on the parametrized policy purely from samples generated from said policy through interactions with the environment. The last line of~\Cref{eq:pg_theorem} still contains a sum over all actions $a$, which we can replace by the sample $A_t \sim \pi$ after multiplying and dividing the terms in the sum by $\pi_{\boldsymbol \theta}(a|S_t)$,
\begin{align}\label{eq:reinforce_derivation}
    \nabla J(\boldsymbol \theta) &\propto \underset{\pi_{\boldsymbol \theta}}{\mathbb{E}}\left[\sum_a \pi_{\boldsymbol \theta}(a|S_t) Q_{\pi}(S_t, a) \frac{\nabla \pi_{\boldsymbol \theta}(a|S_t)}{\pi_{\boldsymbol \theta}(a|S_t)} \right]\nonumber\\
    &= \underset{\pi_{\boldsymbol \theta}}{\mathbb{E}}\left[Q_{\pi}(S_t, A_t) \frac{\nabla \pi_{\boldsymbol \theta}(A_t|S_t)}{\pi_{\boldsymbol \theta}(A_t|S_t)} \right]\nonumber\\
    &= \underset{\pi_{\boldsymbol \theta}}{\mathbb{E}}\left[G_t \frac{\nabla \pi_{\boldsymbol \theta}(A_t|S_t)}{\pi_{\boldsymbol \theta}(A_t|S_t)} \right],
\end{align}
where $G_t$ is the expected return from \Cref{eq:expected_return}. Now, by using the fact that $\nabla \log x = \frac{\nabla x}{x}$, we can write
\begin{equation}\label{eq:pg_final}
    \underset{\pi_{\boldsymbol \theta}}{\mathbb{E}}\left[G_t \frac{\nabla \pi_{\boldsymbol \theta}(A_t|S_t)}{\pi_{\boldsymbol \theta}(A_t|S_t)} \right]
    = \underset{\pi_{\boldsymbol \theta}}{\mathbb{E}}\left[G_t \cdot \nabla \log \pi_{\boldsymbol \theta}(A_t|S_t) \right].
\end{equation}
This equation allows us to estimate the gradient of $J(\boldsymbol \theta)$ by samples from the current policy $\pi_{\boldsymbol \theta}$, and leads us to the following parameter update in each iteration of the algorithm,
\begin{equation}
    \boldsymbol \theta \leftarrow \boldsymbol \theta + \alpha \gamma^t \sum_{k=t+1}^T \gamma^{k-t-1} R_k ~ \nabla \log \pi_{\boldsymbol \theta}(A_t, S_t),
\end{equation}
where $\alpha$ is again the learning rate, $R_k$ is the reward, and $T$ is the length of the episode. Quantum versions of policy gradient based learning have been introduced in~\cite{jerbi2021parametrized,sequeira2022variational}, where the policy is implemented in form of a PQC.

\section{Environments and implementation}\label{sec:environments_implementation}

Our goal is to get insight into the effect of noisy training on quantum RL algorithms. For this, we consider quantum versions of the two main paradigms in RL that have been introduced in previous sections: value-based methods (see \Cref{sec:value_methods}) and policy gradient methods (see \Cref{sec:pg_methods}). As we are interested in the effect of noisy training on models that have otherwise been proven to work well in the noise-free setting, we study models and environments that have been already investigated in this setting before \cite{jerbi2021parametrized,skolik2022quantum,skolik2022equivariant}. In this way, we have evidence that the models and hyperparameters that we choose are suitable for the studied environments, and can focus our efforts on understanding the effect that noise has on the training and performance of these agents. The code that was used to generate the numerical results in this work can be found on Github \cite{noisy_qrl_code}.

\begin{figure}
  \subfloat[HWE-Q]{\includegraphics[scale=1]{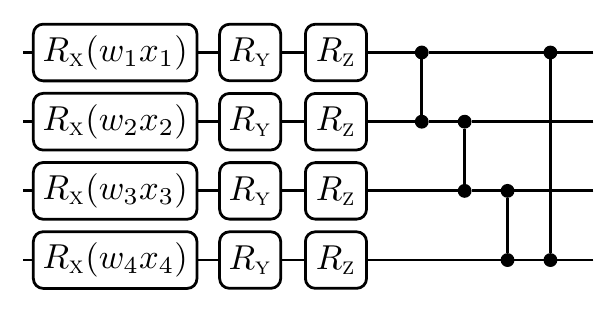}}\\
  \subfloat[HWE-PG]{\includegraphics[scale=1]{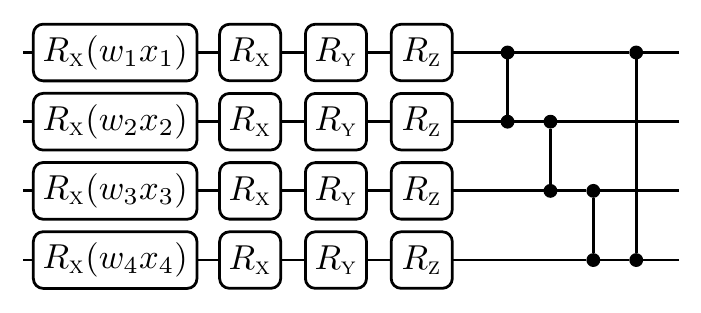}}\\
  \subfloat[EQC]{\includegraphics[scale=1]{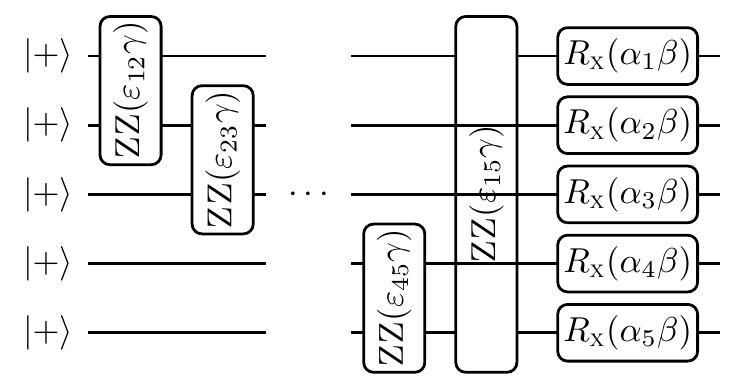}}\quad
  \caption{Parameterised circuits used in this work. (a) Hardware-efficient ansatz for Q-learning in the CartPole environment from~\cite{skolik2022quantum}, (b) hardware-efficient ansatz for policy gradient method in the CartPole environment from~\cite{jerbi2021parametrized}, (c) equivariant quantum circuit for Q-learning and policy gradient method in the TSP environment from~\cite{skolik2022equivariant}. For (a) and (b) we use 5 repetitions of the template shown above, while for (c) we use just one layer.}
\label{fig:ansatzes}
\end{figure}

\subsection{CartPole}

The first environment that we study is a benchmark task from the classical literature and implemented in the OpenAI Gym~\cite{openaiGym}: the CartPole environment. It has been previously studied in classical and quantum RL literature \cite{lillicrap2015continuous,jerbi2021parametrized,lockwood2020reinforcement,skolik2022quantum}. In this environment, the goal is to learn to balance a pole that is attached to a cart that can move left and right on a frictionless track. The state $s$ of the environment is represented by a four dimensional input vector $s\rightarrow \bm{x}=(x_1, x_2, x_3, x_4) \in \mathbb{R}^4$ encoding the position and velocity of the cart, and the velocity and angle of the pole. There are two actions that the agent can perform: moving the cart left or right. The environment is considered as solved when the agent manages to balance the pole for an average of at least 195 time steps for 100 consecutive episodes. We implement noisy training for the CartPole environment using the policy gradient approach introduced in~\cite{jerbi2021parametrized} and the Q-learning approach introduced in~\cite{skolik2022quantum}.

The circuit used for Q-learning in~\cite{skolik2022quantum} consists of five layers of a hardware-efficient ansatz~\cite{kandala2017hardware}, where each circuit layer consists of one parametrized rotation around the $x$-axis per qubit that is used to encode the input states $\bm{x}$, and additional parametrized $y$- and $z$-rotations on each qubit that contain the free parameters to be trained (see~\Cref{fig:ansatzes}(a)). Furthermore, additional trainable parameters multiplying each input feature are used to increase the expressivity of the reuploading quantum circuit \cite{Perez2020Reuploading, Schuld2020Encoding}. Each layer also has a final layer of CZ-gates arranged in a circular topology. The observable for taking the action ``left'' is $O_L = Z_1 Z_2$, where $Z_1$ and $Z_2$ are Pauli-Z operators acting on the first and second qubit, respectively. Similarly, action ``right'' is associated to the observable $O_R = Z_3 Z_4$, defined on the third and fourth qubit. In order to facilitate the function approximation of the optimal Q-function, which has a range of output values beyond that of $Z_i Z_j$ operators, each expectation value is further multiplied with an additional trainable weight, such that the final Q-value for action ``left'' is
\begin{align}
\label{eq:q-values}
    Q(s, L) &= \frac{\expval{O_L}_{s,\bm{\theta}} + 1}{2}\,w_L\\
    & = \frac{\mel{\bm{0}}{U_{\boldsymbol \theta}^{\dagger}(s) O_L U_{\boldsymbol \theta}(s)}{\bm{0}} + 1}{2}\,w_{L},
\end{align}
where $U_{\boldsymbol \theta}(s)$ represents the unitary of the parameterised circuit depending on the trainable parameters $\bm{\theta}$ and the input state $s$, and $w_L$ is the trainable weight corresponding to observable $O_L$. The Q-value for the action ``right'' is defined in a similar manner.

For the policy gradient method, we follow the implementation used in~\cite{jerbi2021parametrized} and made available at~\cite{tfqRlTutorial}, which uses five layers of the same hardware-efficient ansatz as described for Q-learning above, except that each layer has an additional trainable rotation around the $x$-axis on each qubit (see~\Cref{fig:ansatzes}(b)), and the actions observables are defined as $O_L = Z_1 Z_2 Z_3 Z_4$ and $O_R = \mathbb{I} - O_L$. As before, input features are multiplied with an additional trainable parameter each. Since the policy is a probability distribution, a final \textsc{SoftMax} layer is used to map the expectation values $\expval{O_a}_{s,\bm{\theta}} \in [-1,1]$ to the appropriate range $[0,1]$, and so probabilities for each action eventually become
\begin{equation}
\label{eq:Policy_PQC}
    \pi_{\boldsymbol \theta}(a|s) = \frac{e^{\beta\expval{O_a}_{s, \boldsymbol \theta}}}{\sum_{a'} e^{\beta\expval{O_{a'}}_{s, \boldsymbol \theta}}},
\end{equation}
where $\beta \in \mathbb{R}$ is a also a trainable parameter.

\subsection{Traveling Salesperson Problem}

The second environment that we study is more complex and requires introducing the field of neural combinatorial optimization (NCO). NCO is an alternative to the hand-crafted heuristics used in combinatorial optimization, where instead a machine learning model is trained to solve instances of a given combinatorial optimization problem \cite{bello2016neural}. In the case of RL-based NCO, the optimization problem is defined in terms of an environment and the quality of the solution is measured by the reward function. In this work, we study a quantum NCO approach that learns to solve instances of the Traveling Salesperson Problem (TSP), as introduced in~\cite{skolik2022equivariant}. In TSP one is presented with a list of cities in form of a weighted graph, and the goal is to find the tour of minimal length that visits each city in this list exactly once. 

In this environment one episode consists in solving one instance, where the agent selects the cities in the tour in a step-wise fashion. States in this environment are instances of the TSP, in addition to the partial tour at the current time step. The actions are defined in terms of the cities, where in each time step the agent can select one of the cities that is not yet in the tour. The reward is the negative difference in length between the tour at the previous time step and the tour after adding the latest city, as we want to minimize the length of the tour while RL agents try to maximize the expected reward. We evaluate the quality of the tours proposed by the agents in terms of the approximation ratio
\[\frac{c(T)}{c(T^*)},\]
where $c(T)$ is the length of the tour $T$ proposed by the agent, and $c(T^*)$ is the length of the optimal tour $T^*$. The stopping criterion for this environment is an average approximation ratio of at least 1.05 over the past 100 episodes.

To implement a quantum agent for this environment, we follow~\cite{skolik2022equivariant}, where the information of the TSP graph instance is directly encoded into a PQC and each graph node corresponds to one qubit. Each layer in this ansatz consists of one rotation around the x-axis parametrized by $\alpha_i \beta_l$, where $\alpha_i \in \{0, \pi\}$ represents whether city $i$ is already in the current tour ($\alpha_i = 0$), or still available for selection ($\alpha_i = \pi$), and $\beta_l$ is a trainable parameter that is shared across all single-qubit gates in layer $l$. The graph's edges in each layer are represented by a ZZ-gate parametrized by $\varepsilon_{ij} \gamma_l$, where $\varepsilon_{ij}$ is the weight of edge connecting nodes $i$ and $j$, and $\gamma_l$ is a trainable parameter that is shared across all two-qubit gates in layer $l$. Such ansatz is shown in~\Cref{fig:ansatzes}(c). 

In the case of Q-learning, the observables are ZZ-operators that correspond to the edges in the graph, i.e., $Z_i Z_j$ is measured for edge $ij$. For policy gradient agents the observables are the same, but as the policy has to be a probability distribution we again use a final \textsc{SoftMax} layer with a trainable inverse temperature $\beta$, as in~\Cref{eq:Policy_PQC}. The authors of~\cite{jerbi2021parametrized} have shown that using this type of final layer can be highly beneficial for policy gradient training, compared to only using the probability distribution resulting from the quantum state directly. This is due to the fact that the trainable inverse temperature enables the agent to tune its level of exploration of the state space. As the optimal solutions to TSP instances are deterministic, it is favourable in this environment to have a tunable inverse temperature that allows exploration of the large state space early in training, as well as close-to-deterministic decisions towards the end.

\section{Shot noise}

We start our studies with the type of noise that is arguably the simplest to characterize: noise induced by statistical errors that result from the probabilistic nature of quantum measurements. For each circuit evaluation, be it for action selection of the RL agent or for computing parameter updates via the parameter shift rule, we take a fixed number of measurements $M$ and compute the resulting expectation value. The precision of this expectation value depends on $M$ and scales like $\epsilon \sim 1/\sqrt{M}$. 

Variational algorithms often require a very large number of measurements to be executed, and this problem is exacerbated in QML tasks that typically involve separate circuit evaluations for all training data points. For this reason, it is not only important to understand the effect of shot noise on the trainability and performance of QML models, but it is also desirable to develop methods that lead to a smaller shot footprint than simply assigning a fixed number of shots to each circuit evaluation. Depending on knowledge of the algorithm itself, it can be possible to make an informed decision on the number of shots that suffice in each step. In this section, we develop such a method specifically for Q-learning that is a natural extension to the original algorithm.

\subsection{Reducing the number of shots in a Q-learning algorithm}
\label{sec:ucb_inspired_alg}

As described in \Cref{sec:value_methods}, a Q-learning agent selects actions based on the following rule (see \Cref{eq:argmax_qval})
\[a_t = \mathrm{argmax}_a Q_{\pi}(s_t, a; \boldsymbol \theta),\]
that is, it chooses actions according to the largest Q-value.\footnote{In the $\epsilon$-greedy policy (see \Cref{sec:value_methods}) we consider here, the agent picks either the action corresponding to the argmax Q-value, or a random action. As no circuit evaluation is required to pick a random action, we only consider the steps with actual action selection by the agent in this section.} Now, consider a quantum agent that only has access to noisy estimates of the Q-values $\tilde{Q}(s_t, a_t; \boldsymbol \theta)$ resulting from the statistical uncertainty of a measurement process involving a finite number of shots $M$. If the sample size is large enough $M\gg 1$, then by the central limit theorem each noisy Q-value can be described as a random variable
\begin{equation}
\label{eq:noisy_q_val}
\tilde{Q}(s_t, a_t; \boldsymbol \theta) = Q(s_t, a_t; \boldsymbol \theta) + \epsilon\,,
\end{equation}
where $Q(s_t, a_t; \boldsymbol \theta)$ is the true noise-free value, and $\epsilon$ is a random variable sampled from a Gaussian distribution centered in zero $\mu_{\epsilon} = \mathbb{E}[\epsilon]=0$, and with standard deviation inversely proportional to the square root of the number of measurement shots $\sigma_\epsilon = \text{Std}[\epsilon] \sim 1/\sqrt{M}$. Since actions are selected through an argmax function, the perturbation $\epsilon$ will not affect the action selection process as long as the order between the largest and the remaining Q-values remains unchanged. Then, one may ask: is there a minimal number of shots that suffice to reliably distinguish the largest Q-value $Q_{max}$ and the second-largest Q-value $Q_2$? 

When the observables associated to the actions are non-commuting, they have to be estimated independently from each other, and one has the freedom of choosing how to allocate the measurement shots among the observables of interest, possibly in a clever way. In our case, the goal is to estimate which of the observables has the highest Q-value while trying to be shot-frugal, and this task can be related to the theory of multi-armed bandits \cite{slivkins2019introduction}. The multi-armed bandit is a RL problem in which an agent can allocate only a limited amount of resources between a number of choices, e.g., a number of arms on a bandit machine, and is asked to determine which of these choices leads to the highest expected reward. There exists a trade-off between exploration (i.e., trying the different arms) and exploitation (always choosing the arm that appears best according to the current knowledge), and the upper confidence bound (UCB)~\cite{lai1985asymptotically,auer2002using} algorithm shows how to use statistical confidence bounds to allocate exploratory resources. The UCB algorithm could be used in the scenario described above where a number of non-commuting observables have to be estimated, and we want to find the optimal strategy to allocate a fixed budget of measurement shots to the task of identifying the largest Q-value.

However, in the specific implementations of QRL agents based on recent literature that we study in this work~\cite{jerbi2021parametrized,skolik2022quantum,skolik2022equivariant}, only commuting observables are used, hence it is not necessary to apply the UCB procedure to determine which one should be measured more often. Nonetheless, inspired by the UCB algorithm, we can still define a rather general simple heuristic that can be used to reduce the overall number of shots required to train the Q-learning models as those studied in this work. The idea is to use the knowledge about the scaling of the estimation error with respect to the number of measurements (see \Cref{eq:noisy_q_val}), to determine with confidence whether we have taken enough shots to determine the maximum Q-value. 

\begin{algorithm}[t]
\caption{Algorithm to reduce the number of measurements in Q-learning}\label{algorithm:ucb_inspired}
\SetKwInOut{Input}{Input}
\SetKwInOut{Output}{Output}
\Input{$m_{\mathrm{init}}$, $m_{\mathrm{inc}}$, $m_{\mathrm{max}}$}
\Output{$m_{\mathrm{est}}$}

$m_{\mathrm{est}} \gets m_{\mathrm{init}}$\;
\While{$m_{\mathrm{est}} < m_{\mathrm{max}}$}{
    $\tilde{Q}(s_t, a_1) = \langle O_{a_1} \rangle_{m_{\mathrm{est}}}$\;
    $\tilde{Q}(s_t, a_2) = \langle O_{a_2} \rangle_{m_{\mathrm{est}}}$\;
    $\Delta \tilde{Q} = \abs{\tilde{Q}(s_t, a_1) - \tilde{Q}(s_t, a_2)}$\;
    \eIf{$\Delta \tilde{Q} < 2/\sqrt{m_{\mathrm{est}}}$}
        {$m_{\mathrm{est}} \gets m_{\mathrm{est}} + m_{\mathrm{inc}}$\;}
        {\Return $\mathrm{min}(m_{\mathrm{est}}, m_{\mathrm{max}})$\;}
    }
\Return $\mathrm{min}(m_{\mathrm{est}}, m_{\mathrm{max}})$\;
\end{algorithm}

The procedure goes as follows. First, we take a small number of initial measurements $m_{\mathrm{init}}$, for example $m_{\mathrm{init}} = 100$, of all observables to compute the estimates $\tilde{Q}_{m_{\mathrm{init}}}(s_t, a),~\forall~a \in \mathcal{A}$. Based on these values, we compute the absolute difference between the largest and the second largest Q-values. If this difference is larger than twice the estimation error $\epsilon = 2/\sqrt{m_{\mathrm{init}}}$ (as both of the Q-values are noisy), we have found the largest Q-value with high confidence and we stop here. On the other hand, if the difference is smaller, we increment the sample size with additional $m_{\mathrm{inc}}$ measurements each, and recompute the estimated Q-values with the $m_{\mathrm{inc}} + m_{\mathrm{init}}$ shots. We again compute the absolute difference of the two largest Q-values and determine whether the number of measurements suffices based on the error $\epsilon = 2/\sqrt{m_{\mathrm{init}} + m_{\mathrm{inc}}}$. This measure-and-compare scheme is performed until either the two largest Q-values can be distinguished with high confidence, or a fixed shot budget $m_{\mathrm{max}}$ is reached. 

In \Cref{algorithm:ucb_inspired} we provide a description of this procedure, where for the sake of simplicity we describe the case where there are only two possible actions, and we therefore only have to find the larger of two Q-values. However, the scheme can be used for an arbitrary number of Q-values, as it is only important to distinguish between the highest and the second-highest Q-value with high confidence. The algorithm takes as input the number of initial measurements $m_{\mathrm{init}}$, the number of additional measurements in every step $m_{\mathrm{inc}}$, and the maximum number of measurements that are allowed in one run of the shot-allocation algorithm (i.e., finding the largest Q-value) $m_{\mathrm{max}}$. The output is the number of measurements $m_\mathrm{est}$ that are sufficient to find the argmax Q-value with high confidence based on the rules above. The values $\langle O_{a_i} \rangle_{m_{\mathrm{est}}}$ are the expectation values of observables $O_{a_i}$ corresponding to action $a_i$, estimated with $m_{\mathrm{est}}$ shots. Note that the proposed scheme works both for commuting or non-commuting observables, where in the former case one can spare shots by computing the observables from the same set of measurement outcomes. Moreover, note that we ignore the coefficients in the statistics of the Q-values coming from~\Cref{eq:q-values}, when considering the measurement stopping criterion. This choice has no impact on the effectiveness of the proposed method, as it is always found to be very well performing in the presented form. 

While this algorithm can clearly determine the optimal number of shots in the action selection process in a methodical manner, one should check that this will not introduce errors in the remaining parts of the variational Q-learning model, i.e., during the parameter update step. Recall that each parameter update of the model is computed based on the output of the model itself (see \Cref{eq:q_update})
\[Q_{\pi}(s_t, a_t; \boldsymbol \theta) \leftarrow r_{t+1} + \gamma \max_a Q_{\pi}(s_{t+1}, a; \boldsymbol \theta),\]
which means that in the parameter update step we do not need to perform action selection, but instead care about the actual Q-values in order to compute the loss function. The question is now to what precision we need to approximate the Q-values in order to learn a good Q-function. Technically, even the noise-free Q-function is only an approximation of the true Q-function, which is the whole point of doing Q-learning with function approximators. This suggests that there is some leeway to make even the approximate function itself an approximation by taking only as many measurements as are necessary to find the argmax Q-value with high confidence. Indeed, it has been shown in \cite{skolik2022quantum} that even the Q-functions of agents that successfully solve an environment can produce Q-values that are far from the optimal Q-values, and that learning the correct order of Q-values is more important in this setting than approximating the optimal Q-value as precisely as possible. Consequently, when we compute the Q-values that are used to perform parameter updates, we use the same algorithm as that in \Cref{algorithm:ucb_inspired} to determine the number of measurements to take.

\subsection{Numerical results}

\begin{figure}
\includegraphics[width=\columnwidth]{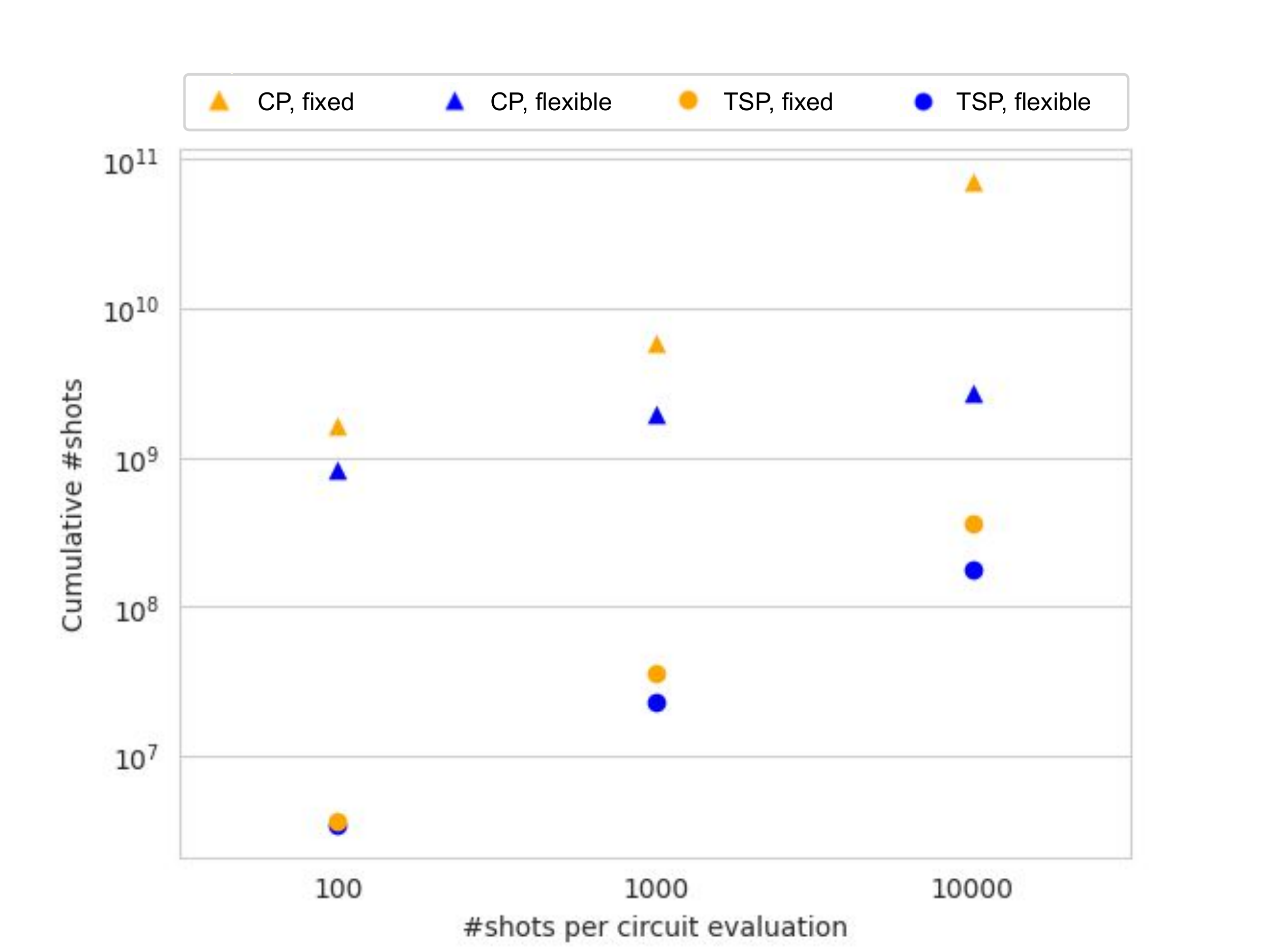}
\caption{Comparison of the cumulative number of shots per observable over a full training run, for the flexible shot allocation technique (blue) and for a standard a fixed measurement scheme using the same number of shots for every circuit evaluation (orange), both for CartPole (triangles) and TSP (circles). Each data point shows the average over ten trained agents.}
\label{fig:ucb_vs_fixed_cumulative}
\end{figure}

\begin{figure*}
  \subfloat[CartPole]{\includegraphics[width=\columnwidth]{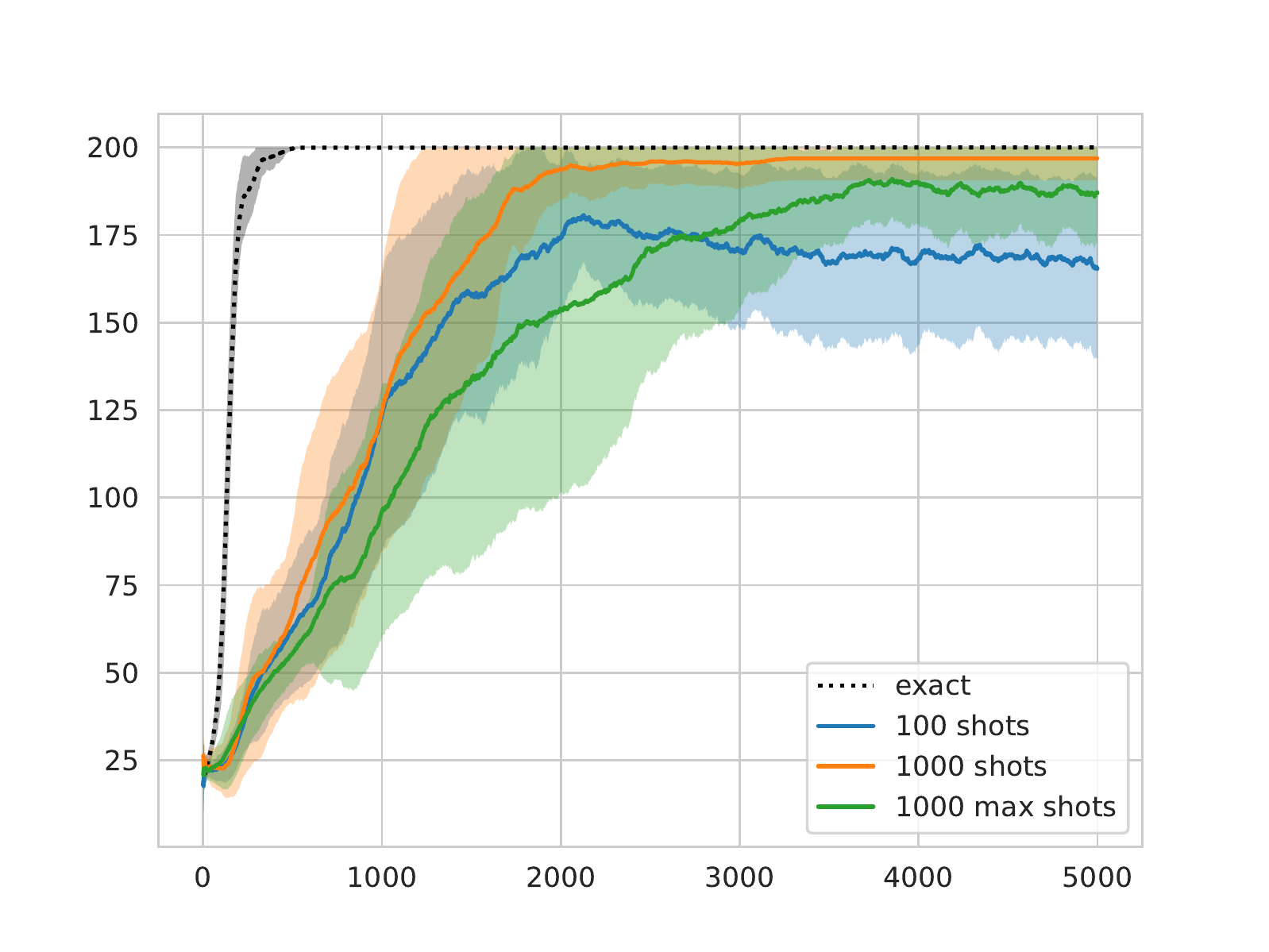}}\quad
  \subfloat[Traveling Salesperson Problem]{\includegraphics[width=\columnwidth]{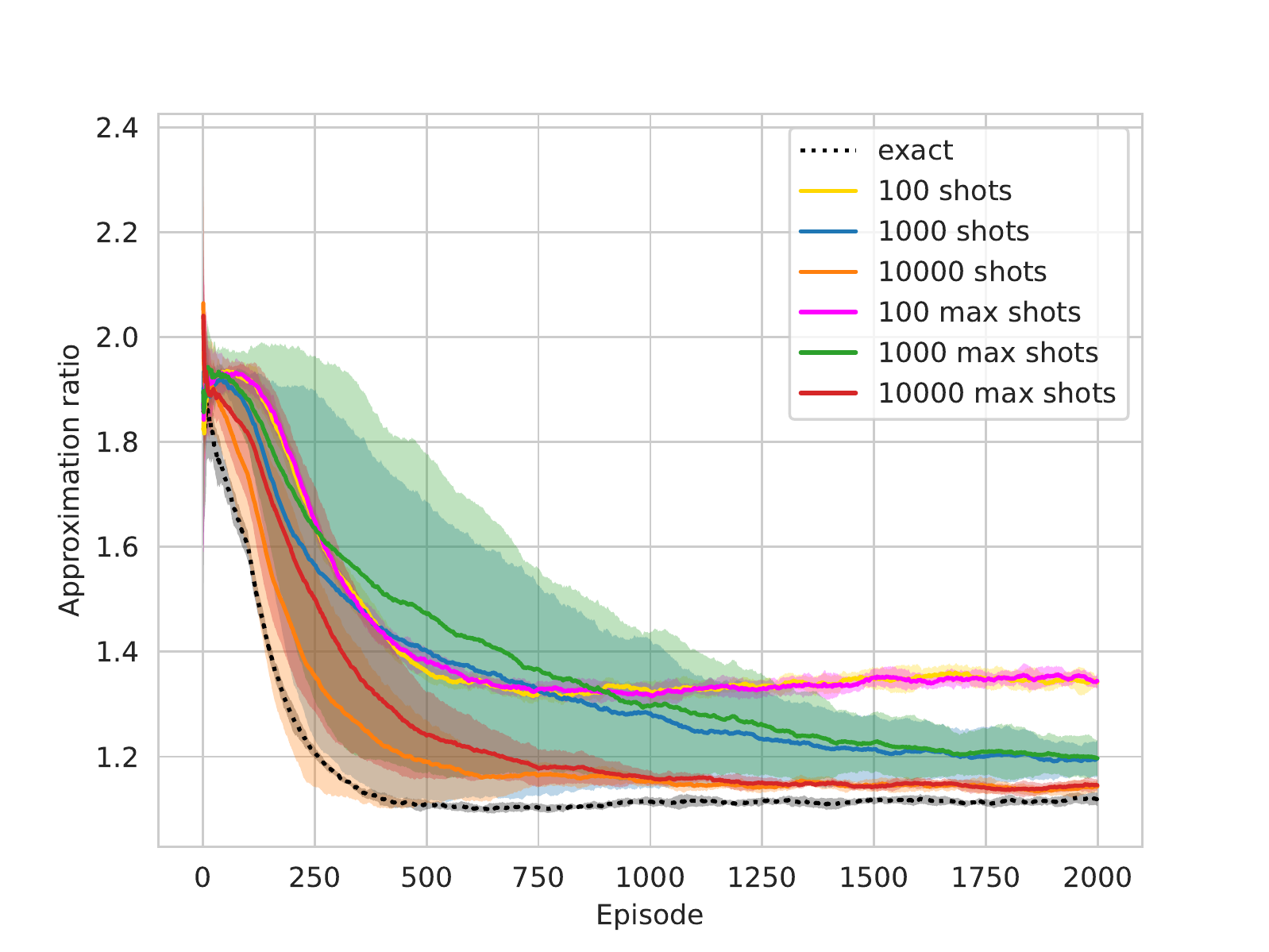}}\quad
  \caption{Comparison of Q-learning with shot noise using the informed shot-allocation method (labeled ``max shots") proposed in this work, and a standard measurement scheme that simply assigns a fixed number of shots to each circuit evaluation (labeled ``shots"). Results are averaged over ten agents for each configuration. (a) Shows results for agents trained in CartPole environment, (b) shows results for agents in the TSP environment.}
\label{fig:shots_ucb_comparison}
\end{figure*}

We now numerically compare the performance of agents in the CartPole and TSP environments in settings where a fixed number of shots is used in each circuit evaluation, and where the number of shots in each step is determined by the algorithm we introduced in \Cref{sec:ucb_inspired_alg}. To give an overview of the number of shots used in one training run under varying hyperparameter settings, we show the average cumulative number of shots for different settings in \Cref{fig:ucb_vs_fixed_cumulative}. For the CartPole environment (triangles), the number of cumulative shots grows quickly with the number of shots in each step in the fixed setting (orange). This is not true for the flexible shot allocation technique (blue), where for values of $m_{\mathrm{max}} \in \{100, 1000, 10000\}$ the cumulative number of shots is relatively similar. As we see in \Cref{fig:shots_ucb_comparison} a), a low number of shots such as 1000 is already sufficient to achieve close to optimal performance in the CartPole environment. Therefore, we focus on comparing settings with 100 and 1000 (maximum) shots per circuit evaluation in that figure. Comparing the cumulative number of shots for $m_{\mathrm{fixed}}=100$ and $m_{\mathrm{max}}=1000$ in \Cref{fig:ucb_vs_fixed_cumulative}, we see that these two configurations use almost the same number of measurements overall. Still, the final performance of the agents trained with the flexible shot allocation technique is almost optimal, while those trained with a fixed number of shots in each circuit evaluation are below a final score of 175 on average. However, as we allow agents to use even less than 100 shots per evaluation with the flexible allocation method of~\Cref{algorithm:ucb_inspired}, performance starts to degrade, so at least 100 shots are required in this setting. To not clutter the figure we show the results for agents that use fewer than 100 shots per circuit evaluation in~\Cref{fig:cp_ucb_worst} in the Appendix.

In the TSP environment, each step in an episode constitutes of a constant and (compared to CartPole) relatively low number of circuit evaluations. We still see that the higher the setting for the (maximum) number of shots is, the bigger the gap in average cumulative number of shots becomes. For agents trained in the TSP environment, shown in~\Cref{fig:shots_ucb_comparison} b), the final performance remains unchanged by the additional noise introduced by the flexible shot allocation technique, and agents reach the same accuracy of those trained with a corresponding but fixed number of shots per circuit evaluation. The only difference between the two approaches is that the agents using the flexible shot allocation method take slightly longer to converge in some cases. Independently from the estimation method used (flexible or fixed), it is clear from~\Cref{fig:shots_ucb_comparison} that it is the number of shots available that plays the major role in determining the performance of the noisy agents, as measured by the proximity to the average approximation ratios reached in the noise-free scenario, namely when agents have access to exact the expectation values ($M\rightarrow\infty$). In this environment, there is a trade-off between delayed convergence due to less precision in the approximation of the Q-function, and using a higher number of shots to arrive at the same final performance.

To summarize, we have seen that Q-learning models can be successfully trained even in the presence of statistical noise introduced by a measurement processes carried out with a limited number of shots. In addition, by leveraging the specifics of the Q-learning algorithm, we introduced an easy-to-implement and effective method that can be used to reduce the number of shots needed to train variational Q-learning agents. How many shots one can save during training with this method depends on the agents' resilience to shot noise, as well as the specific characteristics of the environment. In the CartPole environment, where one bad decision does not lead to immediate failure, the additional noise introduced by estimating expectation values with a low number of measurements and approximating an imprecise Q-function does not affect performance severely. In the TSP environment on the other hand, where one bad choice of the next city in the tour can lead to a much longer path, we observe that the number of measurements has to be relatively high to get close to optimal performance. However, even in this setting we can achieve a reduction in the overall number of measurements by taking an informed approach at when to measure an observable more often.

\section{Coherent noise}\label{sec:gauss_noise_section}
In this section, we turn our attention to coherent noise, that is, errors that preserve the unitary evolution of the quantum circuit but still change its output~\cite{cai2020mitigating}. In our analysis, we model coherent noise as an \textit{over-} or \textit{under-rotation} of the parametrized gates, by adding a random Gaussian perturbation to the variational parameters in the considered circuits. 

This type of error could occur in real quantum devices as a drift in the parameters for example due to an imperfect control of the system or a miscalibration of the hardware, and it is therefore an important component of the overall picture of an imperfect quantum device. Specifically, we assume that the perturbation remains unchanged during the estimation of a given observable, i.e. it does not change considerably between repeated measurements on the same experimental setup. However the perturbation amount changes whenever the experiment is changed, for example due to measuring a different observable, or using the circuit with a different set of parameters. 

Gaussian coherent noise is also an interesting model because it lends itself very well to theoretical analysis, and one can estimate the effect of such an error on the output of a parameterised quantum circuit. In the following, we first proceed with an analytical treatment of the error introduced by Gaussian perturbations on variational circuits, and then proceed with the numerical results for the two environments considered in this work.

\subsection{Effect of Gaussian coherent noise on circuit output}\label{sec:gauss_effect_output}

Consider a general parametrized quantum circuit acting on a system of $n$ qubits, with unitary $U(\bm{\theta}) \in \mathbb{C}^{2n} \times  \mathbb{C}^{2n}$ and parameter vector $\bm{\theta} = (\theta_1, \hdots, \theta_M) \in \mathbb{R}^{M}$. Let $O$ be on observable and $\rho = \dyad{0}$ the initial state of the quantum system, the outcome of the variational circuit is the expectation value
\begin{equation}
\label{eq:function_pqc}
    f(\bm{\theta}) = \expval{\obs}_{\bm{\theta}} =  \Tr[\obs\, U(\bm{\theta}) \rho U^{\dagger}(\bm{\theta})]\, .
\end{equation}
 
Suppose that the parameters are affected by a noise process that adds a perturbation 
\begin{equation}
    \bm{\theta} \rightarrow \bm{\theta} + \delta \bm{\theta}\, ,
\end{equation}
where $\delta\bm{\theta} = (\delta\theta_1,\hdots, \delta\theta_M) \in \mathbb{R}^M$ are \textit{i.i.d.}\ according to a Gaussian distribution $\mathcal{N}(\mu, \sigma)$ with zero mean $\mu=0$ and equal variance $\sigma^2$, namely
\begin{align}
\label{eq:gaussian_distrib}
    & \delta\theta_i \sim \mathcal{N}(0, \sigma^2)\,, & \nonumber\\
    & \mathbb{E}[\delta\theta_i] = 0\,, &\hspace*{-1cm} \forall i \in \{1,\hdots, M\}\,   \\
    & \mathbb{E}[\delta\theta_i\delta\theta_j] = \sigma^2 \delta_{ij}\,. & \nonumber
\end{align}
As discussed earlier, in our analysis in this section and in the numerical simulations in~\cref{sec:res_cp_gaussian}, we assume that the perturbed parameters remain the same during the evaluation of a single expectation value. In a real experiment on quantum hardware, this would mean that for all measurements used to estimate the expectation value, the perturbations stay at least approximately unchanged. Of course, without this assumption, the resulting noise model could not be considered unitary, and one may then resort to a noise channel formulation of Gaussian noise as proposed in~\cite{GentiniNoiseVQA_2020, ito2021universal}. Hence, in the following we restrict our attention to the setting described above.

The effect of Gaussian noise on the circuit can be evaluated by Taylor expanding the circuit around the unperturbed parameters $\bm{\theta}$. For ease of explanation, we hereby report only the main ideas and results, and we refer to Appendix~\ref{app:higher-order} for a complete and detailed derivation of all the calculations performed in this section.

Let $f(\bm{\theta} + \delta\bm{\theta})$ be the function evaluated on the perturbed parameters, its Taylor expansion up to fourth-order reads
\begin{equation}
\label{eq:taylor_expansion}
\begin{aligned}
    f(\bm{\theta}+\delta\bm{\theta}) \approx & f(\bm{\theta}) + \sum_{i=1}^M\frac{\partial f(\bm{\theta})}{\partial\theta_i}\delta\theta_i + \frac{1}{2}\sum_{i,j=1}^{M}\frac{\partial^2 f(\bm{\theta})}{\partial\theta_i\partial\theta_j}\delta\theta_{i}\delta\theta_{j}\\
    & + \frac{1}{3!}\sum_{i,j,k=1}^M\frac{\partial f(\bm{\theta})}{\partial\theta_i\partial\theta_j\partial\theta_k}\delta\theta_{i}\delta\theta_{j}\delta\theta_{k} + \order{\delta\theta^4}\, .
\end{aligned}
\end{equation}

With this expression one can evaluate the expected value of the noisy function $\mathbb{E}[f(\bm{\theta}+\delta\bm{\theta})]$ over the distribution of the Gaussian perturbations, $\mathbb{E}(\cdot) = \mathbb{E}_{\delta\theta_i \sim \mathcal{N}(0, \sigma^2)}(\cdot)$. Since every odd moment of a Gaussian distribution vanishes, using relations~\eqref{eq:gaussian_distrib} in the expansion~\eqref{eq:taylor_expansion} one obtains
\begin{equation}
\label{eq:taylor4}
\begin{aligned}
    \mathbb{E}[f(\bm{\theta}+\delta\bm{\theta})] & \approx f(\bm{\theta}) + \frac{1}{2}\sum_{ij}\frac{\partial f(\bm{\theta})}{\partial \theta_i \partial \theta_j}\mathbb{E}[\delta\theta_i\delta\theta_j] \\
    & \approx f(\bm{\theta}) + \frac{1}{2}\sigma^2\sum_{ij}\frac{\partial f(\bm{\theta})}{\partial \theta_i \partial \theta_j}\delta_{ij} \\
    & \approx f(\bm{\theta}) + \frac{1}{2}\sigma^2 \Tr[H(\bm{\theta})]\, + \order{\sigma^4},
\end{aligned}
\end{equation}
where $\Tr[H(\bm{\theta})]$ denotes the trace of the Hessian matrix
\begin{equation}
    H_{ij}(\bm{\theta}) = \frac{\partial^2 f(\bm{\theta})}{\partial \theta_i \partial \theta_j}\, \quad i,\,j=1\hdots, M.
\end{equation}

Thus, the first non-vanishing correction term caused by the noise is proportional to the noise variance $\sigma^2$, and the Hessian of the parametrized quantum circuit, which conveys geometric information about the curvature of the function landscape around the unperturbed point $\bm{\theta}$.

Higher-order terms in the expansion can be evaluated in a similar way, specifically making use of so-called \textit{Wick's relations} for multivariate normal distributions as shown in Appendix~\ref{app:higher-order}. If all the derivatives of the function $f(\bm{\theta})$ are bounded, as it is the case for parametrized quantum circuits, then it is possible to derive an upper bound on the error induced by the perturbations which only depends on the noise strength $\sigma^2$ and the total number of parameters $M$, as we show in the following. 

Using the parameter shift rule~\cite{Schuld2019Gradients, Mitarai2018Learning}, one can show that any derivative of a parametrized quantum circuit can be expressed as a linear combination of circuit outcomes evaluated at specific points in parameter space~\cite{ito2021universal, CerezoHigherOrder2021}. Let $\bm{\alpha} = (\alpha_1, \hdots, \alpha_M) \in \mathbb{N}^{M}$ be a multi index keeping track of the order of partial derivatives, define the derivative operator
\begin{equation}
    \partial^{\bm{\alpha}}  := \frac{\partial^{|\bm{\alpha}|}}{\partial\theta_{1}^{\alpha_1}\cdots\, \partial\theta_{M}^{\alpha_M}}\, ,
\end{equation}
where $\abs{\bm{\alpha}} := \sum_{i=1}^M \alpha_i$. By nested applications of the parameter shift rule, one can show that
\begin{equation}
\label{eq:multideriv_pqc}
    \partial^{\bm{\alpha}} f(\bm{\theta}) = \frac{1}{2^{\abs{\bm{\alpha}}}} \sum_{i=1}^{2^{\abs{\bm{\alpha}}}} s_m\, f(\bm{\theta}_m)\,,
\end{equation}
where $s_m \in \{\pm 1\}$ are signs, and $\bm{\theta}_{m}$ are parameters obtained shifting the parameter vector $\bm{\theta}$ along different directions. Now, since the measurement outcome of every circuit is bounded by the maximum absolute eigenvalue of the observable, i.e.\ $\abs{f(\bm{\theta})} \leq \|\obs\|_\infty$, consequently it also holds that $\abs{\partial^{\bm{\alpha}} f(\bm{\theta})} \leq \|\obs\|_\infty$ (see Appendix \ref{app:higher-order}). Note that we only consider bounded observables here, like the Pauli operators commonly used in variational RL algorithms \cite{chen2020variational,lockwood2020reinforcement,jerbi2021parametrized,skolik2022quantum}.

Since all the derivatives of the function are bounded, it is possible to bound every term in the Taylor series and then compute an upper bound to the error caused by the perturbation. In fact, defining the absolute (average) error caused by the noise as
\begin{equation}
    \label{eq:gauss_error}
    \varepsilon_{\bm{\theta}} := \abs{\mathbb{E}[f(\bm{\theta }+ \delta \bm{\theta})] - f(\bm{\theta})}\, ,
\end{equation}
one can prove that this is upper bounded by (see Appendix~\ref{app:higher-order})
\begin{equation}
    \label{eq:gaussian_error}
    \varepsilon_{\bm{\theta}} \leq \norm{\obs}_{\infty} \qty( e^{\sigma^2 M /2 } - 1)\, .
\end{equation}
Note that since $\varepsilon_{\bm{\theta}}\leq 2\norm{O}_\infty$ is always true, the bound is informative only as long as $e^{\sigma^2 M /2 } - 1<2$.

This expression only depends on the noise strength $\sigma^2$, the total number of noisy parameters $M$, and the operator norm of the observable $\norm{\obs}_{\infty}$, and it can be used to estimate a \textit{sufficient} condition on the noise strength to guarantee a desired error threshold $\varepsilon_{\bm{\theta}}$. Rearranging \Cref{eq:gaussian_error}, a sufficient condition to have error $\varepsilon_{\bm{\theta}}$ not larger than $\epsilon$, is to have Gaussian perturbations satisfying
\begin{equation}
\label{eq:error_bound}
    \sigma \leq \sqrt{\frac{2}{M}}\log(1+\frac{\epsilon}{\norm{\obs}_{\infty}})\,.
\end{equation}

As the allowable error is small $\epsilon \ll 1$, by approximating the logarithm $\log(1+x) \approx x$, one derives that the perturbations must follow the scaling
\begin{equation}
    \sigma \in \order{\frac{\epsilon}{M^{1/2}\norm{\obs}_\infty}}\, .
\end{equation}
Note that a similar scaling law was recently derived also in~\cite{ito2021universal}, though via a slightly different method based on the moment generating function of the probability distribution characterising the perturbations. 

\begin{figure}
    \centering
    \includegraphics[width=\columnwidth]{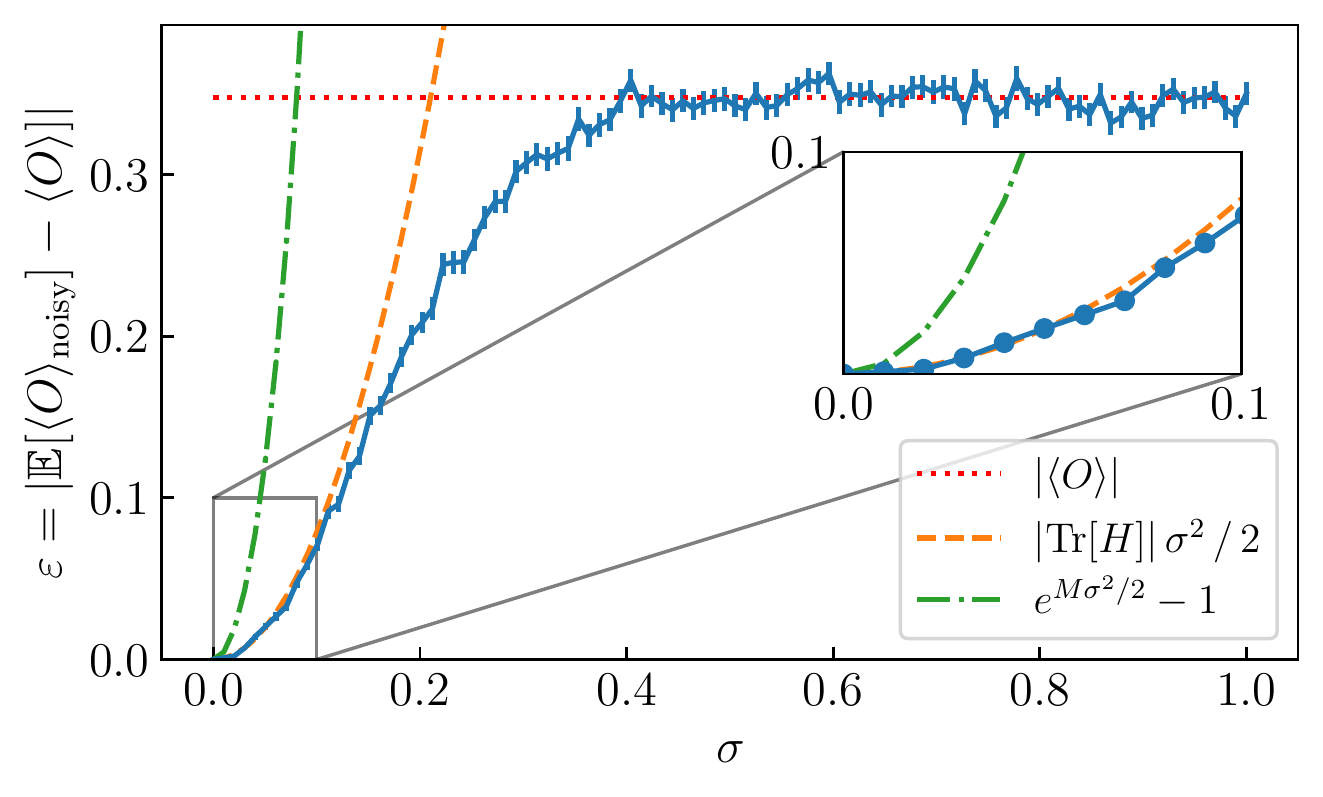}
    \caption{Effect of Gaussian coherent noise on the output of the parametrized quantum circuit shown in \Cref{fig:ansatzes}(b). The plot is obtained by first choosing a parameter vector $\bm{\theta}_0 \in \mathbb{R}^{92}$ corresponding to a the ideal noise-free expectation value $f(\bm{\theta}_0)=\expval{O}$ with $O = Z^{\otimes 4}$. With this baseline fixed, random Gaussian perturbations are added to the angles $\bm{\theta}_{noisy} = \bm{\theta}_0 + \delta\bm{\theta}$, and the resulting noisy expectation vales $\expval{O}_{noisy}$ are computed. Each point in the plot is the average over $N=10^5$ different perturbation vectors sampled from a multivariate Gaussian distribution of a given $\sigma$. The experiments are then repeated for increasing values of the noise strength $\sigma$. The error bars show the statistical error of the mean. For small noise levels, the output of the quantum circuit closely follows the behaviour predicted by \Cref{eq:taylor4}, where the Hessian is evaluated at the unperturbed value $H=H(\bm{\theta}_0)$. When the error is too large the circuit behaves as a random circuit whose output is on average zero, hence the error plateaus to the unperturbed expectation value $\varepsilon = \abs{\expval{O}} = \abs{f(\bm{\theta}_0)} $). The upper bound predicted by \Cref{eq:gaussian_error} is very loose in general, and holds tightly only for very small values of $\sigma\lessapprox 0.01$.}
    \label{fig:gaussian_perturbation}
\end{figure}

To provide an example, assume one is willing to tolerate an error of $\epsilon = 10\%$, that $\norm{\obs}_{\infty} = 1$ as for measuring a Pauli operator and that the PQC consists of $M=100$ noisy parametrized gates, then one can be sure of such accuracy if $\sigma \sim 0.1 / \sqrt{100} = 0.01$. However, we stress again that the scaling \Cref{eq:error_bound} is only a \textit{sufficient} but not necessary condition for achieving an error $\epsilon$. In fact, apart from the requirement of bounded derivatives, \Cref{eq:error_bound} is agnostic with respect to the specifics of the function, and such bound can be quite loose in real instances where a much larger noise level still causes a small error, as shown in \Cref{fig:gaussian_perturbation}. 

In \Cref{fig:gaussian_perturbation}, we report simulation results obtained by simulating the parametrized ansatz depicted in \Cref{fig:ansatzes}(b) subject to Gaussian coherent noise of increasing strength. It is clear that the output of the circuit closely follows the approximation of \Cref{eq:taylor4} given by the Hessian even at moderately large value of the noise $\sigma \lessapprox 0.15$. When the noise is too strong ($\sigma > 0.2$), the circuit becomes essentially random, and the average expectation value when measuring a Pauli operator is zero. This is a consequence of PQCs often behaving like unitary designs upon random initialization of the parameters~\cite{McCleanBarren2018, CerezoBarrenLocalCost2021}, a fact which we discuss in detail in Sec.~\ref{sec:gaussian_noise_resilience}. At last, as discussed earlier, while the upper bound~\eqref{eq:gaussian_error} holds, it is indeed very loose and only holds tightly at small $\sigma \lessapprox 0.01$. 

We now proceed discussing why hardware-efficient parametrized quantum circuits can be resilient to Gaussian coherent noise. Roughly, this is because such circuits are found to behave like random unitaries upon random assignment of the parameters, which implies that the derivatives of such circuits tend to vanish as the system size grows large~\cite{CerezoHigherOrder2021}.  

\subsection{Resilience of Hardware-Efficient ansatzes to Gaussian coherent noise \label{sec:gaussian_noise_resilience}}

The previous analysis showed that Gaussian perturbations induce an error depending on the Hessian of the circuit (see \Cref{eq:taylor4}), so that up to fourth order in the perturbation it holds that
\begin{equation}
\label{eq:hessian_correction}
    \mathbb{E}[f(\bm{\theta}+\delta\bm{\theta})] \approx f(\bm{\theta}) + \frac{1}{2}\sigma^2 \Tr[H(\bm{\theta})]\,.
\end{equation} 

This equation tells us that if the optimization landscape is flat or close to being flat, then the Hessian is small, and so the perturbation will have little effect on the output of the circuit. On the contrary, in the presence of a very curved landscape, noise will have a great impact and the output of the circuit may change sensibly. It is known that the curvature of the optimization landscape produced by a PQC is closely related to the barren plateau phenomenon \cite{McCleanBarren2018,CerezoBarrenLocalCost2021, Holmes2021connecting}, where the variance of the first and second derivative vanishes exponentially in the number of qubits and layers in a random circuit. Additionally, the hardware-efficient ansatz we use for some of the environments in this work is known to suffer from barren plateaus when the system size is large. As the curvature of the optimization landscape of these types of circuits is very flat, it can also be expected that the type of noise induced by the Gaussian perturbations on parameters that we study in this work should not affect circuits that generally produce small first and second order derivatives. While circuits that are in the barren plateau regime are obviously undesirable as they quickly become untrainable, one can consider circuits of the size such that the variance in gradients is relatively small, but the circuit has not yet converged to an approximate 2-design, as shown in \cite{McCleanBarren2018}. We make this statement more formal in the following.

We can use standard results on averages of unitary designs \cite{HuangPredicting2020, HaarPuchala2017} to characterize the Hessian of hardware-efficient circuits, and thus gain insight on their performance under Gaussian noise. We report the main results of our analysis here, full derivations can be found in \Cref{app:gaussian_noise_resilience}. In the following, we suppose that sampling a random value of the parameter vector $\bm{\theta}$ in the parametrized circuit $U(\bm{\theta})$, is equivalent to sampling a unitary from a unitary 2-design, defined as a set of unitary matrices that match the Haar distribution up to the second moment. Also, we consider observables $O$ being Pauli strings, so that $\Tr[O] = 0$ and $\Tr[O^2] = 2^n$. In order to distinguish from the previous notation where averages were computed over the Gaussian distribution of the perturbations, we use $\mathbb{E}_{U}[\cdot]$ and $\text{Var}_{U}[\cdot]$ to denote average values and variances evaluated over the random unitaries.

Then, under reasonable and usual assumptions on parts of the parametrized quantum circuit being 2-designs, it is possible to show that the diagonal elements of the Hessian $H_{ii} = \partial^2 f(\bm{\theta})/\partial\theta_i^2$ satisfy~\cite{CerezoHigherOrder2021} (see also Appendix~\ref{app:gaussian_noise_resilience} for an explicit derivation)
 \begin{equation} 
 \label{eq:haar_Hii}
    \mathbb{E}_{U}\qty[H_{ii}] = 0\,,\quad \text{Var}_{U}\qty[ H_{ii} ]  \in \order{\frac{1}{2^n}}.
 \end{equation}
That is, in addition to first order derivatives, also second order derivatives of random parameterized quantum circuits are found to be zero on average, and with a variance which is exponentially vanishing. 

Starting from the results above, one can calculate the statistics of the trace of the Hessian, for which it holds
\begin{equation}
\label{eq:haar_TrH}
    \mathbb{E}_U\qty[\Tr[H]] = 0\,,\quad \text{Var}_U\qty[\Tr[H]] \lessapprox \frac{M^2}{2^n}\,.
\end{equation}
Furthermore, our numerical simulations suggest that the variance of the trace of the Hessian is actually smaller, and is well captured by the following expression
\begin{equation}
\label{eq:approxtrh}
    \text{Var}_U\qty[\Tr[H]] \approx \frac{M(M+1)}{4(2^n+1)} \approx \frac{1}{4}\frac{M^2}{2^n}\, ,
\end{equation}
a fact which we justify and discuss in \Cref{app:TrHSec}. 

\begin{figure}
    \centering
    \includegraphics[width=\columnwidth]{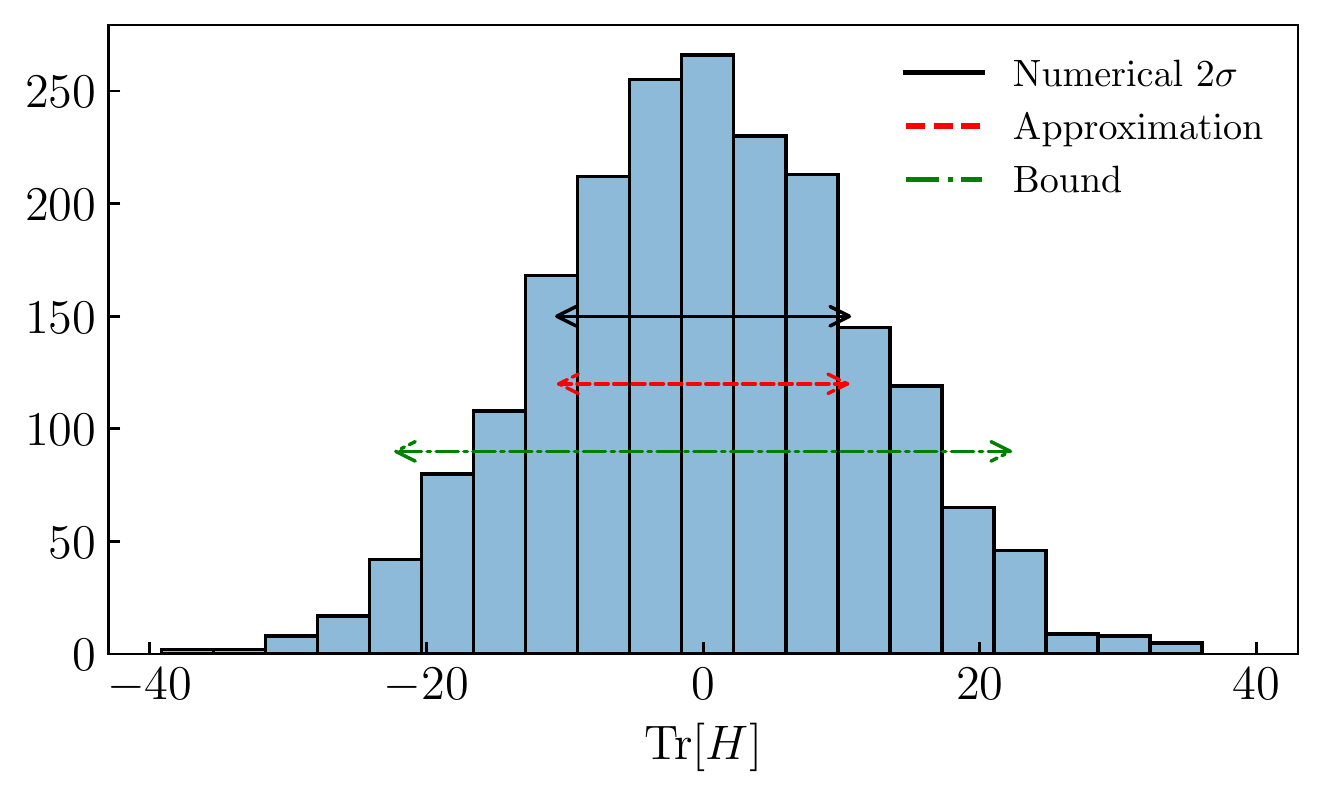}
    \caption{Simulation results of evaluating the trace of the Hessian matrix for the circuit shown in Fig.~\ref{fig:ansatzes}(b) with random assignments of the parameters and $O = Z^{\otimes 4}$. The simulations are performed by sampling $2000$ random parameter vectors $\{\bm{\theta}_m\}_{m=1}^{2000}$ with $\theta_i \sim \text{Unif}[0,2\pi[$ and then evaluating the trace of the corresponding Hessian matrix $\Tr[H(\bm{\theta}_m)]$. These values are used to build the histogram showing the frequency distribution of $\Tr[H]$. The length of the arrows are, respectively: ``Numerical $2\sigma$" (black solid line) twice the  numerical standard deviation, ``Approximation" (dashed red) twice the square root of the approximation in Eq.~\eqref{eq:approxtrh}, ``Bound" (dashed-dotted green) twice the square root of the upper bound in Eq.~\eqref{eq:haar_TrH}.}
    \label{fig:hessians_distribution}
\end{figure}

In \Cref{fig:hessians_distribution} we report simulation results of evaluating the trace of the Hessian matrix for the circuit shown in~\Cref{fig:ansatzes}(b). The histogram represents the frequency of obtaining a given value of the trace of the Hessian $\Tr[H(\bm{\theta})]$ upon random assignments of the parameters. Indeed, there is a very good agreement between the variance obtained via numerical simulations (black solid line), and the one calculated with the approximation~\eqref{eq:approxtrh} (dashed red line).

\begin{figure*}
  \subfloat[training performance]{\includegraphics[scale=0.52]{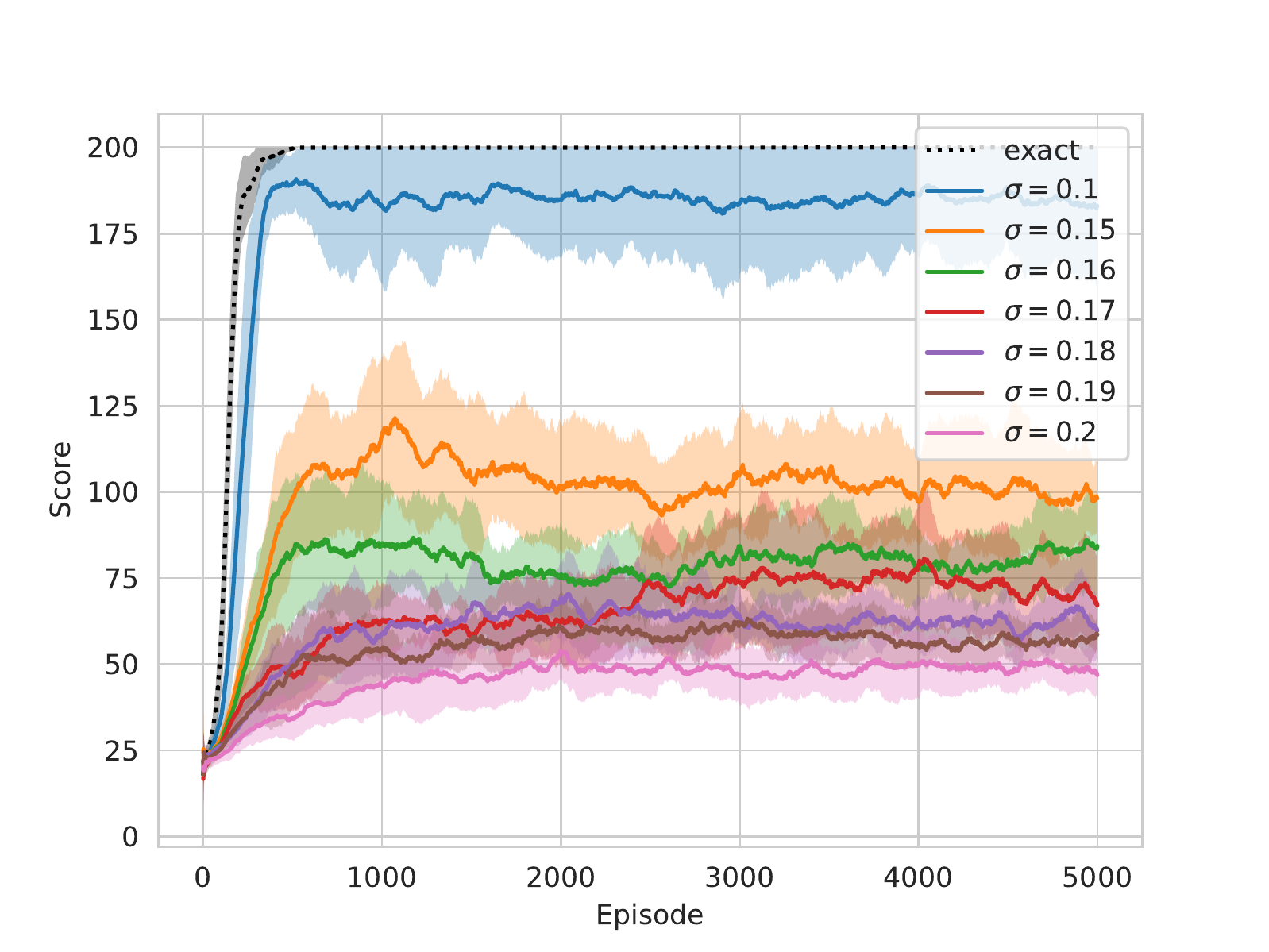}}\quad
  \subfloat[evaluation performance]{\includegraphics[scale=0.48]{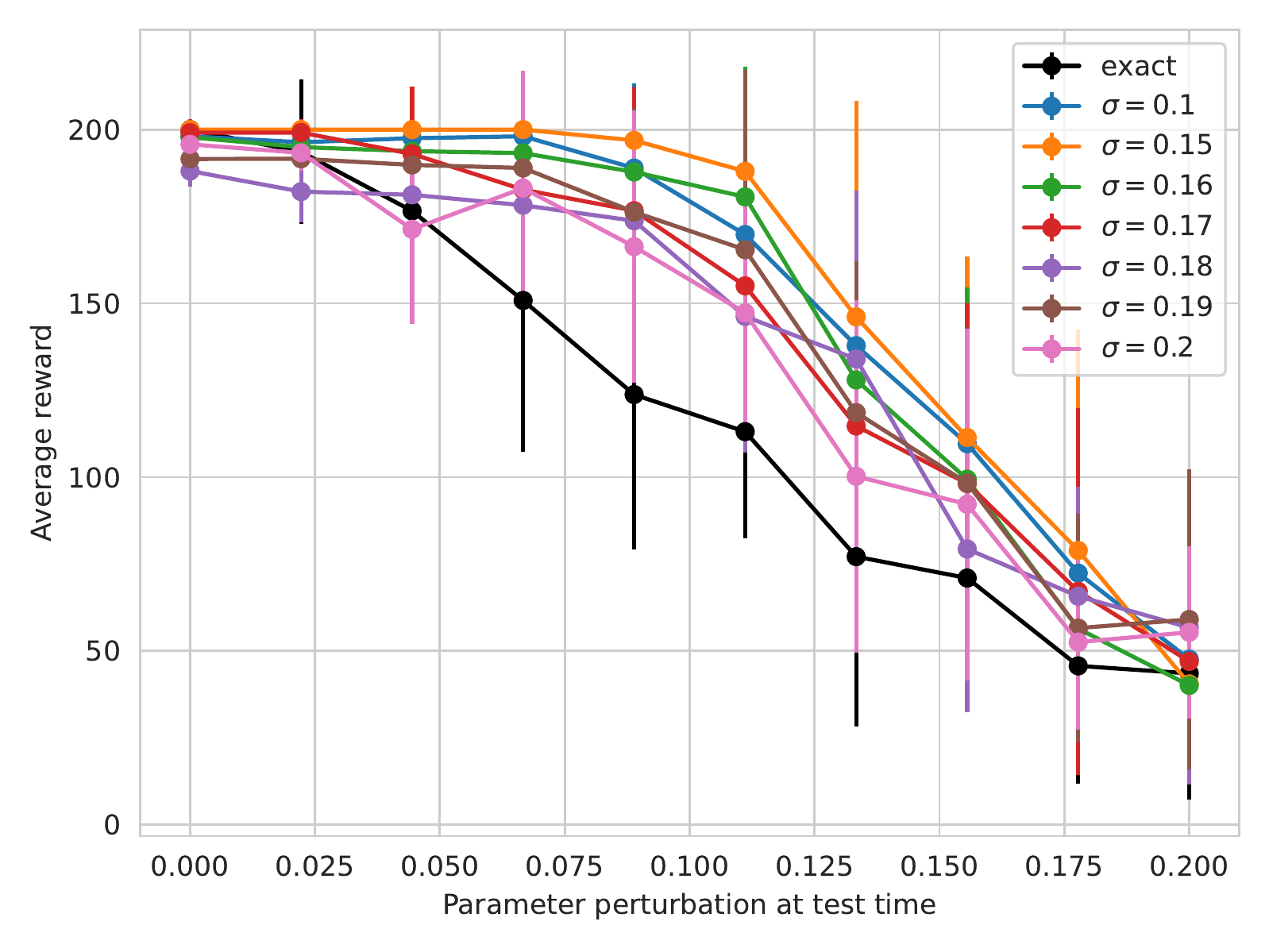}}\quad
  \caption{Q-learning agents on the CartPole environment trained and evaluated at varying perturbations $\sigma$. Panel (a) shows training performance, while panel (b) shows the performance of the same agents after training and evaluated under different perturbation levels than those present during training. Each point is computed as the average score of the 10 agents under the perturbation indicated on the x-axis.}
\label{fig:cp_q_gaussian}
\end{figure*}

\begin{figure*}[ht]
  \subfloat[training performance]{\includegraphics[scale=0.52]{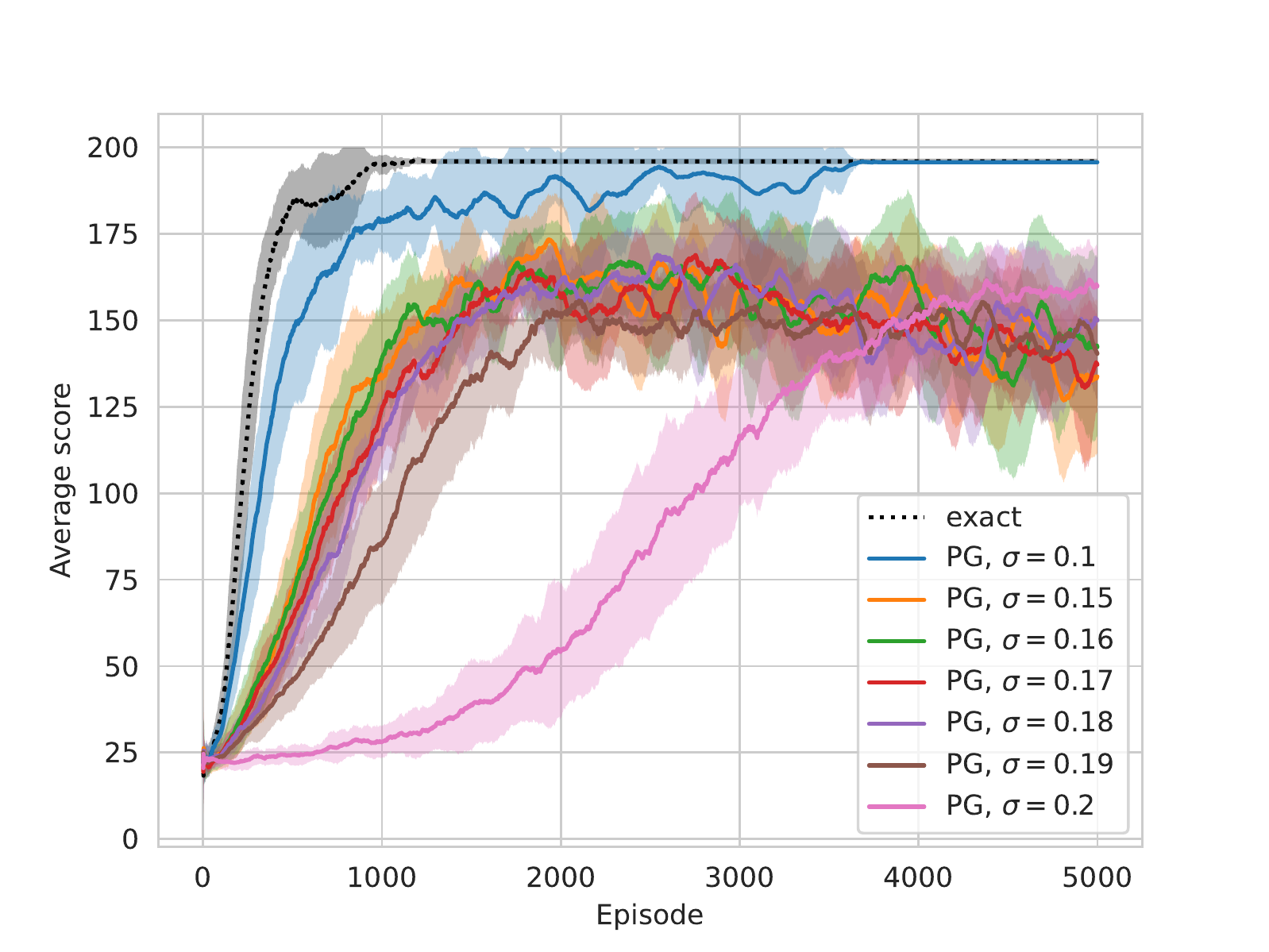}}\quad
  \subfloat[evaluation performance]{\includegraphics[scale=0.48]{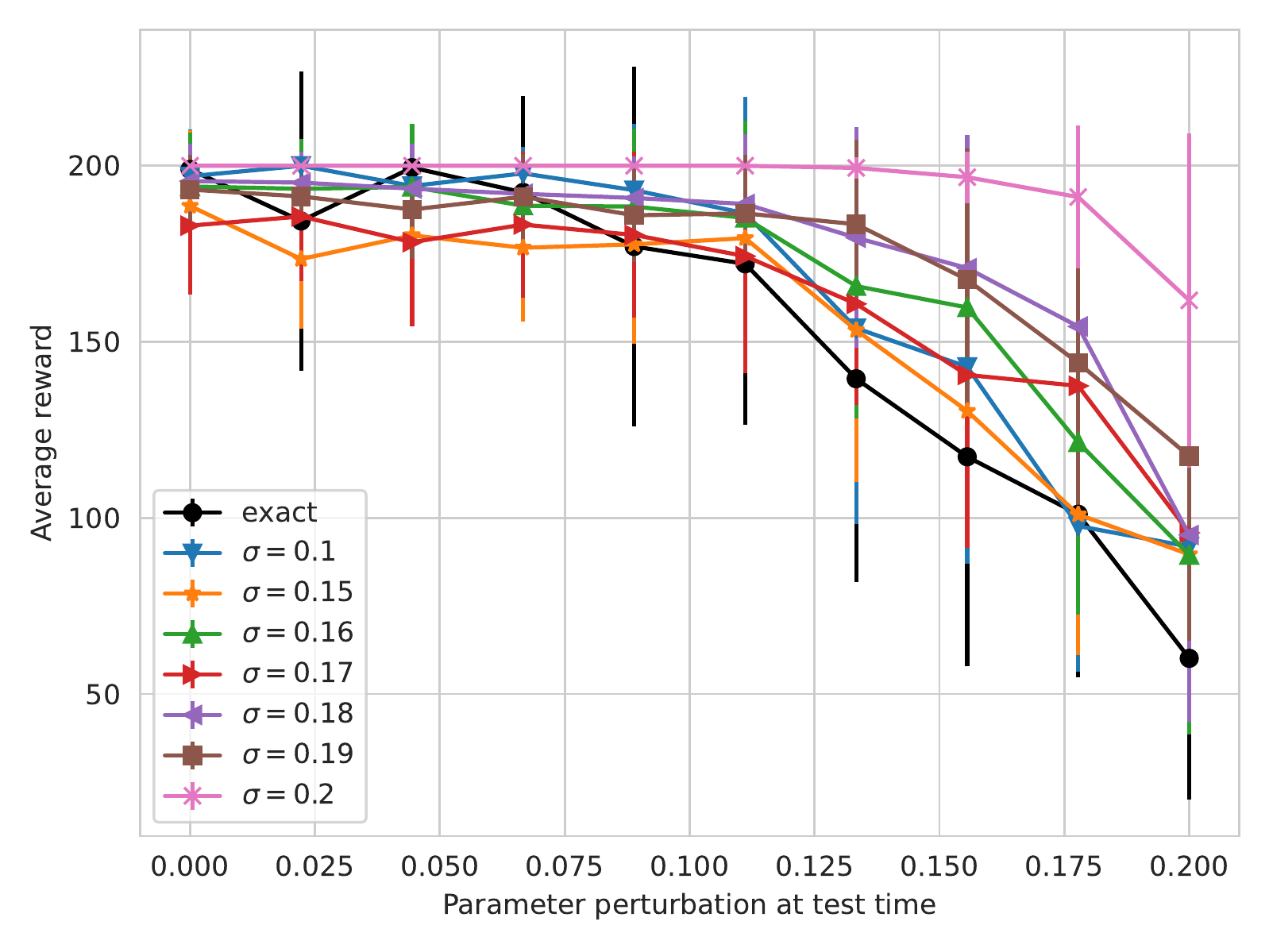}}\quad
  \caption{Policy gradient agents on the CartPole environment trained and evaluated at varying perturbations $\sigma$. Panel (a) shows training performance, while panel (b) shows the performance of the same agents after training and evaluated under different perturbation levels than those present during training. Each point is computed as the average score of the 10 agents under the perturbation indicated on the $x$-axis.}
\label{fig:cp_pg_gaussian}
\end{figure*}

The circuit used has $M=92$ parameters and $n=4$ qubits, and plugging these values in \Cref{eq:approxtrh} yields a standard deviation $\sigma_U = \text{Std}_U\qty[\Tr[H]] \approx 11$. Then, if the behaviour of the PQCs in practical scenarios is well described by its random parameter regime, one expects the trace of the Hessian to be on average zero and in general not much bigger (in absolute value) than $\sigma_U \approx 11$. With this order of magnitude for the trace, the first order correction \Cref{eq:hessian_correction} even with a Gaussian noise level of $\sigma = 0.1$ is very small, as it amounts to 
$$
\abs{\mathbb{E}[f(\bm{\theta}+\delta\bm{\theta})] - f(\bm{\theta})} \approx \frac{1}{2}\sigma^2 \abs{\Tr[H(\bm{\theta})]} \approx 0.05\,.
$$

Summing up, for those PQCs whose cost landscape is close to being flat, then Gaussian perturbations on the variational parameters will have a limited impact on the output of the quantum circuit. 

\begin{figure*}
  \subfloat[pole angle, cart position (PG)]{\includegraphics[scale=0.33]{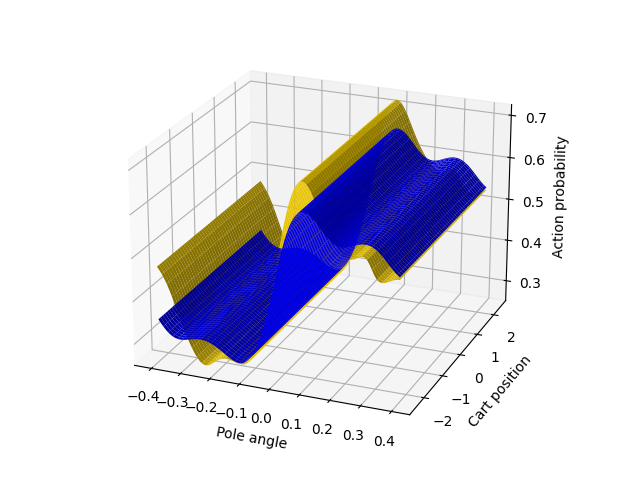}}\quad
  \subfloat[pole angle, cart velocity (PG)]{\includegraphics[scale=0.33]{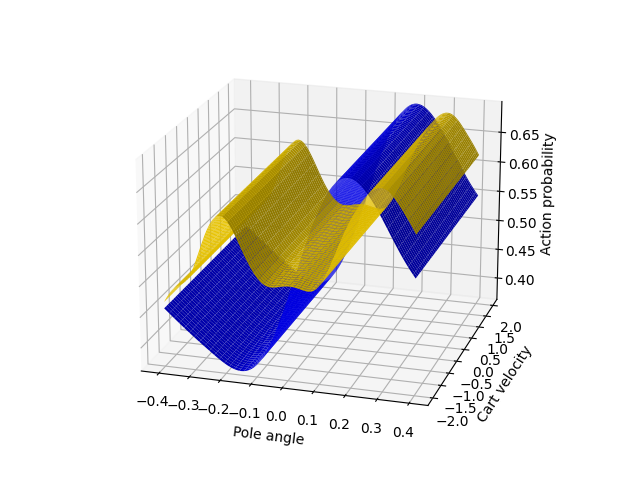}}\quad
  \subfloat[pole angle, pole velocity (PG)]{\includegraphics[scale=0.33]{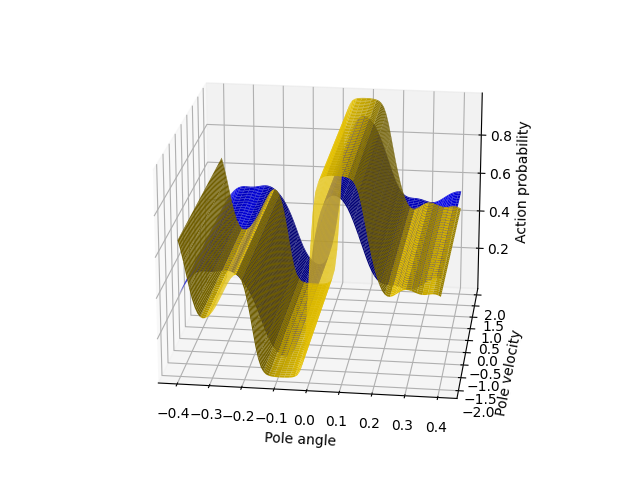}}\quad
  \subfloat[pole angle, cart position (QL)]{\includegraphics[scale=0.33]{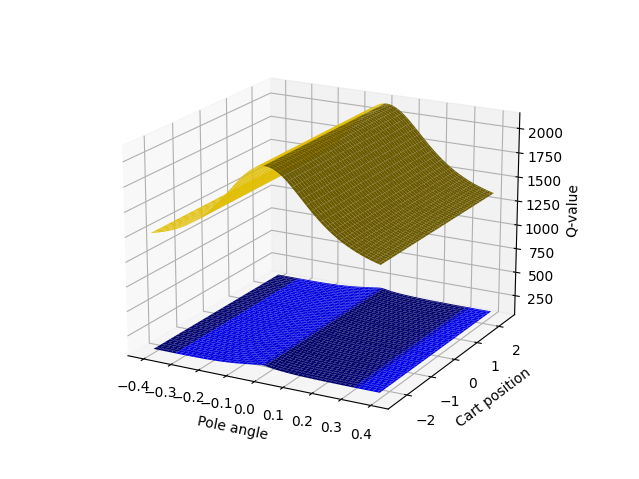}}\quad
  \subfloat[pole angle, cart velocity (QL)]{\includegraphics[scale=0.33]{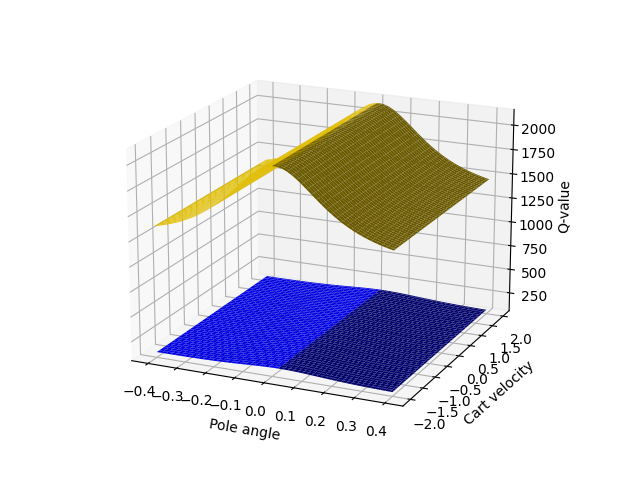}}\quad
  \subfloat[pole angle, pole velocity (QL)]{\includegraphics[scale=0.33]{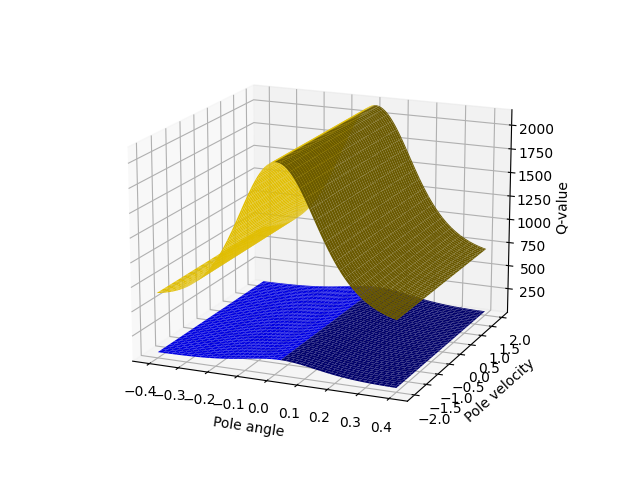}}\quad
  \caption{Comparison of average learned policies (PG) and Q-functions (QL) of agents from~\Cref{fig:cp_q_gaussian} and~\Cref{fig:cp_pg_gaussian}, in the noise-free setting (blue) and with a perturbation level $\sigma=0.2$ (yellow).}
\label{fig:res_cp_policies}
\end{figure*}

\subsection{Numerical results}
\subsubsection{CartPole}\label{sec:res_cp_gaussian}

First, we evaluate the performance of policy gradient and Q-learning algorithms when Gaussian perturbations are applied at each circuit evaluation during training. In \Cref{fig:cp_q_gaussian} (a) and (b), we show the training and evaluation performance, respectively, of Q-learning agents in the CartPole environment with perturbations in the range $\sigma \in \{0, 0.1, 0.15, 0.16, 0.17, 0.18, 0.19, 0.2\}$. 
Only the agent trained with noise level $\sigma=0.1$ learns the environment successfully and remains close to optimal performance. As suggested by our theoretical analysis in \Cref{sec:gauss_effect_output}, performance starts to degrade as we consider higher perturbations of $\sigma > 0.1$, and none of those agents manage to achieve a better performance than a score of 125 on average. In~\Cref{fig:cp_q_gaussian} (b) we evaluate the performance of trained agents when they act in an environment with different perturbation levels than those present when they were trained. Even agents that do not perform well during training achieve close to optimal performance when evaluated in the noise-free setting. This suggests that despite their bad training performance due to the added perturbations, these agents still learn a good Q-function. Notably, the agents trained without noise perform worst when they are evaluated under various levels of perturbations.

Results for agents trained with the policy gradient method are shown in \Cref{fig:cp_pg_gaussian} (a). While again only the agents trained with a perturbation of $\sigma=0.1$ perform well and even reach optimal performance, agents with higher perturbations also largely stay close to optimal performance with a final score of 125 on average. Even the agent trained with a relatively high $\sigma=0.2$ is robust in this setting, even though it requires by far the most training episodes to get to a good score. This positive trend is also visible in~\Cref{fig:cp_pg_gaussian}(b), where we see that all agents achieve close to optimal performance when evaluated with perturbation levels $\sigma \leq 0.1$, which is again in line we our theoretical analysis in \cref{sec:gauss_effect_output}. The difference between agents trained with Gaussian perturbations and those trained without is not as large as in the Q-learning setting, and at evaluation time both algorithms perform similarly. Another observation about the policy gradient agents is that those trained with $\sigma=0.2$ achieve optimal or close to optimal performance in the environment under various perturbation levels at evaluation time, and are the most robust out of all agents trained in this setting. Overall, the policy gradient method shows a larger resilience to Gaussian noise in our experiments for the CartPole environment. It is an open question why this is the case, however, we did not observe better performance of the policy gradient algorithm under noise in general, as results in later sections will show.

\begin{figure*}
  \subfloat[training performance]{\includegraphics[width=\columnwidth]{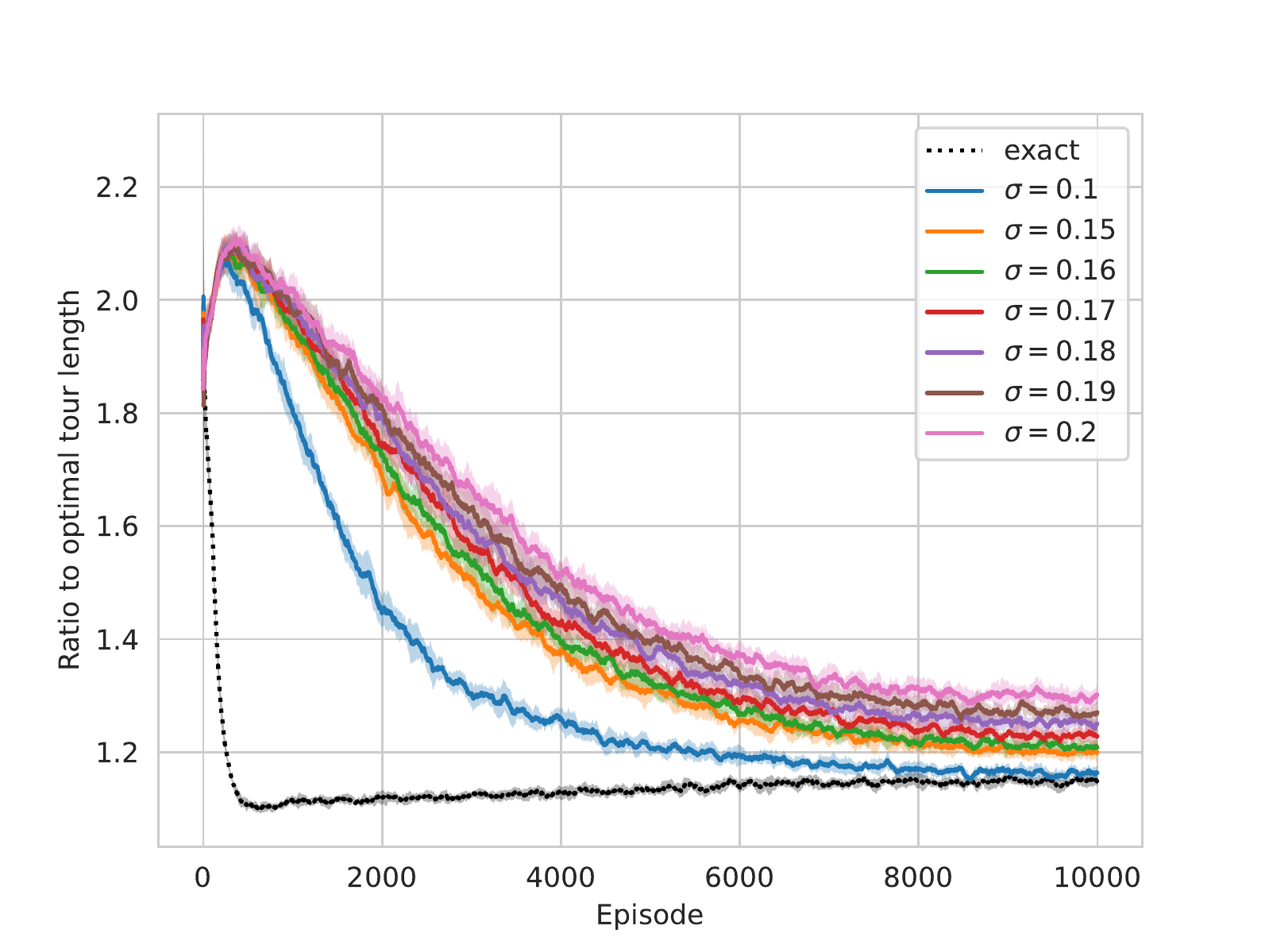}}\quad
  \subfloat[evaluation performance]{\includegraphics[width=\columnwidth]{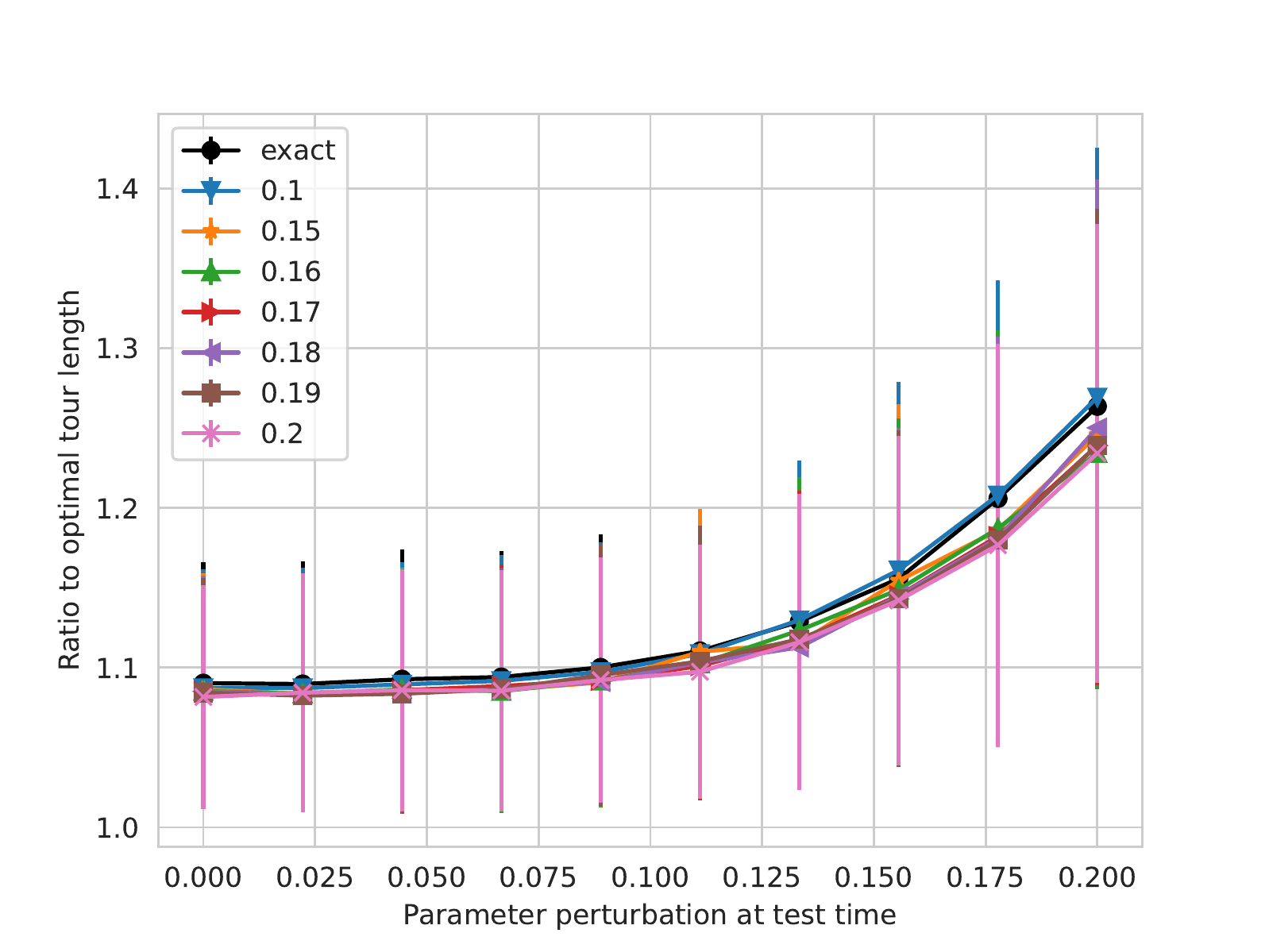}}\quad
  \caption{Training and evaluation of Q-learning agents in the TSP environment under various perturbations $\sigma$. Panel (a) shows the effect of perturbations during training, panel (b) shows results for the same agents evaluated on varying perturbation levels after training, different to those present at training time.}
\label{fig:res_tsp_q_gaussian}
\end{figure*}

\begin{figure*}
  \subfloat[training performance]{\includegraphics[width=\columnwidth]{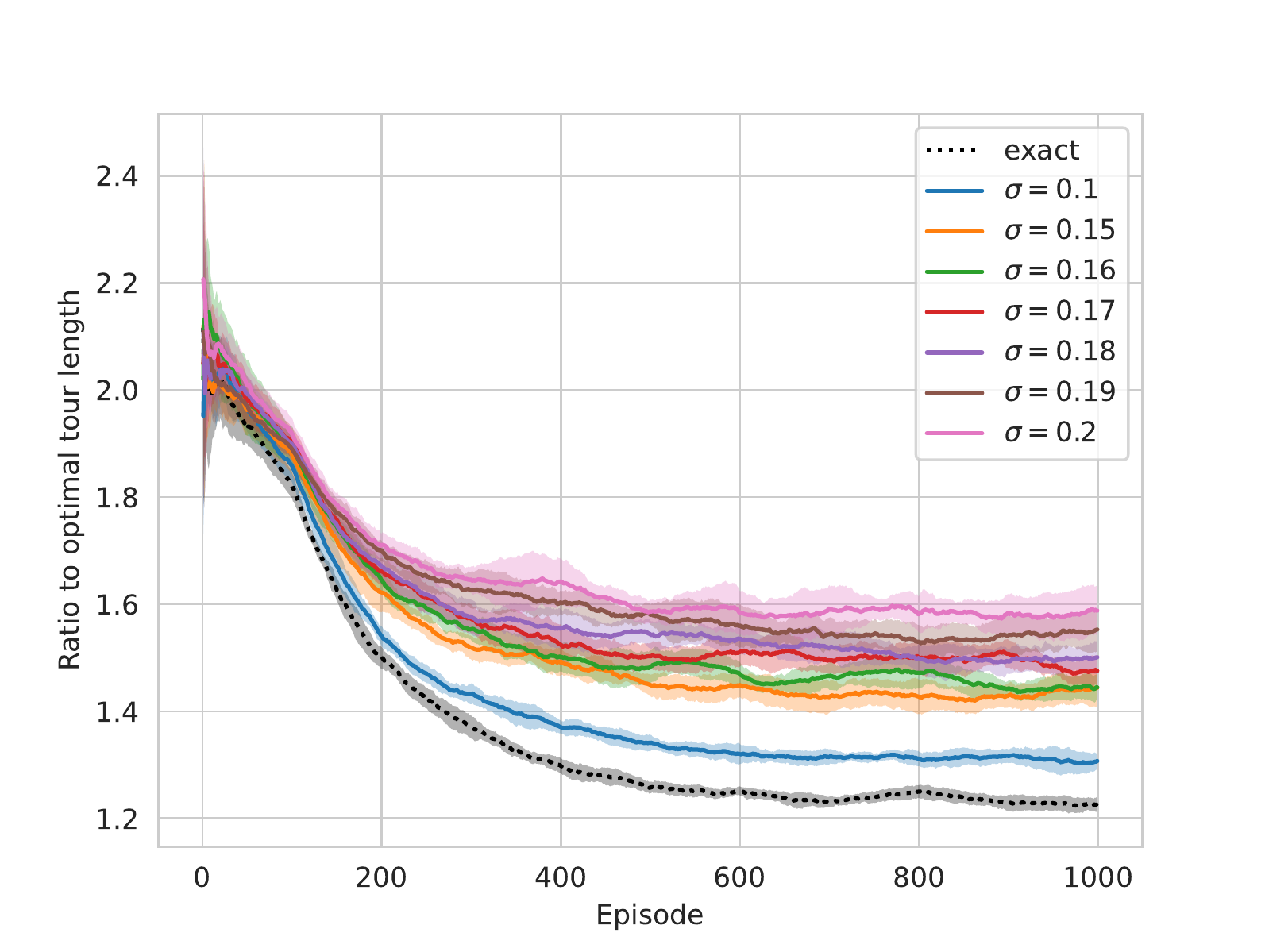}}\quad
  \subfloat[evaluation performance]{\includegraphics[width=\columnwidth]{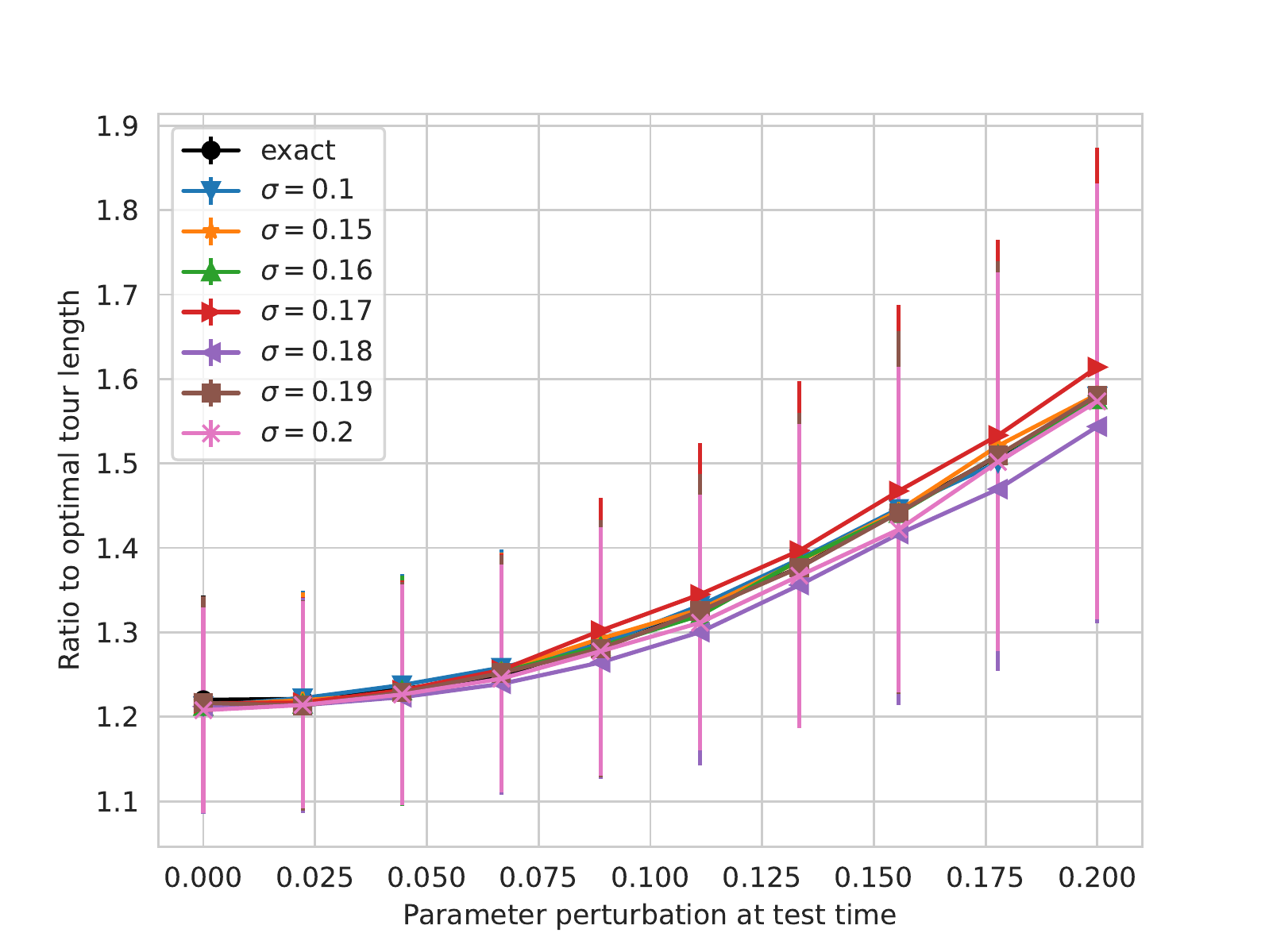}}\quad
  \caption{Training and evaluation of policy gradient agents in the TSP environment under various perturbations $\sigma$. Panel (a) shows the effect of perturbations during training, panel (b) shows results for the same agents evaluated on varying perturbation levels after training, different to those present at training time.}
\label{fig:res_tsp_pg_gaussian}
\end{figure*}

In addition to studying the performance of Q-learning and policy gradient agents at training and evaluation time, we visualize the learned policies and Q-functions of both in the noisy and noise-free setting in \Cref{fig:res_cp_policies}. As learned policies and Q-functions can look different even when training the same agent twice, we show averages of the ten agents shown in \Cref{fig:cp_q_gaussian} and \Cref{fig:cp_pg_gaussian} for both algorithms, and for perturbation levels of $\sigma=0$ (blue) and $\sigma=0.2$ (yellow), respectively. The CartPole environment has four inputs: cart position and velocity, and pole angle and velocity. To visualize the learned policies and Q-functions, we show the probabilities and Q-values for taking the action ``right'' as a function of pairs of state values. The state inputs that are not in the figure are set to zero, and for the sake of clarity we do not apply perturbations to the parameters when visualizing the policy. In \Cref{fig:res_cp_policies} (a)-(c), we see results for policy gradient agents. Overall, it can be seen that the agents trained without perturbations learn smoother policies, hence for most states there is a clear decision on which action to take. Training with perturbations makes the policies slightly more rippled, but they still mostly follow the contours of the policy learned under ideal conditions. 

The approximated Q-functions can be seen in \Cref{fig:res_cp_policies}(d)-(f). One observation we make here is that the range that Q-values take blows up considerably compared to the noise-free setting. This is due to the trainable output weights that the expectation values are multiplied with in the Q-learning setting (see \Cref{sec:environments_implementation}) becoming considerably larger for agents trained in the noisy setting. However, as we can see in the Appendix in \Cref{fig:vis_q_noisy_free}, the shapes of the learned Q-functions of the noise-free and noisy agents are still very similar, which explains why even the agents trained with $\sigma = 0.2$ perform almost optimally when evaluated without perturbations in \Cref{fig:cp_q_gaussian} (b). We also note that the range of Q-values of both the noisy and noise-free agents is much larger than the range of optimal Q-values given in~\cite{skolik2022quantum}. This can be understood as the agent consistently overestimating the expected return, a problem known to arise in classical Q-learning, and which is exacerbated by noise \cite{van2016deep}. However, the authors of \cite{skolik2022quantum} also point out that in the function approximation setting, it is more important to learn the order of Q-values for each state (i.e., preserving that the argmax Q-value corresponds to the optimal action) than learning a close representation of the optimal Q-values.

\subsubsection{TSP}

In this section, we study the performance of Q-learning and policy gradient algorithms with Gaussian coherent noise in the TSP environment. Panels (a) and (b) in \Cref{fig:res_tsp_q_gaussian} show the training and evaluation performance of Q-learning agents in this environment under perturbations in the range $\sigma \in \{0, 0.1, 0.15, 0.16, 0.17, 0.18, 0.19, 0.2\}$. We note that the Q-learning agents trained without noise already converge after 600 episodes on average, but to get an equal runtime in terms of episodes for all settings, we also let them run for 10000 episodes. This unnecessarily long runtime causes the optimizer to leave the local minimum again, which we ignore as an artifact here and consider the lowest average approximation ratio for the comparison with the other models. 

For the TSP environment, we observe that with increasing levels of Gaussian perturbations, convergence of agents is delayed and their final approximation ratio becomes worse compared to the noise-free agents' performance. Still, all agents seem to learn very similar policies despite being trained with different settings of $\sigma$, as we can see by their almost identical performance at evaluation time shown in \Cref{fig:res_tsp_q_gaussian} (b). Despite a drop in performance during training, the final performance of the models on a test set of  previously unseen TSP instances stays almost unaffected by the noise present during training. While we see that agents trained with more noise seem to learn more noise-robust policies as in the case of the CartPole environment, this effect is not as pronounced here. Additionally, we again see that performance of trained models in \Cref{fig:res_tsp_q_gaussian} (b) starts to drop at $\sigma > 0.1$, as indicated by our theoretical analysis in \Cref{sec:gauss_effect_output}. While the policy gradient method shows a certain robustness to noise during training in the CartPole environment, this is not the case for the TSP environment, as we show in \Cref{fig:res_tsp_pg_gaussian} (a). The only agent that gets close in performance to the noise free agent is the one trained with $\sigma=0.1$, while higher perturbations yield agents that are relatively bad with an approximation ratio between 1.4 and 1.6 on average. However, again, all agents seem to learn similar policies as indicated by their test performance in \Cref{fig:res_tsp_pg_gaussian} (b). Similar to CartPole, the agents' performance on the test set under varying perturbation levels closely matches that of the noise-free agents, and again we see a large drop in performance for perturbations that are higher than $\sigma = 0.1$. 

Overall, the Q-learning algorithm performs better in the TSP environment than the policy gradient method. The optimal tour for each TSP instance is deterministic, so using a stochastic policy as in the policy gradient approach introduces an additional source of error, as there is always a non-zero probability to chose a non-optimal action. This leads to an increased susceptibility to the Gaussian perturbations present during the evaluation of the policy gradient algorithm. This is not the case for Q-learning, where choices are made based on the argmax Q-value. Additionally, the ansatz that we use does not separate between data encoding and trainable parameters as described in~\Cref{sec:environments_implementation}. As the optimal tour of a TSP instance does not change upon small perturbations of the edge weights, this leads to a relative robustness of this ansatz used in conjunction with Q-learning to Gaussian coherent noise in this environment.

\section{Incoherent noise}\label{sec:res_incoherent_noise}

The Gaussian perturbation noise that we studied in \Cref{sec:gauss_noise_section} is well-suited to model coherent errors due to imprecision in the control of the quantum device, but it does not reflect noise that results from undesired interactions of the quantum system with its environment. To study the effect of this type of incoherent noise we perform additional experiments  in this section.

We simulate this type of noise with TensorFlow Quantum (TFQ) \cite{Broughton2020TFQ}, where they are implemented through a Monte-Carlo trajectory sampling method \cite{CirqDepolarizing, GoogleNoisySim} that approximates the effect of noise by averaging over state vectors generated from a probabilistic application of the noise channel. This method of simulating noise essentially trades off the overhead in memory needed to store the $2^n \times 2^n$ sized density matrices necessary to simulate incoherent noise, with a runtime overhead. The precision of this approximation is determined by the number of repetitions, which specifies how many ``noisy'' state vectors are used. This adds a stochastic element to the simulation of the noise channels, and we get closer to simulating the exact noise model as the number of trajectories increases. Depending on the environments, we choose the number of trajectories so that it is possible to perform simulations in a reasonable time frame, and specify this number individually for each of the experiments below. We note that the runtime requirements for CartPole when simulating this type of noise are especially high, as the number of time steps in each episode, as well as the number of episodes itself depends strongly on the performance of the agent. In particular, agents that perform neither very well nor very badly, which are exactly the noise configurations we are interested in studying here, take especially long to simulate, as they do not converge early by solving the environment, but still take on the order of 100 time steps in each episode. Therefore we focus our attention mainly on the TSP environment in this section.

\subsection{Depolarizing noise}\label{sec:results_depol}

Depolarization noise affects a quantum state by either replacing it with the completely mixed state with probability $p$, or leaving it untouched otherwise \cite{NielsenChuang}. Let $\rho$ be the density matrix of a qubit, then depolarizing noise is defined by the map
\begin{equation}
\label{eq:depol_error}
    \mathcal{D}_p(\rho) = (1-p) \rho + p \frac{\mathds{1}}{2}.
\end{equation}
We model depolarization noise with Cirq \cite{CirqDepolarizing} and TFQ by appending a layer of local depolarizing channels to every qubit after each time step of the computation, where a time step is defined as the largest set of gates that can be implemented simultaneously. This implementation takes into account the possibility of cross-talk between qubits~\cite{ProctorMirrorRB_2022}. Also, note that while the use of depolarizing channels alone may not be a good approximation of real single qubits errors, it may become a good effective description of the overall noise process for the case where many qubits and layers are used \cite{Vovrosh2021GlobalDepolarizing}.

In our simulations, we assume that both single- and two-qubits gates are noisy, and consist of a composition of the ideal gates followed by \textit{local} depolarizing channels of equal probability $p$, acting independently on each qubit. In particular, the application of a depolarizing noise channel is implemented by performing one out of four actions at each circuit execution (trajectory): do nothing with probability $1-p$, or apply at random one of the three Pauli operators with probability $p$, and then average over the results. We remark that the average gate error of single-qubit gates in currently available superconducting quantum computing hardware is of the order of $r\lessapprox 0.01$, with gate fidelities exceeding $>99\%$. Finally, we note that one can relate the depolarisation strength $p$ to the average gate error $r$ over single qubit Cliffords, as measured by Randomized Benchmarking (RB)~\cite{MagnesanRB_2012, ProctorMirrorRB_2022, McKay3QubRB_2019} and commonly reported for quantum devices~\cite{AndersonQuantinuumQEC_2022, IBMQuantum}, via $r = p/2$. However, our circuits do not only use Cliffords, and moreover the RB's estimates for the gate error depend on the basis gates available on the device. Therefore, one should consider our simulations with depolarizing noise of strength $p$ as a proxy for a quantum device whose average error rate $r$ is of the same order of magnitude of $p$. While a single-qubit error noise model may not be accurate enough to closely mimic the behaviour of a real quantum device, it gives us the possibility to study the effect of single-qubit errors separately, before we go on to study a noise model that also includes two-qubit gate errors in \cref{sec:res_custom_noise}. 

\begin{figure}
\includegraphics[scale=0.5]{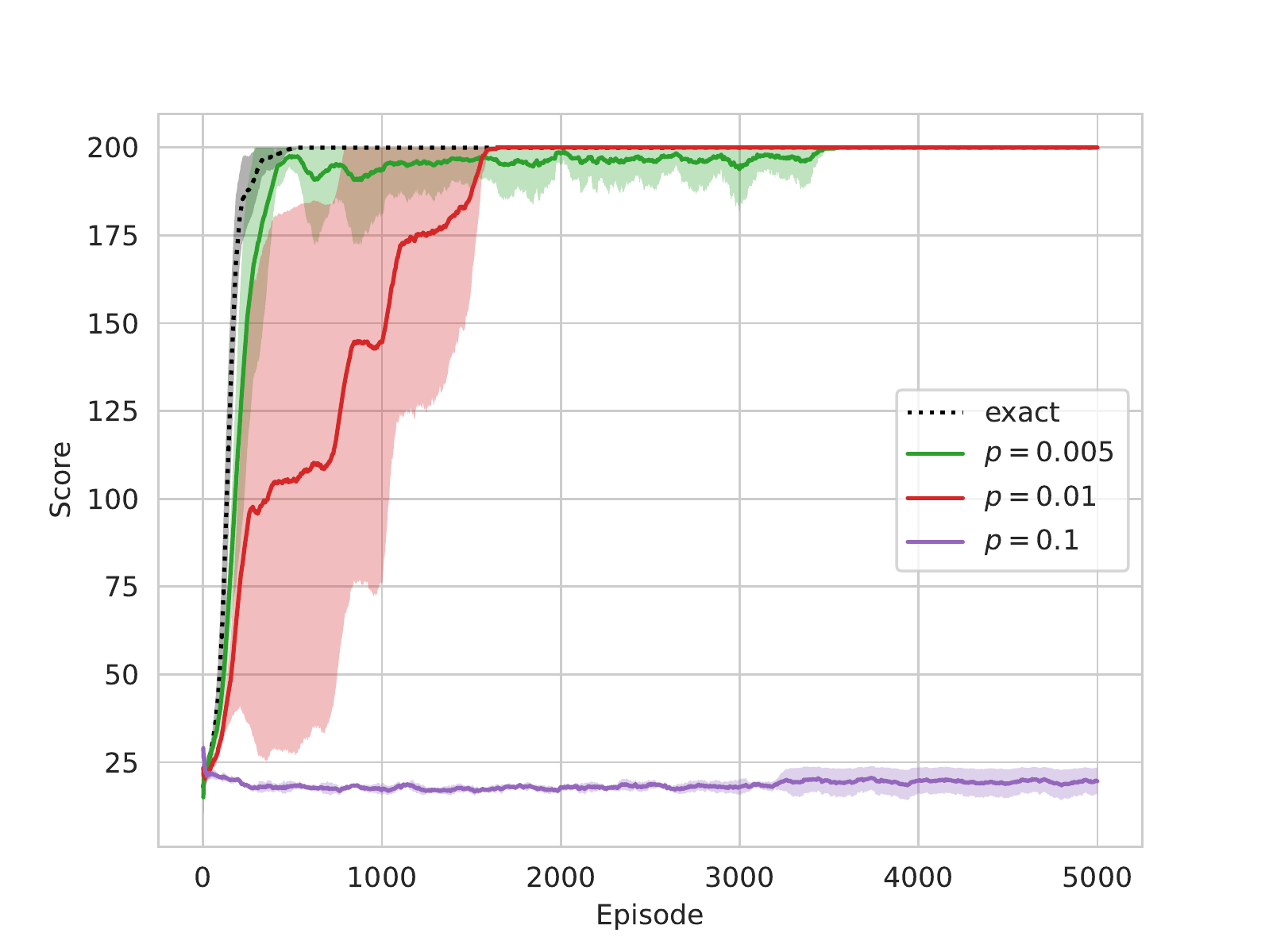}
\caption{Q-learning agents trained with varying probabilities $p$ of depolarization errors, and five layers of the circuit depicted in \Cref{fig:ansatzes} a). Noise is simulated with 100 Monte Carlo trajectories. The noisy curves are averaged over 5 agents, the exact one is averaged over 10 agents as in previous figures.}
\label{fig:res_cp_depolarizing}
\end{figure}

As mentioned above, simulating incoherent noise has high runtime requirements, so in the following we limit our studies to: (\textit{i}) Q-learning in the CartPole environment, and (\textit{ii}) the policy gradient method in the TSP environment. We pick these settings as they were the ones that were more sensitive to Gaussian coherent noise in our studies in~\Cref{sec:gauss_noise_section}, and in that sense represent the worst case instances from the previous section. To simulate the noisy quantum circuits, we use the Monte Carlo sampling as described above, where the number of trajectories used depends on the environment. As the CartPole environment requires a very high number of environment interactions (the better the agent, the more circuit evaluations are required per episode), we use 100 trajectories in this setting. In the TSP environment, the number of steps in each episode is constant and therefore we can use a higher number of 1000 trajectories and still perform simulations in a timely manner.

\begin{figure}
\includegraphics[scale=0.5]{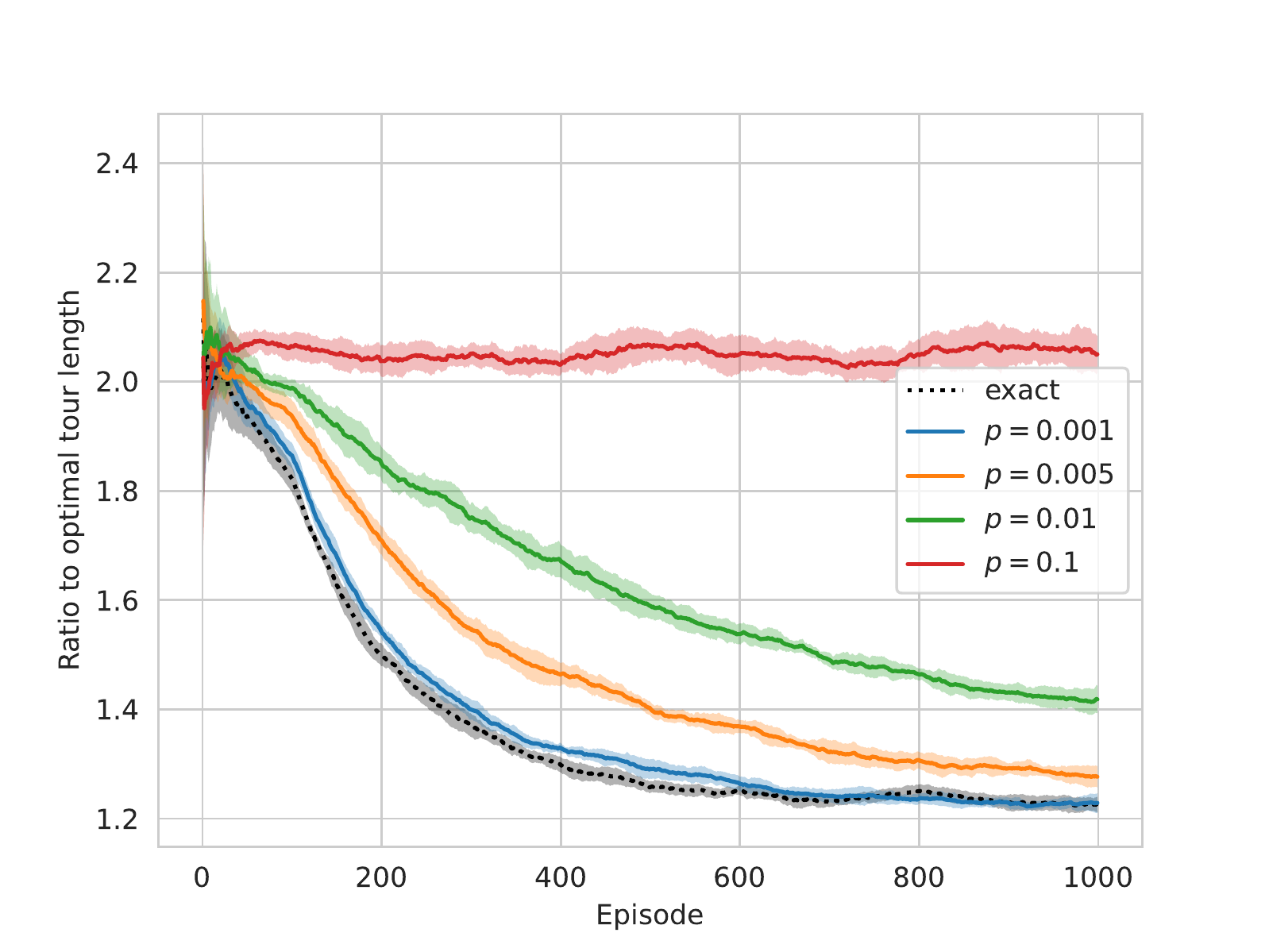}
\caption{Policy gradient agents trained in the TSP environment with varying probabilities $p$ of depolarization error, with one layer of the circuit depicted in \Cref{fig:ansatzes} c). Noise is simulated with 1000 Monte Carlo trajectories. All curves are averaged over 10 agents.}
\label{fig:res_tsp_depolarizing}
\end{figure}

\Cref{fig:res_cp_depolarizing} shows results of Q-learning agents trained in the CartPole environment with various error probabilities $p$. Agents with a realistic error probability of up to $p=0.01$ still solve the environment in less than 2000 episodes on average. Agents trained with error probability $p=0.005$ reach higher scores almost as quickly as agents trained in the noise-free setting, but stay somewhat unstable until they solve the environment after 3500 episodes on average. When the noise probability is increased to $p=0.1$, we see that agents fail to make any learning progress at all. 

\Cref{fig:res_tsp_depolarizing} shows the performance of the policy gradient method under one-qubit depolarization errors in the TSP environment. In this setting, agents trained with error probability $p=0.01$, as is a realistic assumption on current devices, perform noticeably worse than agents in the noise-free setting with a drop in approximation ration of around 0.2 on average. Only when we consider an error probability of $p=0.001$ do we get performance that is almost exactly the same as that in the noise-free case. Similar to the results of the Q-learning agent in the CartPole environment, agents trained with an error probability of $p=0.1$ show no meaningful learning progress. 

\subsection{Noise model based on current hardware}\label{sec:res_custom_noise}

After studying the effect of single-qubit depolarization errors in \Cref{sec:results_depol}, we now study the performance of the Q-learning algorithm in the TSP environment in the presence of a more realistic noise model that captures the behaviour of a near-term superconductive quantum device. The error sources we incorporate into this noise model are the following: single-qubit and two-qubit depolarization errors, single qubit amplitude damping error, and measurement noise. While hardware providers like IBM and Google offer the possibility of simulating noise models of specific devices, we do not want to take device-specific factors like qubit topology and native gate sets into account in this work, as the performance in these settings also depends strongly on the quality of the circuit compiled to the native gate set and qubit connectivity \cite{pelofske2022quantum}. Instead, we define a custom noise model based on gate fidelities published by hardware vendors, but do not take the above details into account. To determine realistic settings for the error probability of each noise source, we use calibration data published by IBM \cite{ibmq_experience} at the time of writing. The noise model used in our simulation is specified as follows:

\begin{itemize}
    \item \textbf{Depolarization error:}
    Single qubit depolarization channels with $p=0.001$ are applied after every single qubit gate. Two-qubit depolarization errors, defined by properly adjusting the definition in~\Cref{eq:depol_error}, with $p_2=0.01$ are applied after every two-qubit gate on the corresponding pair of qubits.
    
    \item \textbf{Amplitude damping error:} 
    Amplitude damping channels with decay parameter $\gamma = 0.003$ are applied after each single- and two-qubit gate on the corresponding qubits. Such a decay rate is valid for real devices having single qubit gate durations of $t=35\mathrm{ns}$, and average qubit decay times $T_1 \approx 100 \mu \mathrm{s}$, which correspond to a decay parameter of $\gamma = 1 - \exp(-t/T1) \approx 0.0003$. 
    
    \item \textbf{Measurement noise}
    Measurement errors are modeled by appending a bit-flip channel with probability $p=0.01$ to every qubit right before the measurement process.
\end{itemize}
We recall that the circuit ansatz for the TSP environment is the one depicted in~\Cref{fig:ansatzes}(c), where input information about the edge weights of the TSP instance is encoded by means of two-qubit gates. We therefore chose to study this ansatz in the context of a noise model that incorporates two-qubit errors, as we expect that these types of errors will affect performance of an ansatz that encodes crucial information in two-qubit gates more severely. Additionally, it is hard to perform simulations in this setting for the CartPole environment in a reasonable amount of time, as discussed above. For these reasons, we restrict our attention to the TSP environment in this section.

\begin{figure}
\includegraphics[scale=0.5]{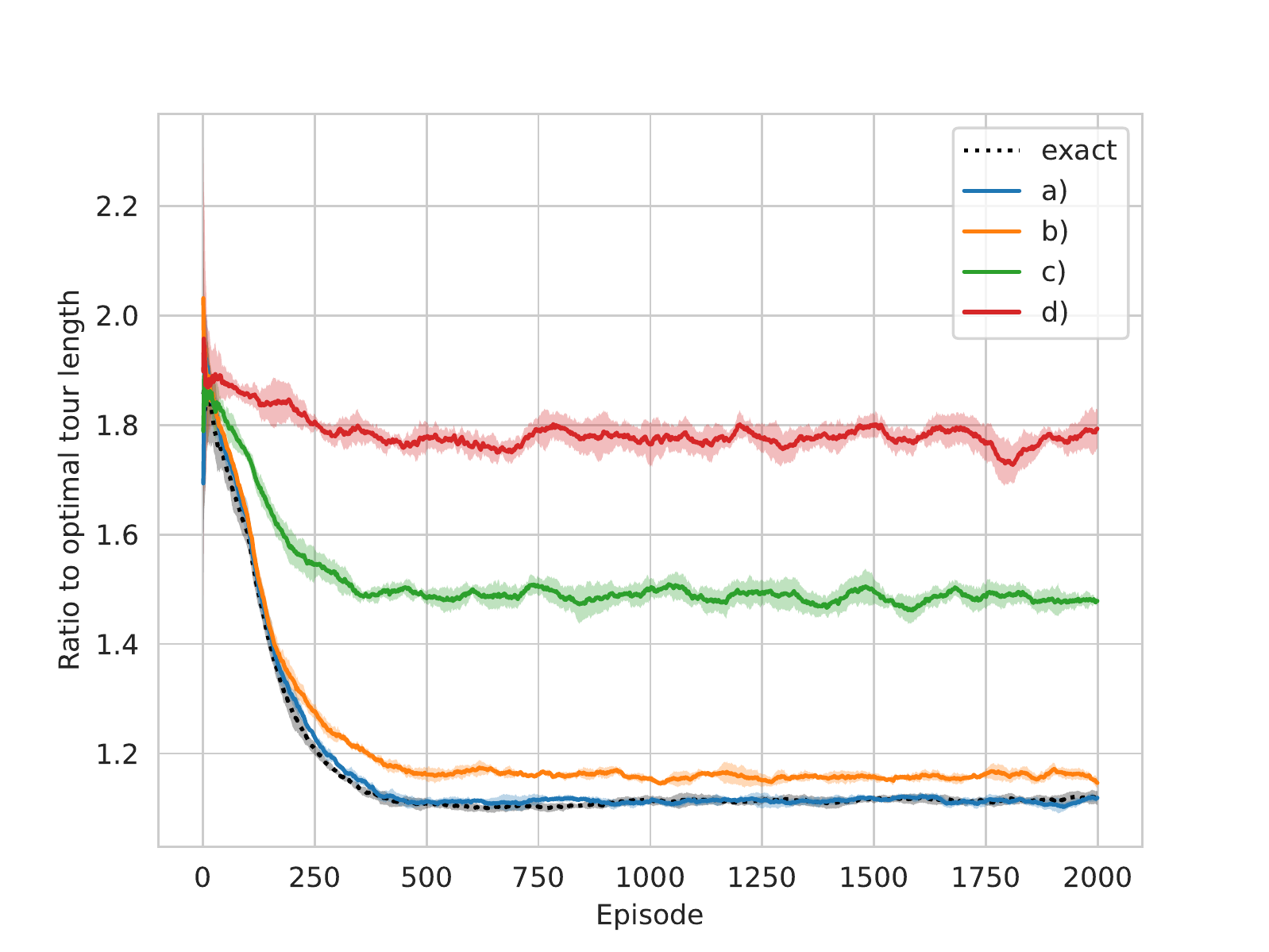}
\caption{Q-learning agents trained in the TSP environment with one layer of the circuit depicted in \Cref{fig:ansatzes} c) and custom noise model, using 1000 Monte Carlo trajectories. The labels indicate the custom noise configurations defined in~\Cref{tab:noise_models}, results are averaged over five agents in each curve, except for the exact curve which is averaged over ten agents as done in previous figures.}
\label{fig:tsp_custom_noise}
\end{figure}

\Cref{fig:tsp_custom_noise} shows results averaged over five Q-learning agents in the TSP environment for each of the error probability configurations of the custom noise model described above. We show the specific error probabilities used for the simulations in \Cref{tab:noise_models}. Configuration a) corresponds to error probabilities that are consistent with those present on current quantum hardware as described above. Based on this, we specify three other error probabilities b) - d) by increasing the error on varying error sources. We note that while the error probabilities themselves in configuration a) are consistent with those on current hardware, our simulation is only an approximation of this error due to the Monte Carlo trajectory sampling method described in \Cref{sec:res_incoherent_noise}. To perform simulations in a reasonable time frame, we use 1000 trajectories for each circuit evaluation. The circuit that we simulate has 145 gates (counting a ZZ-gate as two CNOTs and one Z gate), and for small error probabilities the chance of applying each of the noise channels is relatively small. This means that in each trajectory, a relatively small number of noise channels is applied. Hence we expect that the results in \Cref{fig:tsp_custom_noise} are slightly better than what we would get if the exact noise model was simulated (i.e., in the limit of a large number of trajectories, or by considering the full density matrix).

Looking at the results in~\Cref{fig:tsp_custom_noise}, we see that for configuration a) (blue), the performance of the agents matches those of the noise-free ones (dotted black) almost exactly, and the noise model based on realistic error strengths of current devices does not affect training. We see a slight drop in performance when we increase the error probability of the amplitude damping channels from 0.0003 to 0.03 (orange), as described in \Cref{tab:noise_models}, column b). For configuration c), we also increase the other remaining error sources' probabilities, which leads to a considerable drop in performance. In configuration d), we assume extremely high error probabilities for each of the noise channels, which leads to a complete failure of the agents to make any meaningful learning progress in this environment. 

\begin{table}
    \centering
    \begin{tabular}{lllll}
    \toprule
    Error source & \textbf{a)} & b) & c) & d) \\ \midrule
    Depolarization (1Q)  & \textbf{0.001} & 0.001 & 0.01 & 0.1 \\ 
    Depolarization (2Q)  & \textbf{0.01}  & 0.01 & 0.1 & 0.2 \\ 
    Amplitude damping    & \textbf{0.0003} & 0.03 & 0.03 & 0.1 \\ 
    Bitflip (measurement) & \textbf{0.01} & 0.01 & 0.1 & 0.1 \\
    \bottomrule
    \end{tabular}
    \caption{Error strengths for the configurations of the custom noise model used in~\Cref{fig:tsp_custom_noise}. Depolarization (1Q) indicates the single qubit depolarising channel applied after each single-qubit gate, and similarly for 2Q for two-qubit gates. Configuration a) in bold is based on error rates published by IBM at the time of writing, as described in the main text.}
    \label{tab:noise_models}
\end{table}

\section{Conclusions}

Our goal in this work was to evaluate the resilience of variational RL algorithms to various types of noise that are present on real quantum hardware. First, we investigated shot noise, which results from the probabilistic nature of quantum measurements. We introduced a method to reduce the number of shots to train a Q-learning agent, motivated by the specific structure of the underlying RL algorithm. Our shot allocation technique enables a more shot-frugal training of variational Q-learning models with little or no effect on the final performance of the agents. 

After considering shot noise, we moved on to study the effect of Gaussian coherent errors that can arise on real hardware due to miscalibration of the device, or imprecise pulse sequences that implement the parameterised gates in the quantum circuit. We gave an analytic expression for how this type of noise affects the output of a quantum RL agent, and provided a bound on the standard deviation of the Gaussian error that elucidates the tolerable magnitude of the error on the output of a quantum model. We confirm this bound in our simulations, where we study the effect of various levels of Gaussian perturbations on the performance of training policy gradient and Q-learning agents in two different environments. For one of these environments, we find that agents trained with higher noise probabilities also learn more robust policies and Q-functions, in the sense that under evaluation of different perturbation levels, these agents achieve optimal or close to optimal performance more often.

Finally, we studied incoherent noise that emerges in real hardware due to undesired interactions of the qubits with the surrounding environment, as the device is not completely shielded from external effects. To this end, we consider single-qubit depolarization errors, as well as a custom noise model that combines single- and two qubit depolarization errors, amplitude damping errors, and bitflip (measurement) errors. For the latter, we perform simulations with realistic error probabilities for each of the noise channels, in line with data published for IBM devices at the time of writing. 

Overall, we find that the effect of noise on training variational RL algorithms for Q-learning and the policy gradient method depends strongly on the strength of the noise, as well as the type of noise itself. For some cases, like decoherence errors with realistic error probabilities of current devices, the drop in performance is relatively small. On the other hand, we find that large Gaussian perturbations as well as errors induced by the probabilistic nature of quantum measurements can affect performance in highly detrimental ways. Additionally, we find that for Gaussian coherent noise agents that are trained with higher perturbations learn more noise-robust policies in some cases, similar to results in classical literature, where noise is used as a regularization technique.

While our results were performed in a regime that is still efficiently simulable on classical computers, it is an interesting question for future work to consider the implications of noise-robustness of large-scale quantum models in light of recent results which show that in certain settings, the outputs of noisy quantum circuits can be efficiently approximated classically \cite{stilck2021limitations,gao2018efficient}. This raises the question to what extent an inherent noise-robustness of hybrid variational quantum machine learning affects the possibility to achieve a quantum advantage with these types of models.

On the practical side, the optimization procedures that we used in this work were the same as those commonly used to train models in noise-free simulations and are not tailored to account for quantum hardware specific noise. This raises the question on how optimization methods that are tailored for the special characteristics of variational quantum models could further improve the performance of these types of models in a noisy setting. For the optimization of PQC parameters in the combinatorial optimization or quantum chemistry setting, it is known that some optmization methods, like simultaneous perturbation stochastic approximation (SPSA), actually become better with noise. It is an interesting area of future research to design quantum-specific optimization routines for machine learning that address or even combat specific types of noise, for example leveraging effective quantum error mitigation techniques~\cite{Mitiq, RussoQEM2022, WangQEM_QML_2021}. Our work motivates the study of these types of optimization methods, as well as continued efforts to find learning tasks where variational RL algorithms can potentially provide an advantage.

\acknowledgements
AS is funded by the German Ministry for Education and Research (BMB+F) in the project QAI2-Q-KIS under grant 13N15587. This work was also supported by the Dutch Research Council (NWO/OCW), as part of the Quantum Software Consortium programme (project number 024.003.037). CM acknowledges support by the National Research Centre for HPC, Big Data and Quantum Computing (ICSC).

\section*{Data and code availability}

The code that was used to generate the numerical results in this work can be found on GitHub (\url{https://github.com/askolik/noisy_qrl}), along with the data set containing the TSP instances studied in this work and their optimal solutions.

\section*{Author contributions}
AS conceived the idea for this work and conducted the numerical experiments. SM performed analytical study on the effect of Gaussian noise and provided decoherence noise model. AS and VD proposed shot allocation algorithm. AS and SM wrote the first version of the manuscript, all authors contributed to the final editing. 

\section*{Competing interests}

The authors declare no competing interests.

\bibliography{bibliography}

\providecommand{\noopsort}[1]{}\providecommand{\singleletter}[1]{#1}%
\begin{thebibliography}{10}

\bibitem{Bharti2021NIQSReview}
Kishor Bharti, Alba Cervera-Lierta, Thi~Ha Kyaw, Tobias Haug, Sumner
  Alperin-Lea, Abhinav Anand, Matthias Degroote, Hermanni Heimonen, Jakob~S.
  Kottmann, Tim Menke, Wai-Keong Mok, Sukin Sim, Leong-Chuan Kwek, and Al\'an
  Aspuru-Guzik.
\newblock Noisy intermediate-scale quantum algorithms.
\newblock {\em Rev. Mod. Phys.}, 94:015004, Feb 2022.

\bibitem{CerezoPQC2021Review}
M.~Cerezo, Andrew Arrasmith, Ryan Babbush, Simon~C. Benjamin, Suguru Endo,
  Keisuke Fujii, Jarrod~R. McClean, Kosuke Mitarai, Xiao Yuan, Lukasz Cincio,
  and et~al.
\newblock Variational quantum algorithms.
\newblock {\em Nature Reviews Physics}, 3(9):625–644, 2021.

\bibitem{ManginiQNN}
S.~Mangini, F.~Tacchino, D.~Gerace, D.~Bajoni, and C.~Macchiavello.
\newblock Quantum computing models for artificial neural networks.
\newblock {\em Europhysics Letters}, 134(1):10002, 2021.

\bibitem{GentiniNoiseVQA_2020}
Laura Gentini, Alessandro Cuccoli, Stefano Pirandola, Paola Verrucchi, and
  Leonardo Banchi.
\newblock Noise-resilient variational hybrid quantum-classical optimization.
\newblock {\em Phys. Rev. A}, 102:052414, Nov 2020.

\bibitem{jim1996analysis}
Kam-Chuen Jim, C~Lee Giles, and Bill~G Horne.
\newblock An analysis of noise in recurrent neural networks: convergence and
  generalization.
\newblock {\em IEEE Transactions on neural networks}, 7(6):1424--1438, 1996.

\bibitem{noh2017regularizing}
Hyeonwoo Noh, Tackgeun You, Jonghwan Mun, and Bohyung Han.
\newblock Regularizing deep neural networks by noise: Its interpretation and
  optimization.
\newblock {\em Advances in Neural Information Processing Systems}, 30, 2017.

\bibitem{graves2011practical}
Alex Graves.
\newblock Practical variational inference for neural networks.
\newblock {\em Advances in neural information processing systems}, 24, 2011.

\bibitem{graves2013speech}
Alex Graves, Abdel-rahman Mohamed, and Geoffrey Hinton.
\newblock Speech recognition with deep recurrent neural networks.
\newblock In {\em 2013 IEEE international conference on acoustics, speech and
  signal processing}, pages 6645--6649. Ieee, 2013.

\bibitem{balda2020adversarial}
Emilio~Rafael Balda, Arash Behboodi, and Rudolf Mathar.
\newblock Adversarial examples in deep neural networks: An overview.
\newblock {\em Deep Learning: Algorithms and Applications}, pages 31--65, 2020.

\bibitem{xie2017mitigating}
Cihang Xie, Jianyu Wang, Zhishuai Zhang, Zhou Ren, and Alan Yuille.
\newblock Mitigating adversarial effects through randomization.
\newblock {\em arXiv preprint arXiv:1711.01991}, 2017.

\bibitem{gilmer2019adversarial}
Justin Gilmer, Nicolas Ford, Nicholas Carlini, and Ekin Cubuk.
\newblock Adversarial examples are a natural consequence of test error in
  noise.
\newblock In {\em International Conference on Machine Learning}, pages
  2280--2289. PMLR, 2019.

\bibitem{jaeckle2021generating}
Florian Jaeckle and M~Pawan Kumar.
\newblock Generating adversarial examples with graph neural networks.
\newblock In {\em Uncertainty in Artificial Intelligence}, pages 1556--1564.
  PMLR, 2021.

\bibitem{goodfellow2014explaining}
Ian~J Goodfellow, Jonathon Shlens, and Christian Szegedy.
\newblock Explaining and harnessing adversarial examples.
\newblock {\em arXiv preprint arXiv:1412.6572}, 2014.

\bibitem{srivastava2014dropout}
Nitish Srivastava, Geoffrey Hinton, Alex Krizhevsky, Ilya Sutskever, and Ruslan
  Salakhutdinov.
\newblock Dropout: a simple way to prevent neural networks from overfitting.
\newblock {\em The journal of machine learning research}, 15(1):1929--1958,
  2014.

\bibitem{WangNoiseinducedBarrenPlateaus2021}
Samson Wang, Enrico Fontana, M.~Cerezo, Kunal Sharma, Akira Sone, Lukasz
  Cincio, and Patrick~J. Coles.
\newblock Noise-induced barren plateaus in variational quantum algorithms.
\newblock {\em Nature Communications}, 12(1):6961, November 2021.

\bibitem{zeng2021simulating}
Jinfeng Zeng, Zipeng Wu, Chenfeng Cao, Chao Zhang, Shi-Yao Hou, Pengxiang Xu,
  and Bei Zeng.
\newblock Simulating noisy variational quantum eigensolver with local noise
  models.
\newblock {\em Quantum Engineering}, 3(4):e77, 2021.

\bibitem{farhi2014quantum}
Edward Farhi, Jeffrey Goldstone, and Sam Gutmann.
\newblock A quantum approximate optimization algorithm.
\newblock {\em arXiv preprint arXiv:1411.4028}, 2014.

\bibitem{alam2019analysis}
Mahabubul Alam, Abdullah Ash-Saki, and Swaroop Ghosh.
\newblock Analysis of quantum approximate optimization algorithm under
  realistic noise in superconducting qubits.
\newblock {\em arXiv preprint arXiv:1907.09631}, 2019.

\bibitem{harrigan2021quantum}
Matthew~P Harrigan, Kevin~J Sung, Matthew Neeley, Kevin~J Satzinger, Frank
  Arute, Kunal Arya, Juan Atalaya, Joseph~C Bardin, Rami Barends, Sergio Boixo,
  et~al.
\newblock Quantum approximate optimization of non-planar graph problems on a
  planar superconducting processor.
\newblock {\em Nature Physics}, 17(3):332--336, 2021.

\bibitem{LaRose2020Robust}
Ryan LaRose and Brian Coyle.
\newblock Robust data encodings for quantum classifiers.
\newblock {\em Phys. Rev. A}, 102:032420, Sep 2020.

\bibitem{liu2022noise}
Junyu Liu, Frederik Wilde, Antonio~Anna Mele, Liang Jiang, and Jens Eisert.
\newblock Noise can be helpful for variational quantum algorithms.
\newblock {\em arXiv preprint arXiv:2210.06723}, 2022.

\bibitem{wang2020reinforcement}
Jingkang Wang, Yang Liu, and Bo~Li.
\newblock Reinforcement learning with perturbed rewards.
\newblock In {\em Proceedings of the AAAI conference on artificial
  intelligence}, volume~34, pages 6202--6209, 2020.

\bibitem{huang2017adversarial}
Sandy Huang, Nicolas Papernot, Ian Goodfellow, Yan Duan, and Pieter Abbeel.
\newblock Adversarial attacks on neural network policies.
\newblock {\em arXiv preprint arXiv:1702.02284}, 2017.

\bibitem{kos2017delving}
Jernej Kos and Dawn Song.
\newblock Delving into adversarial attacks on deep policies.
\newblock {\em arXiv preprint arXiv:1705.06452}, 2017.

\bibitem{yu2018towards}
Yang Yu.
\newblock Towards sample efficient reinforcement learning.
\newblock In {\em IJCAI}, pages 5739--5743, 2018.

\bibitem{chen2020variational}
Samuel Yen-Chi Chen, Chao-Han~Huck Yang, Jun Qi, Pin-Yu Chen, Xiaoli Ma, and
  Hsi-Sheng Goan.
\newblock Variational quantum circuits for deep reinforcement learning.
\newblock {\em IEEE Access}, 8:141007--141024, 2020.

\bibitem{lockwood2020reinforcement}
Owen Lockwood and Mei Si.
\newblock Reinforcement learning with quantum variational circuit.
\newblock {\em Proceedings of the AAAI Conference on Artificial Intelligence
  and Interactive Digital Entertainment}, 16(1):245--251, Oct. 2020.

\bibitem{jerbi2021parametrized}
Sofiene Jerbi, Casper Gyurik, Simon Marshall, Hans Briegel, and Vedran Dunjko.
\newblock Parametrized quantum policies for reinforcement learning.
\newblock {\em Advances in Neural Information Processing Systems},
  34:28362--28375, 2021.

\bibitem{skolik2022quantum}
Andrea Skolik, Sofiene Jerbi, and Vedran Dunjko.
\newblock Quantum agents in the gym: a variational quantum algorithm for deep
  q-learning.
\newblock {\em Quantum}, 6:720, 2022.

\bibitem{lan2021variational}
Qingfeng Lan.
\newblock Variational quantum soft actor-critic.
\newblock {\em arXiv preprint arXiv:2112.11921}, 2021.

\bibitem{wu2020quantum}
Shaojun Wu, Shan Jin, Dingding Wen, and Xiaoting Wang.
\newblock Quantum reinforcement learning in continuous action space.
\newblock {\em arXiv preprint arXiv:2012.10711}, 2020.

\bibitem{sequeira2022variational}
Andr{\'e} Sequeira, Luis~Paulo Santos, and Lu{\'\i}s~Soares Barbosa.
\newblock Variational quantum policy gradients with an application to quantum
  control.
\newblock {\em arXiv preprint arXiv:2203.10591}, 2022.

\bibitem{lockwood2021playing}
Owen Lockwood and Mei Si.
\newblock Playing atari with hybrid quantum-classical reinforcement learning.
\newblock In {\em NeurIPS 2020 Workshop on Pre-registration in Machine
  Learning}, pages 285--301. PMLR, 2021.

\bibitem{franz2022uncovering}
Maja Franz, Lucas Wolf, Maniraman Periyasamy, Christian Ufrecht, Daniel~D
  Scherer, Axel Plinge, Christopher Mutschler, and Wolfgang Mauerer.
\newblock Uncovering instabilities in variational-quantum deep q-networks.
\newblock {\em arXiv preprint arXiv:2202.05195}, 2022.

\bibitem{ito2021universal}
Kosuke Ito, Wataru Mizukami, and Keisuke Fujii.
\newblock Universal noise-precision relations in variational quantum
  algorithms.
\newblock {\em arXiv preprint arXiv:2106.03390}, 2021.

\bibitem{CerezoHigherOrder2021}
M~Cerezo and Patrick~J Coles.
\newblock Higher order derivatives of quantum neural networks with barren
  plateaus.
\newblock {\em Quantum Science and Technology}, 6(3):035006, jun 2021.

\bibitem{sutton2018reinforcement}
Richard~S Sutton and Andrew~G Barto.
\newblock {\em Reinforcement learning: An introduction}.
\newblock MIT press, 2018.

\bibitem{mnih2015human}
Volodymyr Mnih, Koray Kavukcuoglu, David Silver, Andrei~A Rusu, Joel Veness,
  Marc~G Bellemare, Alex Graves, Martin Riedmiller, Andreas~K Fidjeland, Georg
  Ostrovski, et~al.
\newblock Human-level control through deep reinforcement learning.
\newblock {\em nature}, 518(7540):529--533, 2015.

\bibitem{skolik2022equivariant}
Andrea Skolik, Michele Cattelan, Sheir Yarkoni, Thomas B{\"a}ck, and Vedran
  Dunjko.
\newblock Equivariant quantum circuits for learning on weighted graphs.
\newblock {\em arXiv preprint arXiv:2205.06109}, 2022.

\bibitem{noisy_qrl_code}
Andrea Skolik and Stefano Mangini.
\newblock Code that was used for training of noisy quantum agents.
\newblock \url{https://github.com/askolik/noisy_qrl}, 2022.

\bibitem{openaiGym}
Openai gym.
\newblock \url{https://github.com/openai/gym/wiki}.
\newblock Accessed: 06-09-2022.

\bibitem{lillicrap2015continuous}
Timothy~P Lillicrap, Jonathan~J Hunt, Alexander Pritzel, Nicolas Heess, Tom
  Erez, Yuval Tassa, David Silver, and Daan Wierstra.
\newblock Continuous control with deep reinforcement learning.
\newblock {\em arXiv preprint arXiv:1509.02971}, 2015.

\bibitem{kandala2017hardware}
Abhinav Kandala, Antonio Mezzacapo, Kristan Temme, Maika Takita, Markus Brink,
  Jerry~M Chow, and Jay~M Gambetta.
\newblock Hardware-efficient variational quantum eigensolver for small
  molecules and quantum magnets.
\newblock {\em Nature}, 549(7671):242--246, 2017.

\bibitem{Perez2020Reuploading}
Adrián Pérez-Salinas, Alba Cervera-Lierta, Elies Gil-Fuster, and José~I.
  Latorre.
\newblock Data re-uploading for a universal quantum classifier.
\newblock {\em Quantum}, 4:226, Feb 2020.

\bibitem{Schuld2020Encoding}
Maria Schuld, Ryan Sweke, and Johannes~Jakob Meyer.
\newblock Effect of data encoding on the expressive power of variational
  quantum-machine-learning models.
\newblock {\em Phys. Rev. A}, 103:032430, Mar 2021.

\bibitem{tfqRlTutorial}
Tensorflow quantum rl tutorial.
\newblock
  \url{https://www.tensorflow.org/quantum/tutorials/quantum_reinforcement_learning}.
\newblock Accessed: 06-09-2022.

\bibitem{bello2016neural}
Irwan Bello, Hieu Pham, Quoc~V Le, Mohammad Norouzi, and Samy Bengio.
\newblock Neural combinatorial optimization with reinforcement learning.
\newblock {\em arXiv preprint arXiv:1611.09940}, 2016.

\bibitem{slivkins2019introduction}
Aleksandrs Slivkins et~al.
\newblock Introduction to multi-armed bandits.
\newblock {\em Foundations and Trends{\textregistered} in Machine Learning},
  12(1-2):1--286, 2019.

\bibitem{lai1985asymptotically}
Tze~Leung Lai, Herbert Robbins, et~al.
\newblock Asymptotically efficient adaptive allocation rules.
\newblock {\em Advances in applied mathematics}, 6(1):4--22, 1985.

\bibitem{auer2002using}
Peter Auer.
\newblock Using confidence bounds for exploitation-exploration trade-offs.
\newblock {\em Journal of Machine Learning Research}, 3(Nov):397--422, 2002.

\bibitem{cai2020mitigating}
Zhenyu Cai, Xiaosi Xu, and Simon~C Benjamin.
\newblock Mitigating coherent noise using pauli conjugation.
\newblock {\em npj Quantum Information}, 6(1):1--9, 2020.

\bibitem{Schuld2019Gradients}
Maria Schuld, Ville Bergholm, Christian Gogolin, Josh Izaac, and Nathan
  Killoran.
\newblock Evaluating analytic gradients on quantum hardware.
\newblock {\em Phys. Rev. A}, 99:032331, Mar 2019.

\bibitem{Mitarai2018Learning}
K.~Mitarai, M.~Negoro, M.~Kitagawa, and K.~Fujii.
\newblock Quantum circuit learning.
\newblock {\em Phys. Rev. A}, 98:032309, Sep 2018.

\bibitem{McCleanBarren2018}
Jarrod~R. McClean, Sergio Boixo, Vadim~N. Smelyanskiy, Ryan Babbush, and
  Hartmut Neven.
\newblock Barren plateaus in quantum neural network training landscapes.
\newblock {\em Nat. Commun.}, 9(1):4812, 2018.

\bibitem{CerezoBarrenLocalCost2021}
M.~Cerezo, Akira Sone, Tyler Volkoff, Lukasz Cincio, and Patrick~J. Coles.
\newblock Cost function dependent barren plateaus in shallow parametrized
  quantum circuits.
\newblock {\em Nat. Commun.}, 12(1), 2021.

\bibitem{Holmes2021connecting}
Zo\"e Holmes, Kunal Sharma, M.~Cerezo, and Patrick~J. Coles.
\newblock Connecting ansatz expressibility to gradient magnitudes and barren
  plateaus.
\newblock {\em PRX Quantum}, 3:010313, 2022.

\bibitem{HuangPredicting2020}
Hsin-Yuan Huang, Richard Kueng, and John Preskill.
\newblock Predicting many properties of a quantum system from very few
  measurements.
\newblock {\em Nature Physics}, 16(10):1050--1057, October 2020.

\bibitem{HaarPuchala2017}
Z.~Puchała and J.A. Miszczak.
\newblock Symbolic integration with respect to the haar measure on the unitary
  groups.
\newblock {\em Bulletin of the Polish Academy of Sciences: Technical Sciences},
  65(No 1):21--27, 2017.

\bibitem{van2016deep}
Hado Van~Hasselt, Arthur Guez, and David Silver.
\newblock Deep reinforcement learning with double q-learning.
\newblock In {\em Proceedings of the AAAI conference on artificial
  intelligence}, volume~30, 2016.

\bibitem{Broughton2020TFQ}
Michael Broughton, Guillaume Verdon, Trevor McCourt, Antonio~J Martinez,
  Jae~Hyeon Yoo, Sergei~V Isakov, Philip Massey, Ramin Halavati, Murphy~Yuezhen
  Niu, Alexander Zlokapa, et~al.
\newblock Tensorflow quantum: A software framework for quantum machine
  learning.
\newblock {\em arXiv preprint arXiv:2003.02989}, 2020.

\bibitem{CirqDepolarizing}
Google Inc.
\newblock Documentation of depolarizing channel in cirq.
\newblock \url{https://quantumai.google/reference/python/cirq/depolarize},
  2022.

\bibitem{GoogleNoisySim}
Sergei~V. Isakov, Dvir Kafri, Orion Martin, Catherine~Vollgraff Heidweiller,
  Wojciech Mruczkiewicz, Matthew~P. Harrigan, Nicholas~C. Rubin, Ross Thomson,
  Michael Broughton, Kevin Kissell, Evan Peters, Erik Gustafson, Andy C.~Y. Li,
  Henry Lamm, Gabriel Perdue, Alan~K. Ho, Doug Strain, and Sergio Boixo.
\newblock Simulations of quantum circuits with approximate noise using qsim and
  cirq, 2021.

\bibitem{NielsenChuang}
Michael~A. Nielsen and Isaac~L. Chuang.
\newblock {\em Quantum computation and quantum information}.
\newblock Cambridge University Press, Cambridge, UK, 2010.

\bibitem{ProctorMirrorRB_2022}
Timothy Proctor, Stefan Seritan, Kenneth Rudinger, Erik Nielsen, Robin
  Blume-Kohout, and Kevin Young.
\newblock Scalable randomized benchmarking of quantum computers using mirror
  circuits.
\newblock {\em Phys. Rev. Lett.}, 129:150502, Oct 2022.

\bibitem{Vovrosh2021GlobalDepolarizing}
Joseph Vovrosh, Kiran~E. Khosla, Sean Greenaway, Christopher Self, M.~S. Kim,
  and Johannes Knolle.
\newblock Simple mitigation of global depolarizing errors in quantum
  simulations.
\newblock {\em Phys. Rev. E}, 104:035309, Sep 2021.

\bibitem{MagnesanRB_2012}
Easwar Magesan, Jay~M. Gambetta, and Joseph Emerson.
\newblock Characterizing quantum gates via randomized benchmarking.
\newblock {\em Phys. Rev. A}, 85:042311, Apr 2012.

\bibitem{McKay3QubRB_2019}
David~C. McKay, Sarah Sheldon, John~A. Smolin, Jerry~M. Chow, and Jay~M.
  Gambetta.
\newblock Three-qubit randomized benchmarking.
\newblock {\em Phys. Rev. Lett.}, 122:200502, May 2019.

\bibitem{AndersonQuantinuumQEC_2022}
C.~Ryan-Anderson, N.~C. Brown, M.~S. Allman, B.~Arkin, G.~Asa-Attuah,
  C.~Baldwin, J.~Berg, J.~G. Bohnet, S.~Braxton, N.~Burdick, J.~P. Campora,
  A.~Chernoguzov, J.~Esposito, B.~Evans, D.~Francois, J.~P. Gaebler, T.~M.
  Gatterman, J.~Gerber, K.~Gilmore, D.~Gresh, A.~Hall, A.~Hankin, J.~Hostetter,
  D.~Lucchetti, K.~Mayer, J.~Myers, B.~Neyenhuis, J.~Santiago, J.~Sedlacek,
  T.~Skripka, A.~Slattery, R.~P. Stutz, J.~Tait, R.~Tobey, G.~Vittorini,
  J.~Walker, and D.~Hayes.
\newblock Implementing fault-tolerant entangling gates on the five-qubit code
  and the color code, 2022.

\bibitem{IBMQuantum}
Ibmquantum.
\newblock \url{https://quantum-computing.ibm.com/}, 2022.

\bibitem{pelofske2022quantum}
Elijah Pelofske, Andreas B{\"a}rtschi, and Stephan Eidenbenz.
\newblock Quantum volume in practice: What users can expect from nisq devices.
\newblock {\em arXiv preprint arXiv:2203.03816}, 2022.

\bibitem{ibmq_experience}
{IBM Quantum Experience}.
\newblock {IBM Quantum Experience}.
\newblock
  \url{https://quantum-computing.ibm.com/services/resources?tab=systems}, 2022.

\bibitem{stilck2021limitations}
Daniel Stilck~Fran{\c{c}}a and Raul Garcia-Patron.
\newblock Limitations of optimization algorithms on noisy quantum devices.
\newblock {\em Nature Physics}, 17(11):1221--1227, 2021.

\bibitem{gao2018efficient}
Xun Gao and Luming Duan.
\newblock Efficient classical simulation of noisy quantum computation.
\newblock {\em arXiv preprint arXiv:1810.03176}, 2018.

\bibitem{Mitiq}
Ryan LaRose, Andrea Mari, Sarah Kaiser, Peter~J. Karalekas, Andre~A. Alves,
  Piotr Czarnik, Mohamed El~Mandouh, Max~H. Gordon, Yousef Hindy, Aaron
  Robertson, Purva Thakre, Misty Wahl, Danny Samuel, Rahul Mistri, Maxime
  Tremblay, Nick Gardner, Nathaniel~T. Stemen, Nathan Shammah, and William~J.
  Zeng.
\newblock Mitiq: {A} software package for error mitigation on noisy quantum
  computers.
\newblock {\em {Quantum}}, 6:774, August 2022.

\bibitem{RussoQEM2022}
Vincent Russo, Andrea Mari, Nathan Shammah, Ryan LaRose, and William~J. Zeng.
\newblock Testing platform-independent quantum error mitigation on noisy
  quantum computers, 2022.

\bibitem{WangQEM_QML_2021}
Samson Wang, Piotr Czarnik, Andrew Arrasmith, M.~Cerezo, Lukasz Cincio, and
  Patrick~J. Coles.
\newblock Can error mitigation improve trainability of noisy variational
  quantum algorithms?, 2021.

\bibitem{HuembeliHessian2021}
Patrick Huembeli and Alexandre Dauphin.
\newblock Characterizing the loss landscape of variational quantum circuits.
\newblock {\em Quantum Science and Technology}, 6(2):025011, feb 2021.

\bibitem{FukudaRTNI2019}
Motohisa Fukuda, Robert König, and Ion Nechita.
\newblock {RTNI}{\textemdash}a symbolic integrator for haar-random tensor
  networks.
\newblock {\em Journal of Physics A: Mathematical and Theoretical},
  52(42):425303, sep 2019.

\bibitem{KeenerStatisticsBook}
Robert W.~Keener (auth.).
\newblock {\em Theoretical Statistics: Topics for a Core Course}.
\newblock Springer Texts in Statistics. Springer, 1 edition, 2010.

\bibitem{Pennylane}
Ville Bergholm, Josh Izaac, Maria Schuld, Christian Gogolin, Shahnawaz Ahmed,
  Vishnu Ajith, M.~Sohaib Alam, Guillermo Alonso-Linaje, B.~AkashNarayanan, Ali
  Asadi, Juan~Miguel Arrazola, Utkarsh Azad, Sam Banning, Carsten Blank,
  Thomas~R Bromley, Benjamin~A. Cordier, Jack Ceroni, Alain Delgado, Olivia
  Di~Matteo, Amintor Dusko, Tanya Garg, Diego Guala, Anthony Hayes, Ryan Hill,
  Aroosa Ijaz, Theodor Isacsson, David Ittah, Soran Jahangiri, Prateek Jain,
  Edward Jiang, Ankit Khandelwal, Korbinian Kottmann, Robert~A. Lang, Christina
  Lee, Thomas Loke, Angus Lowe, Keri McKiernan, Johannes~Jakob Meyer, J.~A.
  Montañez-Barrera, Romain Moyard, Zeyue Niu, Lee~James O'Riordan, Steven Oud,
  Ashish Panigrahi, Chae-Yeun Park, Daniel Polatajko, Nicolás Quesada, Chase
  Roberts, Nahum Sá, Isidor Schoch, Borun Shi, Shuli Shu, Sukin Sim, Arshpreet
  Singh, Ingrid Strandberg, Jay Soni, Antal Száva, Slimane Thabet, Rodrigo~A.
  Vargas-Hernández, Trevor Vincent, Nicola Vitucci, Maurice Weber, David
  Wierichs, Roeland Wiersema, Moritz Willmann, Vincent Wong, Shaoming Zhang,
  and Nathan Killoran.
\newblock Pennylane: Automatic differentiation of hybrid quantum-classical
  computations, 2018.

\end{thebibliography}
\bibliographystyle{unsrt}

\onecolumngrid

\appendix

\section{Additional results for flexible vs. fixed number of shots in Q-learning}

\begin{figure}[h]
\includegraphics[scale=0.5]{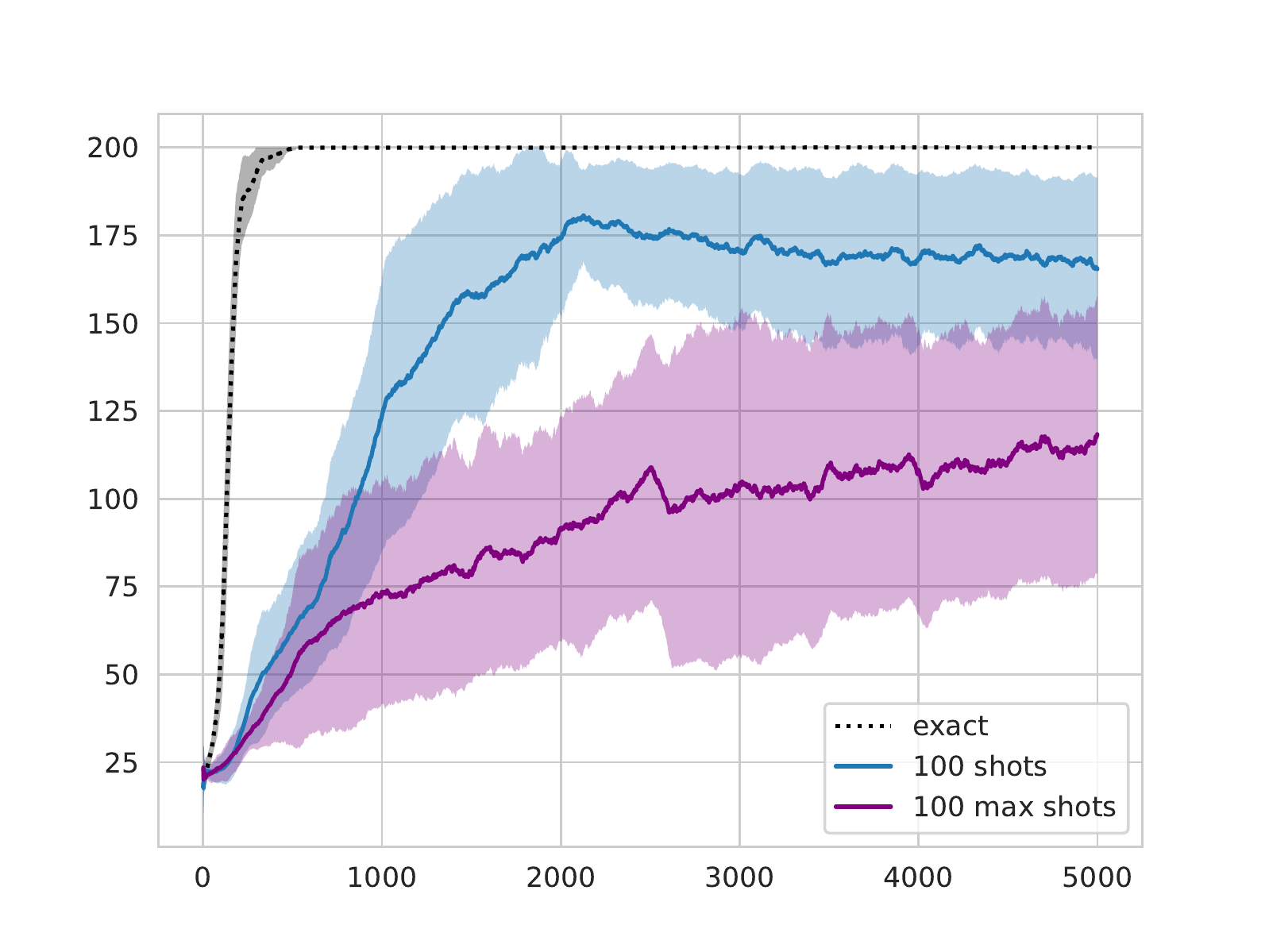}
\caption{Performance of agents trained with a fixed number of 100 shots (blue) and $m_{\mathrm{max}} = 100$ with flexible shot allocation (purple), compared to model trained without shot noise (black dotted curve).}
\label{fig:cp_ucb_worst}
\end{figure}

\section{Gaussian Noise Analysis \label{app:higher-order}}

In this Appendix we perform the noise analysis of a scalar function whose parameters are corrupted by independently distributed Gaussian perturbations. Let $f: \mathbb{R}^{M} \rightarrow \mathbb{R}$ be the function under investigation, whose parameters $\bm{\theta} = (\theta_1, \hdots, \theta_{M})\in \mathbb{R}^M$ are corrupted by a Gaussian noise $\theta_i \rightarrow \theta_i + \delta\theta_i$ with zero mean and variance $\sigma^2$, i.e.
\begin{equation}
\label{eq:gauss_prop}
    \begin{aligned}
        &\delta\theta_i \sim \mathcal{N}(0, \sigma^2) \quad \forall i=1,\hdots, M\,,\\
        &\mathbb{E}[\delta\theta_i] = 0\,,\\
        &\mathbb{E}[\delta\theta_i\delta\theta_j] = \sigma^2\delta_{ij}\,.     
    \end{aligned}
\end{equation}

Since the perturbations are independently distributed and Gaussian, all higher order moments can be evaluated starting from two points correlators of the form $\mathbb{E}[\delta\theta_{i}\delta\theta_{j}]$, as dictated by \textit{Wick's formulas} for multivariate normal distributions
\begin{equation}
\label{eq:wick_integrals}
    \begin{aligned}
        & \mathbb{E}[\delta\theta_{i_1}\cdots \delta\theta_{i_{2n+1}}] = 0\,,\\ 
        & \mathbb{E}[\delta\theta_{i_1}\cdots \delta\theta_{i_{2n}}] = \sum_{\cal{P}}\mathbb{E}[\delta\theta_{k_{1}}\delta\theta_{k_{2}}]\cdots \mathbb{E}[\delta\theta_{k_{2n-1}}\delta\theta_{k_{2n}}]\, ,
    \end{aligned}
\end{equation}
where with $\cal{P}$ we denote all the possible distinct $(2n-1)!!$ pairings of the $n$ variables, as these can be used to express all higher order even moments in terms of products of second moments. Note that all the terms involving an odd number of perturbations $\delta\theta_i$ vanish, and only even moments of remain. For example, expression~\eqref{eq:wick_integrals} for the fourth-order moment ($n=4$) amounts to
\begin{equation}
\label{eq:fourthorder}
    \begin{aligned}
     \mathbb{E}[\delta\theta_{i}\delta\theta_{j}\delta\theta_{k}\delta\theta_{m}] & = \mathbb{E}[\delta\theta_{i}\delta\theta_{j}]\,\mathbb{E}[\delta\theta_{k}\delta\theta_{m}] +\, \mathbb{E}[\delta\theta_{i}\delta\theta_{k}]\,\mathbb{E}[\delta\theta_{j}\delta\theta_{m}] +\, \mathbb{E}[\delta\theta_{i}\delta\theta_{m}]\,\mathbb{E}[\delta\theta_{j}\delta\theta_{k}] \\
     & = \sigma^4\big(\delta_{ij}\delta_{km}+\delta_{ik}\delta_{jm}+\delta_{im}\delta_{jk}\big)\, .
    \end{aligned}
\end{equation}

We now proceed considering the multi dimensional Taylor expansion of the function $f(\bm{\theta}+\delta\bm{\theta})$ around the noise-free point. Up to arbitrary order, this reads 
\begin{equation}
\label{eq:fullTaylor_expansion}
    f(\bm{\theta}+\delta\bm{\theta}) = f(\bm{\theta}) + \sum_{i=1}^{M}\frac{\partial f(\bm{\theta})}{\partial\theta_i}\delta\theta_i + \frac{1}{2!}\sum_{i,j=1}^{M}\frac{\partial^2 f(\bm{\theta})}{\partial\theta_i\partial\theta_j}\delta\theta_i\delta\theta_j + \frac{1}{3!}\sum_{i,j,k=1}^{M}\frac{\partial^3 f(\bm{\theta})}{\partial\theta_i\partial\theta_j\partial\theta_k}\delta\theta_i\delta\theta_j\delta\theta_k + \hdots\, .
\end{equation}
where we used the equal sign because we are considering the full Taylor series, and we assume that this converges to the true function (this statement can be made precise by showing that the reminder term of the expansion goes to zero as the order of expansion goes to infinity).

Before proceeding, we simplify the notation to make the calculation of the Taylor expansion easier to follow. First, we denote the partial derivatives with respect to parameter $\theta_i$ as $\partial_i := \partial /\partial\theta_i$, and similarly for higher order derivatives, for example $\partial_{ij} = \partial^2 /\partial{\theta_i}\partial{\theta_j}$. Also, we suppress the explicit dependence of the function on $\bm{\theta}$, using the short-hand $f$ instead of $f(\bm{\theta})$. At last, we make use of Einstein' summation notation where repeated indexes imply summation. 

With this setup, using Eqs.~\eqref{eq:gauss_prop},~\eqref{eq:wick_integrals} and~\eqref{eq:fourthorder} in~\eqref{eq:fullTaylor_expansion}, one can evaluate the expectation value of the function over the perturbations' distributions as
\begin{equation}
    \begin{aligned}
    \mathbb{E}[f(\bm{\theta}+\delta\bm{\theta})] &= f(\bm{\theta}) + \partial_i f\, \mathbb{E}[\delta\theta_i] + \frac{1}{2}\partial_{ij}f\,\mathbb{E}[\delta\theta_i\delta\theta_j] + \frac{1}{3!}\partial_{ijk} f\,\mathbb{E}[\delta\theta_i\delta\theta_j\delta\theta_k] + \frac{1}{4!}\partial_{ijkm} f\,\mathbb{E}[\delta\theta_i\delta\theta_j\delta\theta_k\delta\theta_m] +\, \hdots \\
    & = f(\bm{\theta}) + \frac{\sigma^2}{2}\partial_{ij}f\,\delta_{ij} + \frac{\sigma^4}{4!}\partial_{ijkm} f\,\qty( \delta_{ij}\delta_{km}+\delta_{ik}\delta_{jm}+\delta_{im}\delta_{jk}) +\,\hdots \\
    & = f(\bm{\theta}) + \frac{\sigma^2}{2}\sum_{i}\frac{\partial^2 f}{\partial\theta_i^2} +\, \frac{\sigma^4}{4!}3\sum_{ij}\frac{\partial^4 f}{\partial\theta_i^2\partial\theta_j^2}\, +\,\hdots
    \end{aligned}
\end{equation}
where in the last line we simplified the fourth order term as
\begin{eqnarray*}
     \mathbb{E}[f^{(4)}] &=&\frac{\sigma^4}{4!}\partial_{ijkm}f\,  \big(
     \delta_{ij}\delta_{km}+\delta_{ik}\delta_{jm}+\delta_{im}\delta_{jk}\big) \\ 
     &=& \frac{\sigma^4}{4!}\bigg(
     \sum_{ik}\frac{\partial^4 f}{\partial\theta_i^2\partial\theta_k^2} + \sum_{ij}\frac{\partial^4 f}{\partial\theta_i^2\partial\theta_j^2} + \sum_{im}\frac{\partial^4 f}{\partial\theta_i^2\partial\theta_m^2}
     \bigg) \\
     &=& \frac{\sigma^4}{4!}\, 3\, \sum_{ij}\frac{\partial^4 f}{\partial\theta_i^2\partial\theta_j^2}\, .
\end{eqnarray*}

Since the expectation values involving an odd number of perturbations vanish, only the even order terms survive, and these can be expressed as
\begin{equation}
    \mathbb{E}[f^{(2n)}]= \frac{\sigma^{2n}}{(2n)!}(2n-1)!!\sum_{i_1,\hdots, i_{n}}\frac{\partial^{2n}{f(\bm{\theta})}}{\partial\theta_{i_{1}}^2 \hdots\, \partial\theta_{i_{n}}^2}\, .
\end{equation}
where the coefficient $(2n-1)!!$ is the number of distinct pairings of $2n$ objects, which comes from Eq.~\eqref{eq:gauss_prop}.

Thus, the full Taylor series can be formally written as
\begin{eqnarray}
    \mathbb{E}[f(\bm{\theta}+\delta\bm{\theta})] &=& f(\bm{\theta}) + \sum_{n=1}^{\infty} \frac{\sigma^{2n}}{(2n)!}(2n-1)!!\sum_{i_1,\hdots, i_{n}=1}^{M}\frac{\partial^{2n}{f(\bm{\theta})}}{\partial\theta_{i_{1}}^2 \hdots\, \partial\theta_{i_{n}}^2} \label{eq:full_expansion1}\\
    &=& f(\bm{\theta}) + \frac{\sigma^2}{2}\Tr[H(\bm{\theta})] + \sum_{n=2}^{\infty} \frac{\sigma^{2n}}{(2n)!}(2n-1)!!\sum_{i_1,\hdots, i_{n}=1}^{M}\frac{\partial^{2n}{f(\bm{\theta})}}{\partial\theta_{i_{1}}^2 \hdots\, \partial\theta_{i_{n}}^2}\label{eq:full_expansion2}\, ,
\end{eqnarray}
where we introduced the Hessian matrix $H(\bm{\theta})$, whose elements are given by $[H(\bm{\theta})]_{ij} = \partial_{ij}f(\bm{\theta})$, and we see that this term represent the first non-vanishing correction to the function caused by the perturbation. 

Our goal is to bound the absolute error
\begin{eqnarray}
\label{eq:gauss_errorApp}
    \varepsilon_{\bm{\theta}} &:=& \abs{\mathbb{E}[f(\bm{\theta}+\delta\bm{\theta})] - f(\bm{\theta})} =  \abs{\sum_{n=1}^{\infty} \frac{\sigma^{2n}}{(2n)!}(2n-1)!!\sum_{i_1,\hdots, i_{n}=1}^{M}\frac{\partial^{2n}{f(\bm{\theta})}}{\partial\theta_{i_{1}}^2 \hdots\, \partial\theta_{i_{n}}^2}}
\end{eqnarray}
caused by the gaussian noise, and we can do that by using the property that all the derivatives of most PQC (Parametrized Quantum Circuit) are bounded. In fact, for those circuits for which a \textit{parameter-shift} rule holds~\cite{Schuld2019Gradients, Mitarai2018Learning}, one can show that any derivative of the function $f(\bm{\theta}) = \expval{O} = \Tr[O\, U(\bm{\theta})\dyad{0} U^\dagger(\bm{\theta})]$ obeys 
\begin{equation}
    \abs{\frac{\partial^{\alpha_1+\hdots\alpha_M}\, f(\bm{\theta})}{\partial\theta_1^{\alpha_1}\hdots\partial\theta_M^{\alpha_M}}} \leq \norm{O}_{\infty}\, ,
\end{equation}
where $\norm{O}_\infty$ is the infinity norm of the observable, namely its largest absolute eigenvalue. We give a proof of this below in Sec.~\ref{ssec:param_shift}. 

Plugging this in Eq.~\eqref{eq:gauss_errorApp}, we can obtain an upper bound to the error $\varepsilon_{\bm{\theta}}$ as desired. Indeed, remembering that for even numbers the double factorial can be expressed as $(2n-1)!! = (2n)!/(2^n n!)$, it holds
\begin{eqnarray}
\label{eq:pert_error}
    \varepsilon_{\bm{\theta}} &=& \abs{\sum_{n=1}^{\infty} \frac{\sigma^{2n}}{(2n)!}(2n-1)!!\sum_{i_1,\hdots, i_{n}=1}^M \frac{\partial^{2n}{f(\bm{\theta})}}{\partial\theta_{i_{1}}^2 \hdots\, \partial\theta_{i_{n}}^2}} \leq \sum_{n=1}^{\infty} \frac{\sigma^{2n}}{(2n)!}(2n-1)!!\sum_{i_1,\hdots, i_{n}=1}^M \underbrace{\abs{\frac{\partial^{2n}{f(\bm{\theta})}}{\partial\theta_{i_{1}}^2 \hdots\, \partial\theta_{i_{n}}^2}}}_{\leq \|\obs\|_\infty} \nonumber\\
    &\leq& \sum_{n=1}^{\infty} \frac{\sigma^{2n}}{(2n)!}(2n-1)!!\, \|\obs\|_\infty\, M^n =  \|\obs\|_\infty \sum_{n=1}^{\infty} \frac{1}{(2n)!}\frac{(2n)!}{2^n n!}(\sigma^2 M)^n = \|\obs\|_\infty \sum_{n=1}^\infty \frac{(M\sigma^2 / 2)^n}{n!} \nonumber\\
    &=& \|\obs\|_\infty \qty(e^{\sigma^2 M/2} - 1) \nonumber\\
    &&  \implies \varepsilon_{\bm{\theta}} = \abs{\mathbb{E}[f(\bm{\theta} + \delta\bm{\theta})] - f(\bm{\theta})} \leq \|\obs\|_\infty \qty(e^{M\sigma^2/2} - 1)\, ,
\end{eqnarray}
where in the last line we used the definition of the exponential function $e^{x} = \sum_{n=0}^{\infty} \frac{x^n}{n!}$. 

One can see that the noise variance $\sigma^2$ must scale as the inverse of the number of parameters $\sigma^2 \in \order{M^{-1}}$ in order to have small deviations induced by the noise. Also, note that since the difference between the noise-free function $f(\bm{\theta})$ and its perturbed version $f(\bm{\theta}+\delta\bm{\theta})$ cannot be larger than twice the maximum eigenvalue of $O$, $\abs{f(\bm{\theta}+\delta\bm{\theta})-f(\bm{\theta})} \leq \abs{f(\bm{\theta}+\delta\bm{\theta})}+\abs{f(\bm{\theta})} = 2 \norm{O}_{\infty}$, the bound~\eqref{eq:pert_error} is informative only as long as $\exp[M\sigma^2/2]-1 < 2$.

It is worth noticing that an identical procedure can be used to bound the average error obtained by approximating the perturbed function with its first non-vanishing correction given by the Hessian. Indeed, starting from Eq.~\eqref{eq:full_expansion2} are repeating the same calculation from above, one obtains
\begin{equation}
    \abs{\mathbb{E}[f(\bm{\theta} + \delta\bm{\theta})]- f(\bm{\theta}) - \frac{\sigma^2}{2}\Tr[H(\bm{\theta})]} \leq \|\obs\|_\infty \qty(e^{M\sigma^2/2} - 1 - \frac{M\sigma^2}{2})\, .
\end{equation}

\subsection{\label{ssec:param_shift}Parameter-Shift rule and bounds to the derivatives} Let $f(\bm{\theta}) = \Tr[O\, U(\bm{\theta})\dyad{0} U^\dagger(\bm{\theta})]$ be the expectation value of an observable $O$ on the parametrized state $\ket{\psi(\bm{\theta})} = U(\bm{\theta})\ket{0}$ obtained with a parametrized quantum circuit $U(\bm{\theta})$. When the variational parameters $\bm{\theta} \in \mathbb{R}^M$ enter in the quantum circuit via rotation gates of the form $V(\theta_i) = \exp[-i \theta_i P / 2]$ with $P^2 = \mathds{1}$ being Pauli operators, then the \textit{parameter-shift} rule can be used to evaluate gradients of the expectation value as~\cite{Schuld2019Gradients, Mitarai2018Learning}
\begin{equation}
    \frac{\partial f(\bm{\theta})}{\partial \theta_i} = \frac{1}{2}\qty(f\qty(\bm{\theta} + \frac{\pi}{2}\bm{e_i}) - f\qty(\bm{\theta} - \frac{\pi}{2}\bm{e_i}))\, ,
\end{equation}
where $\bm{e}_i$ is the unit vector with zero entries and a one in the $i-$th position corresponding to angle $\theta_i$. Similarly, by applying the parameter-shift rule twice one can express second order derivatives as follows using four evaluations of the circuit~\cite{ito2021universal, HuembeliHessian2021}
\begin{eqnarray}
    \frac{\partial^2 f(\bm{\theta})}{\partial\theta_i \partial\theta_j} &=& \frac{1}{2}\qty[\frac{\partial}{\partial \theta_i}f\qty(\bm{\theta} + \frac{\pi}{2}\bm{e_j}) - \frac{\partial}{\partial \theta_i}f\qty(\bm{\theta} - \frac{\pi}{2}\bm{e_j})] \\
    &=& \frac{1}{4}\qty[f\qty(\bm{\theta} + \frac{\pi}{2}\bm{e_j} + \frac{\pi}{2}\bm{e_i}) - f\qty(\bm{\theta} + \frac{\pi}{2}\bm{e_j} - \frac{\pi}{2}\bm{e_i}) - f\qty(\bm{\theta} - \frac{\pi}{2}\bm{e_j} + \frac{\pi}{2}\bm{e_i}) + f\qty(\bm{\theta} - \frac{\pi}{2}\bm{e_j} - \frac{\pi}{2}\bm{e_i})]\, .
\end{eqnarray}

In particular, for the diagonal elements $i=j$, one has
\begin{eqnarray}
\label{eq:diag_hessian}
    \frac{\partial^2 f(\bm{\theta})}{\partial\theta_i^2} &=& \frac{1}{4}\qty[f\qty(\bm{\theta} + \pi\bm{e_i}) - 2f\qty(\bm{\theta}) + f\qty(\bm{\theta} - \pi\bm{e_i})]\nonumber\\
    &=& \frac{1}{2}\qty[f\qty(\bm{\theta} + \pi\bm{e_i}) - f(\bm{\theta})]\, ,
\end{eqnarray}
where we used the fact that $f\qty(\bm{\theta} + \pi\bm{e_i}) = f\qty(\bm{\theta} -\pi\bm{e_i})$. This last equality can be seen intuitively from the $2\pi$ periodicity of the rotation gates or by direct evaluation. In fact, let $U(\bm{\theta}) = U_2\,\exp[-i \theta_i P_i/2]\, U_{1}$ be a factorization of the parametrized unitary where we isolated the dependence on the parameter $\theta_i$ to be shifted. Then, since $\exp[-i\, 2\pi P /2] = \cos{\pi}\,\mathbb{I} - i\sin{\pi}\,P = -\mathbb{I}$, one has
\begin{equation}
    \begin{aligned}
        \ket{\psi(\bm{\theta}-\pi\bm{e}_i)} &= U_2\,\exp[-i (\theta_i - \pi) P_i/2]\, U_{1}\ket{0} \\
        & =  U_2\,\exp[-i (\theta_i -\pi) P_i/2] \underbrace{- \exp[-i\, 2\pi\, P_i / 2]}_{\mathbb{I}} U_{1}\ket{0} \\
        & =  -U_2\,\exp[-i (\theta_i -\pi + 2\pi) P_i/2] U_{1}\ket{0} \\
        & =  -\ket{\psi(\bm{\theta}+\pi\bm{e}_i)} \, ,
    \end{aligned}
\end{equation}
and thus $\mel{\psi(\bm{\theta}-\pi\bm{e}_i)}{O}{\psi(\bm{\theta}-\pi\bm{e}_i)} = \mel{\psi(\bm{\theta}+\pi\bm{e}_i)}{O}{\psi(\bm{\theta}+\pi\bm{e}_i)}$. 

Hence, using Eq.~\eqref{eq:diag_hessian} it is possible to estimate the diagonal elements of the Hessian matrix with just two different evaluations of the quantum circuit. 

By repeated application of the parameter-shift rule one can also evaluate arbitrary higher-order derivatives as linear combination of circuit evaluations~\cite{ito2021universal, CerezoHigherOrder2021}. Let $\bm{\alpha} = (\alpha_1, \hdots, \alpha_M) \in \mathbb{N}^{M}$ be a multi-index keeping track of the orders of derivatives, and let $\abs{\bm{\alpha}} = \sum_{i=1}^M \alpha_i$. Then
\begin{equation}
\label{eq:higher_order}
    \partial^{\bm{\alpha}}f(\bm{\theta}) := \frac{\partial^{\abs{\bm{\alpha}}}\, f(\bm{\theta})}{\partial\theta_1^{\alpha_1}\hdots\partial\theta_M^{\alpha_M}} = \frac{1}{2^\abs{\bm{\alpha}}} \sum_{m=1}^{2^{\abs{\bm{\alpha}}}} s_m f(\tilde{\bm{\theta}}_m)\,,
\end{equation}
where $s_m \in \{\pm 1\}$ are signs, and $\tilde{\bm{\theta}}_m$ are angles obtained by accumulation of shifts along multiple directions. 

Since the output of any circuit evaluation is bounded by the infinity norm (i.e the largest absolute eigenvalue) of the observable $\|O\|_\infty = \max\{\abs{o_i}\,, \, \obs = \sum_{i} o_i \dyad{o_i}\}$
\begin{equation}
    \abs{f(\bm{\theta})} = \abs{\Tr[\obs\, \rho(\bm{\theta})]} \leq \|\obs\|_{\infty} \|\rho(\bm{\theta})\|_1 = \|\obs\|_{\infty} \quad \forall \bm{\theta} \in \mathbb{R}^{M}\, ,
\end{equation}
then one can bound the sum in Eq.~\eqref{eq:higher_order} simply as
\begin{eqnarray}
\label{eq:deriv_bound}
     \abs{\partial^{\bm{\alpha}}f(\bm{\theta})} \leq \frac{1}{2^{\abs{\bm{\alpha}}}} \sum_{m=1}^{2^{\abs{\bm{\alpha}}}} \abs{f(\bm{\tilde{\theta}}_m)} \leq \|\obs\|_\infty\, .
\end{eqnarray}

\subsection{Average value of the Hessian of random PQCs}
\label{app:gaussian_noise_resilience}

In this section we derive the formulas~\eqref{eq:haar_Hii} and~\eqref{eq:haar_TrH} for the expected value of the Hessian as shown in the main text. Consider a system of $n$ qubits and a parametrized quantum circuit with unitary $U(\bm{\theta}) \in \mathcal{U}(2^n)$, where $\mathcal{U}(2^n)$ is the group of unitary matrices of dimension $2^n$. Given a set of parameter vectors $\{\bm{\theta}_1, \bm{\theta}_2, \hdots, \bm{\theta}_K\}$, one can construct the corresponding set of unitaries $\mathbb{U} = \{U_1, U_2, \hdots, U_K\}$, with $U_i = U(\bm{\theta}_i)$ and clearly $\mathbb{U} \in \mathcal{U}(2^n)$.

It is now well known that sampling a parametrized quantum circuit from a random assignment of the parameters is approximately equal to drawing a random unitary from the Haar distribution, a phenomenon which is at the root of the insurgence of Barren Plateaus (BPs)~\cite{Holmes2021connecting, McCleanBarren2018, CerezoBarrenLocalCost2021}. Specifically, it is numerically observed that parametrized quantum circuits behave like unitary 2-designs, that is averaging over unitaries $U_i$ sampled from $\mathbb{U}$ yields the same result of averaging over Haar-random unitaries, up until second order moments. 

As standard in the literature regarding BPs, in the following we assume that the considered parametrized unitaries (and parts of them) are indeed 2-designs, and so we make use of the following relations for integration over random unitaries~\cite{HaarPuchala2017, FukudaRTNI2019, HuangPredicting2020, Holmes2021connecting, CerezoBarrenLocalCost2021}
\begin{gather}
\mathbb{E}_{U}[U A U^\dagger] = \int d\mu(U)\, U A U^\dagger = \frac{\mathds{1}\Tr[A]}{2^n}\, \label{eq:1haar}\\
\mathbb{E}_{U}[A U B U^\dagger C U D U^\dagger] = \frac{\Tr[BD]\Tr[C]A + \Tr[B]\Tr[D]AC}{2^{2n}-1} - \frac{\Tr[BD]AC + \Tr[B]\Tr[C]\Tr[D]A}{2^{n}(2^{2n}-1)}\label{eq:2harr}
\end{gather}

\subsubsection{Statistics of the Hessian}
Let $f(\bm{\theta}) = \Tr[O U(\bm{\theta})\dyad{0}U(\bm{\theta})^\dagger]$ and assume that the observable $O$ is such that $\Tr[O] = 0$ and $\Tr[O^2] = 2^n$, as is the case of measuring a Pauli string. As shown in Eq.~\eqref{eq:diag_hessian}, diagonal elements of the Hessian matrix $H$ can be calculated as
\begin{equation}
\label{eq:hess_diag}
    H_{ii} = \frac{\partial^2f(\bm{\theta})}{\partial\theta_i^2} =  \frac{1}{2}[f(\bm{\theta}+\pi\bm{e}_i)-f(\bm{\theta})]\, .
\end{equation}

For simplicity, from now on drop the explicit dependence on the parameter vector $\bm{\theta}$ when not explicitly needed. The variational parameters enter the quantum circuit via Pauli rotations $e^{-i\theta_i P_i/2}$ with $P_i = P_i^\dagger$ and $P_i^2 = \mathds{1}$, and so the shifted unitary $U(\bm{\theta}+\pi\bm{e}_i)$ can be rewritten as
\begin{equation}
\label{eq:URLdef}
    U(\bm{\theta}+\pi\bm{e}_i)  = U_L e^{-i\pi P_i/2} U_R = -i\, U_L P_i U_R\,,
\end{equation}
where $U_L$ and $U_R$ form a bipartition of the circuit at the position of the shifted angle, so that $U(\bm{\theta}) = U_LU_R$.

Assuming that the set of unitaries $\mathbb{U}_L$ generated by $U_L$ is at least a 1-design, one has that
\begin{align}
    \mathbb{E}_{U_L}[f(\bm{\theta}+\pi\bm{e}_i)] &= \mathbb{E}_{U_L}\qty[\Tr[\obs\, U_L P_i U_R\dyad{0}U_R^\dagger P_i U_L^\dagger]] \\
    &= \Tr\qty[\obs\, \mathbb{E}_{U_L}\qty[U_L P_i U_R\dyad{0}U_R^\dagger P_i U_L^\dagger]]\\
    &= \Tr\qty[\obs\, \frac{\Tr[P_i U_R\dyad{0}U_R^\dagger P_i]\mathds{1}}{2^n}] = \frac{\Tr[O]}{2^n} = 0\, ,
\end{align}
where in the first line we exchanged the trace and the expectation value since both are linear operations, and in the second line we made use of Eq.~\eqref{eq:1haar} for the first moment of the Haar distribution. Similarly, one can show that if $\mathbb{U}_R$ forms a 1-design, then averaging over it yields the same result, namely $\mathbb{E}_{U_R}[f(\bm{\theta}+\pi\bm{e}_i)] = 0$. The same calculation for $f(\bm{\theta})$ shows that $\mathbb{E}_{U_R}[f(\bm{\theta})]=\mathbb{E}_{U_L}[f(\bm{\theta})] = 0$. 

Thus, for every diagonal element of the Hessian, if either $\mathbb{U}_L$ or $\mathbb{U_R}$ is a 1-design (that is Eq.~\eqref{eq:1haar} hold), then its expectation value vanishes
\begin{equation}
\label{eq:Hiiexp}
    \mathbb{E}_{U_R,U_L}[H_{ii}] = 0 \quad \forall i\quad \text{if either $\mathbb{U}_L$ or $\mathbb{U_R}$ is a 1-design}.
\end{equation}

The variance of the diagonal elements can be calculated in a similar manner, even though the calculation is more involved. Substituting Eq.~\eqref{eq:hess_diag} in the defition of the variance, one obtains
\begin{align}
\label{eq:Hiivar}
    \Var[H_{ii}] &:= \mathbb{E}[H_{ii}^2] - \mathbb{E}[H_{ii}]^2 = \mathbb{E}[H_{ii}^2]\nonumber\\
    &=\frac{1}{4}\qty[\E[f(\bm{\theta}+\pi\bm{e}_i)^2] + \E[f(\bm{\theta})^2] - 2\E[f(\bm{\theta}+\pi\bm{e}_i)f(\bm{\theta})]]\,.
\end{align}

In order to use Eq.~\eqref{eq:2harr} for second moment integrals, we can rewrite these expectation values as follow
\begin{align}
  \E[f(\bm{\theta}+\pi\bm{e}_i)^2] &= \E\qty[\Tr[\obs\, U_L P_i U_R\dyad{0}U_R^\dagger P_i U_L^\dagger]^2]\nonumber\\
  &=\E\qty[\Tr[\obs\, U_L P_i U_R\dyad{0}U_R^\dagger P_i U_L^\dagger]\mel{0}{U_R^\dagger P_i U_L^\dagger\obs U_L P_i U_R}{0}]\nonumber\\
  &= \E\qty[\Tr\qty[\obs\, U_L P_i U_R\dyad{0}U_R^\dagger P_i U_L^\dagger\obs U_L P_i U_R\dyad{0}U_R^\dagger P_i U_L^\dagger]]\nonumber\\
  &= \Tr\qty[\E[\obs\, U_L P_i U_R\dyad{0}U_R^\dagger P_i U_L^\dagger\obs U_L P_i U_R\dyad{0}U_R^\dagger P_i U_L^\dagger]]\,,
\end{align}
and similarly for the remaining two terms. Assuming that the set of unitaries $\mathbb{U}_L$ generated by $U_L$ is a 2-design, then
\begin{align}
    \E_{U_L}[f(\bm{\theta}+\pi\bm{e}_i)^2] &= \Tr\qty[\E_{U_L}[\obs\, U_L  \underbrace{P_i U_R\dyad{0}U_R^\dagger P_i}_{B} U_L^\dagger\obs U_L \underbrace{P_i U_R\dyad{0}U_R^\dagger P_i}_{B} U_L^\dagger]]\\
    &= \Tr\qty[\frac{\Tr[B^2]\Tr[O]O + \Tr[B]^2 O^2}{2^{2n}-1} - \frac{\Tr[B^2]O^2 + \Tr[B]^2\Tr[O]O}{2^{n}(2^{2n}-1)}]\\
    &= \frac{\Tr[O]^2 + \Tr[O^2]}{2^{2n}-1} - \frac{\Tr[O^2] + \Tr[O]^2}{2^{n}(2^{2n}-1)} = \frac{1}{2^n+1}\,,
\end{align}
where in the second line we made use of Eq.~\eqref{eq:2harr}, and the third line the used that $\Tr[B]=\Tr[B^2]=1$ since $B = P_i U_R\dyad{0}U_R^\dagger P_i$ is a projector, and that $\Tr[O]=0$ and $\Tr[O^2]=2^n$. Similarly, one can show that integration over $\mathbb{U}_R$ yields the same result. Also, the same calculation leads to $\E_{U_L}[f(\bm{\theta})^2] = \E_{U_R}[f(\bm{\theta})^2] = 1/(2^{n}+1)$. Thus, if either $\mathbb{U}_L$ or $\mathbb{U_R}$ is a 2-design then
\begin{equation}
\label{eq:f2_haar}
    \mathbb{E}_{U_R,U_L}[f(\bm{\theta})^2] = \mathbb{E}_{U_R,U_L}[f(\bm{\theta}+\pi\bm{e}_i)^2] =  \frac{1}{2^n+1} \quad \forall i\quad \text{if either $\mathbb{U}_L$ or $\mathbb{U_R}$ is a 2-design}.
\end{equation}

Now we evaluate the correlation term $\E[f(\bm{\theta}+\pi\bm{e}_i)f(\bm{\theta})]$. If $\mathbb{U}_L$ is a 2-design, then
\begin{align}
\label{eq:Ul2des}
\E_{U_L}[f(\bm{\theta}+\pi\bm{e}_i)f(\bm{\theta})] &= \Tr\qty[\E_{U_L}\qty[\obs\, U_L P_i U_R\dyad{0}U_R^\dagger U_L^\dagger\obs U_L U_R\dyad{0}U_R^\dagger P_i U_L^\dagger]]\nonumber\\
 &= \Tr\qty[\frac{\Tr[P_i U_R\dyad{0}U_R^\dagger]^2 O^2}{2^{2n}-1} - \frac{O^2}{2^{n}(2^{2n}-1)}]\nonumber \\
  &= \frac{1}{2^{2n}-1}\qty[2^n\,\Tr[P_i U_R\dyad{0}U_R^\dagger]^2 - 1]\,.
\end{align}    
While if $\mathbb{U}_R$ is a 2-design instead it holds
\begin{align}
\label{eq:Ur2des}
\E_{U_R}[f(\bm{\theta}+\pi\bm{e}_i)f(\bm{\theta})] &= \Tr\qty[\obs\, U_L P_i \,\E_{U_R}\qty[U_R\dyad{0}U_R^\dagger U_L^\dagger\obs U_L U_R\dyad{0}U_R^\dagger] P_i U_L^\dagger]\nonumber\\
 &= \Tr\qty[\obs\, U_L P_i\,\frac{(2^n-1)U_L^\dagger O U_L}{2^n(2^{2n}-1)}\, P_i U_L^\dagger]\nonumber\\
  &= \frac{1}{2^n(2^{n}+1)}\Tr[\obs\, U_L P_i U_L^\dagger O U_L P_i U_L^\dagger]\,.
\end{align}  

If both of them are 2-designs, then continuing from Eq.~\eqref{eq:Ur2des}, one obtains
\begin{align}
\label{eq:Url2des}
\E_{U_L,U_R}[f(\bm{\theta}+\pi\bm{e}_i)f(\bm{\theta})] &= \frac{1}{2^n(2^{n}+1)}\Tr\qty[\E_{U_L}\qty[\obs\, U_L P_i U_L^\dagger O U_L P_i U_L^\dagger]]\nonumber\\
&= \frac{1}{2^n(2^{n}+1)}\Tr\qty[\frac{\Tr[P_i]^2 O^2+\Tr[P_i^2]\Tr[O]O}{2^{2n}-1}-\frac{\Tr[P_i^2]O^2+\Tr[P_i]^2\Tr[O]O}{2^n(2^{2n}-1)}]\nonumber\\
&= -\frac{1}{2^n(2^{n}+1)}\frac{\Tr[P_i^2]\Tr[O^2]}{2^n(2^{2n}-1)} = -\frac{1}{(2^{n}+1)(2^{2n}-1)} \in \order{2^{-3n}}
\end{align}  

Finally, plugging Eqs.~\eqref{eq:Ul2des},~\eqref{eq:Ur2des} and~\eqref{eq:Url2des} in Eq.~\eqref{eq:Hiivar}, one has $\forall i=1,\hdots, M$ 
\begin{align}
    \Var_{U_L,U_R}[H_{ii}] &= \frac{1}{2}\E[f(\bm{\theta})^2]-\frac{1}{2}\E[f(\bm{\theta}+\pi\bm{e}_i)f(\bm{\theta})]\nonumber \\
    &= \frac{1}{2(2^n+1)} - \frac{1}{2} 
    \begin{cases}
    \label{eq:Hiivartot}
        \displaystyle{\frac{1}{2^{2n}-1}\qty[2^n\,\Tr[P_i U_R\dyad{0}U_R^\dagger]^2 - 1]} \quad &\forall i\,, \text{if $\mathbb{U}_L$ 2-design}\\[10pt]
        \displaystyle{\frac{1}{2^n(2^{n}+1)}\Tr[\obs\, U_L P_i U_L^\dagger O U_L P_i U_L^\dagger]} \quad &\forall i\,, \text{if $\mathbb{U}_R$ 2-design}\\[10pt]
        \displaystyle{-\frac{1}{(2^{n}+1)(2^{2n}-1)}} \quad &\forall i\,, \text{if $\mathbb{U}_L, \mathbb{U}_R$ 2-designs}\\
    \end{cases} 
\end{align}
where $\mathbb{U}_R = \mathbb{U}_R^{(i)}$ and $\mathbb{U}_L = \mathbb{U}_L^{(i)}$ are defined as in Eq.~\eqref{eq:URLdef} and actually depend on the index $i$ of the parameter. 

Not surprisingly, as it happens for first order derivatives, also second order derivatives of PQCs are found to be exponentially vanishing~\cite{CerezoHigherOrder2021, Holmes2021connecting}, as from Eq.~\eqref{eq:Hiivartot} one can check that $\Var[H_{ii}] \in \order{2^{-n}}$.

\subsubsection{Statistics of the trace of the Hessian}
\label{app:TrHSec}
The average value of the trace of the Hessian is easily found to be zero using Eq.~\eqref{eq:Hiiexp}, in fact
\begin{equation}
    \E_{U_R, U_L}[\Tr[H]] = \sum_{i=1}^M \E_{U^{(i)}_R, U^{(i)}_L}[H_{ii}] = 0\,,
\end{equation}
where we assume that for every parameter $i$ either $\mathbb{U}_R^{(i)}$ or $\mathbb{U}_L^{(i)}$ is a 1-design. The variance of the trace is instead
\begin{align}
\label{eq:VarTrh1}
\Var_{U_R, U_L}[\Tr[H]] &= \Var\qty[\sum_{i=1}^M H_{ii}] =  \sum_{i=1}^M \Var[H_{ii}] + 2\sum_{i<j}^M \text{Cov}[H_{ii}H_{jj}]\,.
\end{align}
We can upper bound this quantity using the covariance inequality~\cite{KeenerStatisticsBook} $$\abs{\text{Cov}[H_{ii}, H_{jj}]} \leq \sqrt{\Var[H_{ii}]\Var[H_{jj}]} \approx \Var[H_{ii}]\,,$$ were we assumed that $\Var[H_{ii}] \approx \Var[H_{jj}]\,\forall i,j$. Using that $\Var[H_{ii}] \in \order{2^{-n}}$ one finally has
\begin{align}
\label{eq:varbound}
\Var_{U_R, U_L}[\Tr[H]] \leq \sum_{i=1}^M \Var[H_{ii}] + 2\sum_{i<j}^M \Var[H_{ii}] \in \order{\frac{M^2}{2^n}}\,.
\end{align}

Alternatively, one can obtain a tighter yet qualitative approximation by explicitly considering the nature of the sums in Eq.~\eqref{eq:VarTrh1}. First, by using Eq.~\eqref{eq:hess_diag}, the covariance term is explicitly
\begin{equation}
    \text{Cov}[H_{ii},H_{jj}] = \E[H_{ii}H_{jj}] = \frac{1}{4}\E[(f_i-f)(f_j-f)] = \frac{1}{4}\E[f^2] + \frac{1}{4}\E[f_if_j] - \frac{1}{4}\E[f_if] - \frac{1}{4}\E[f_jf]\,,
\end{equation}
where for ease of notation we defined $f_{i,j} = f(\bm{\theta}+\pi\bm{e}_{i,j})$ and $f=f(\bm{\theta})$. Note that except for the first term which is always positive, all remaining correlations terms can be both positive and negative. Also, all of these terms are bounded from above by the same quantity, as via Cauchy-Schwarz it follows
\begin{equation}
\label{eq:covine}
    \abs{\E[f_if_j]} \leq \sqrt{\E[f_i^2]\E[f_j^2]} = \frac{1}{2^n+1} \quad\text{and}\quad \abs{\E[f_if]} \leq \sqrt{\E[f_i^2]\E[f^2]} = \frac{1}{2^n+1}\,,
\end{equation}
where we have used $E[f^2]=E[f_i^2]=1/(2^n+1)$ from Eq.~\eqref{eq:f2_haar}. Then, the variance can be written as
\begin{align}
\label{eq:trhapprox1}
\Var_{U_R, U_L}[\Tr[H]] &= \sum_{i=1}^M \Var[H_{ii}] + 2\sum_{i<j}^M \E[H_{ii}H_{jj}]\nonumber \\
&= \sum_{i=1}^M \frac{\E[f^2]- \E[f_if]}{2} + 2\sum_{i<j}^M\frac{\E[f^2]+\E[f_if_j] - \E[f_if] - \E[f_jf]}{4} \nonumber\\
&= \frac{1}{2}\qty(\sum_{i=1}^M+\sum_{i<j}^M)\E[f^2] - \frac{1}{2}\qty(\sum_{i=1}^M\E[f_if] + \sum_{i<j}^M\E[f_if]+\sum_{i<j}^M\E[f_jf])+\frac{1}{2}\sum_{i<j}^M\E[f_if_j]\nonumber\\
&= \frac{M(M+1)}{4}\E[f^2] - \underbrace{\frac{M}{2}\sum_{i=1}^M\E[f_if]+\frac{1}{2}\sum_{i<j}^M\E[f_if_j]}_{\Delta}\,.
\end{align}

\begin{figure}
  \centering
\begin{tikzpicture}
    \node (a) {\includegraphics[width=0.45\textwidth]{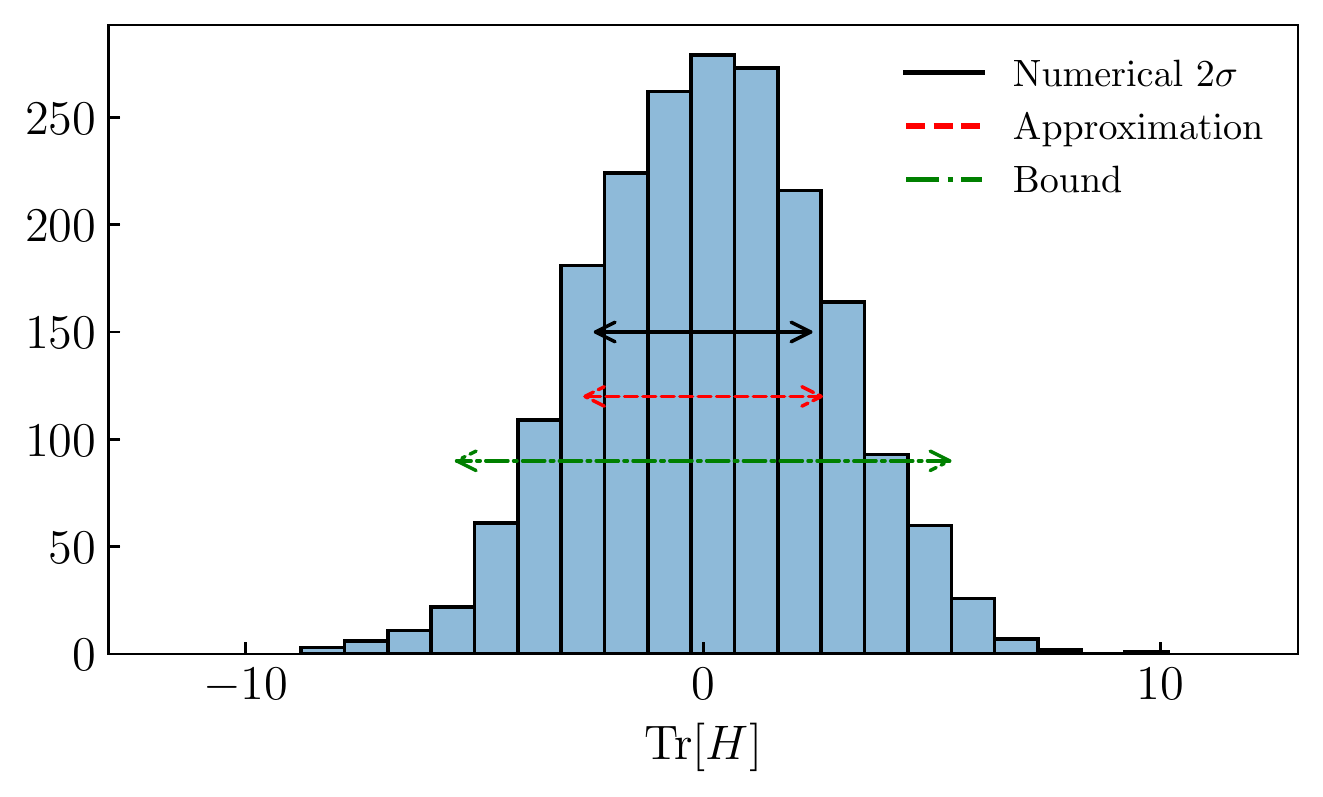}};
    \node[yshift=1cm, xshift=-2.25cm] (a.center)  {\includegraphics[width=0.1\textwidth]{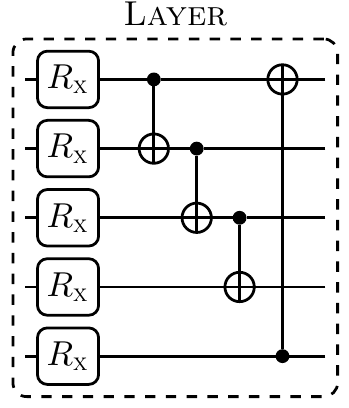}};
    \node[right=of a] (b) {\includegraphics[width=0.45\textwidth]{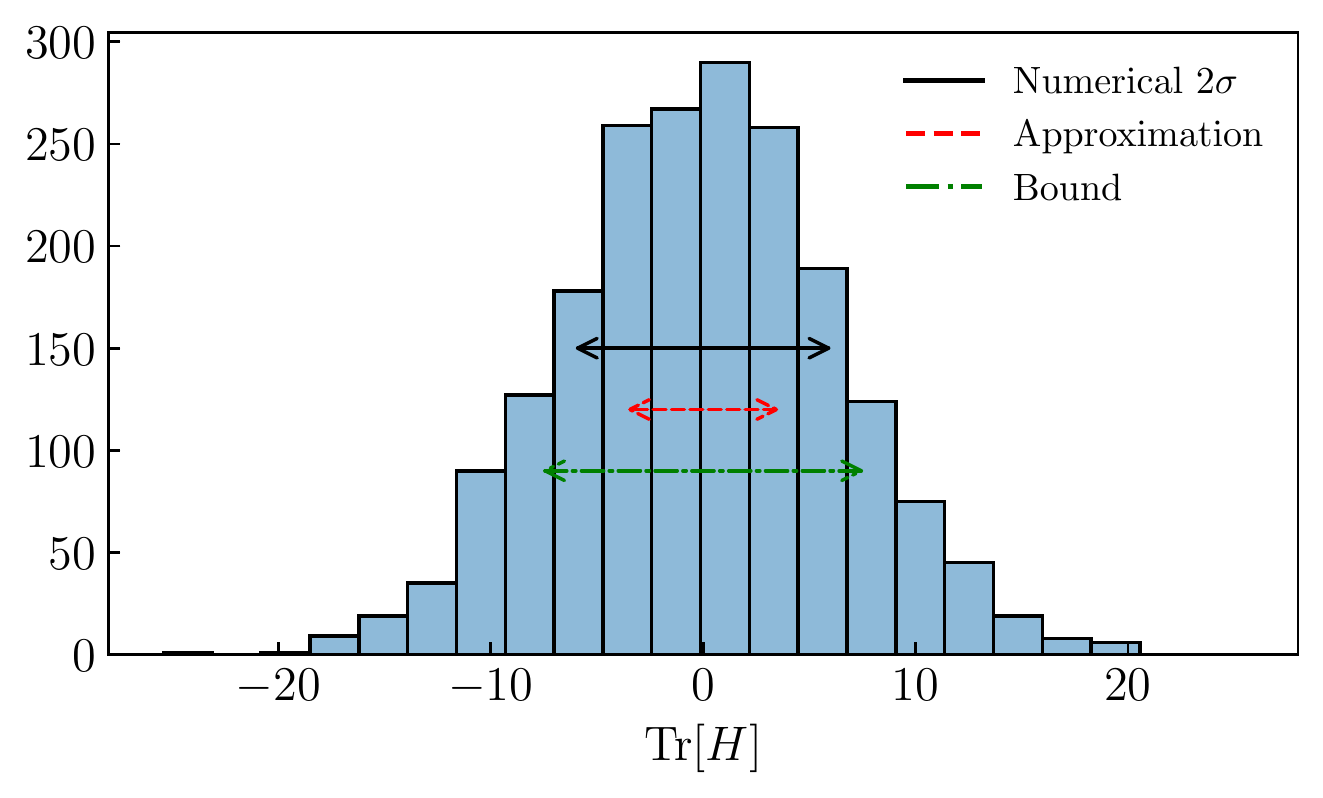}};
    \node[yshift=1cm, xshift=-2.cm] at (b.center) {\includegraphics[width=0.13\textwidth]{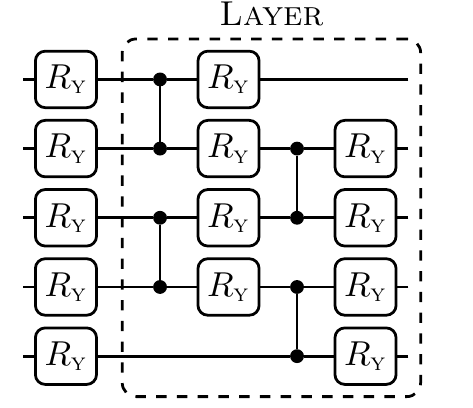}};
\end{tikzpicture}
    \caption{Simulation results of evaluating the trace of the Hessian matrix for two different hardware-efficient ansatzes with random values of the parameters. The plot on the left is obtained using the layer template shown in the figure for $n=6$ qubits and $l = 6$ layers. The plot on the right instead with $n=5$ and $l=5$ layers of the template shown in the corresponding inset. The simulations are performed by sampling $2000$ random parameter vectors $\bm{\theta}_m$ with $\theta_i \sim \text{Unif}[0,2\pi[$, evaluating the trace of the Hessian matrix $\Tr[H(\bm{\theta})]$, and then building the histogram to show its frequency distribution. In both experiments the measured observable is $Z^{\otimes n}$. The length of the arrows are respectively: ``Numerical $2\sigma$" (black solid line) twice the  numerical standard deviation, ``Approximation" (dashed red) twice the square root of the approximation in Eq.~\eqref{eq:Trhapprox}, ``Bound" (dashed-dotted green) twice the square root of the upper Bound in Eq.~\eqref{eq:varbound}.
    These parametrized circuits correspond to the templates \texttt{BasicEntanglinLayer} and \texttt{Simplified2Design} defined in Pennylane~\cite{Pennylane}, and used for example in~\cite{CerezoBarrenLocalCost2021} to study barren plateaus.}
    \label{fig:HessExp}
\end{figure}

\paragraph*{Numerical simulations---} In addition to Figure~\ref{fig:hessians_distribution} in the main text, in Figure~\ref{fig:HessExp} we report numerical evidence for the trace of the Hessian for two common hardware-efficient parametrized quantum circuit ansatzes. The histograms represent the frequency of obtaining a given value of the trace of the Hessian $\Tr[H(\bm{\theta})]$ upon random assignments of the parameters. The length of the arrows are, respectively: ``Numerical $2\sigma$" (black solid line) twice the  statistical standard deviation computed from the numerical results, ``Approximation" (dashed red) twice the square root of the Eq.~\eqref{eq:trhapprox1} with $\Delta=0$, ``Bound" (dashed-dotted green) twice the square root of the upper Bound in Eq.~\eqref{eq:varbound}.

All simulations confirm the bound~\eqref{eq:varbound}, and, more interestingly, both the circuit on the left of Fig.~\ref{fig:HessExp} and the one in Fig.~\ref{fig:hessians_distribution} in the main text, have a numerical variance which is very well approximated by Eq.~\eqref{eq:trhapprox1} with $\Delta = 0$. We conjecture this is due to the fact that all correlation terms in Eq.~\eqref{eq:trhapprox1} are roughly of the same order of magnitude (see Eq.~\eqref{eq:covine}), and can be either positive and negative, depending on the parameter and the specifics of the ansatz. Thus, one can expect the whole contribution to either vanish $\Delta \approx 0$, or be negligible with respect to the leading term. If this is the case, then substituting $\E[f^2] = 1/(2^n+1)$, the variance of the Hessian is approximately
\begin{equation}
\label{eq:Trhapprox}
    \Var_{U_R, U_L}[\Tr[H]] \approx \frac{M(M+1)}{4}\E[f^2] = \frac{M(M+1)}{4(2^n+1)} \approx \frac{1}{4}\frac{M^2}{2^n}\,,
\end{equation}
which is four times smaller then the upper bound Eq.~\eqref{eq:varbound}, but clearly has the same scaling. While we numerically verified it also at other number of qubits, more investigations are needed to understand if and when this approximation holds, and we leave a detailed study of this phenomenon for future work.

\section{Visualization of CartPole policies obtained with Q-learning}
\label{apdx:policy_visualizations}

\begin{figure}[h]
 \subfloat[$\sigma=0$]{\includegraphics[scale=0.5]{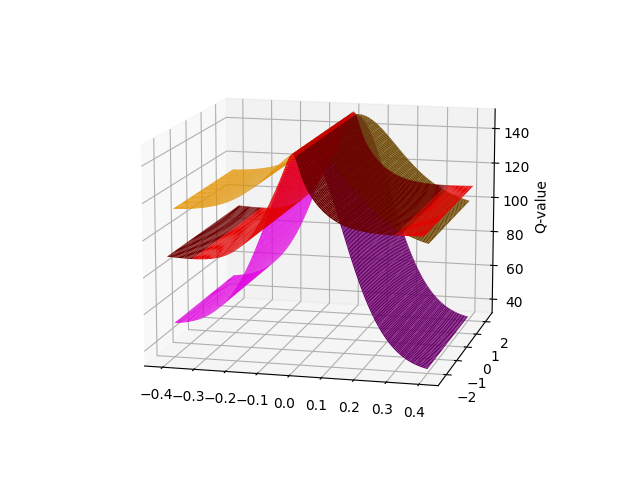}}\quad
 \subfloat[$\sigma=0.2$]{\includegraphics[scale=0.5]{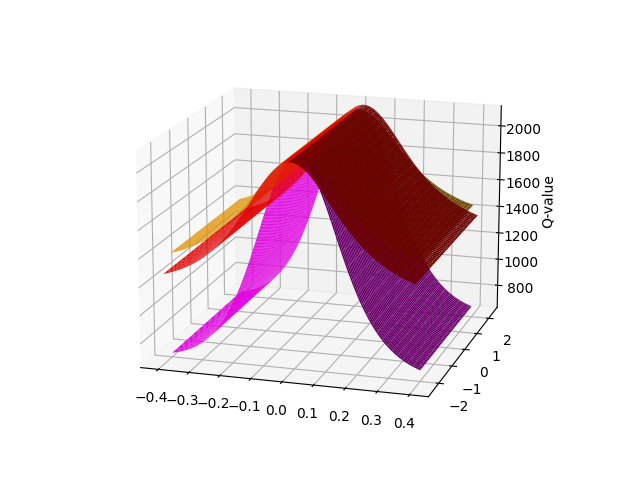}}\quad
  \caption{Visualization of the Q-functions learned in the noise-free (a) and noisy (b) settings. The red surface shows Q-values for pole angle and cart position, orange for pole angle and cart velocity, and magenta for pole angle and pole velocity.}
  \label{fig:vis_q_noisy_free}
\end{figure}

\end{document}